\documentclass{article}
\usepackage[utf8]{inputenc}
\usepackage{comment}

\pdfoutput=1
\usepackage{amssymb}
\usepackage{amsmath}
\usepackage{setspace}
\usepackage{slashed}
\usepackage{mathtools}
\usepackage{amsfonts}
\usepackage{tikz}
\usepackage{xcolor}
\usepackage{physics}
\usepackage{comment}
\usepackage{subcaption}
\usepackage[percent]{overpic}
\usepackage{cite}
\usepackage{physics}
\usepackage{moresize}
\usepackage{dsfont}
\usepackage{dsdshorthand}
\usepackage{bbm}
\usepackage{booktabs}

\definecolor{darkgreen}{rgb}{0,0.5,0}
\definecolor{darkblue}{rgb}{0,0,0.6}
\definecolor{purple}{rgb}{0.4,.2,0.7}

\usepackage[colorlinks=true,citecolor=darkgreen,linkcolor=black,urlcolor=purple]{hyperref}

\usepackage{pdfsync}

\makeatletter
\newcommand*{\defeq}{\mathrel{\rlap{%
                     \raisebox{0.3ex}{$\m@th\cdot$}}%
                     \raisebox{-0.3ex}{$\m@th\cdot$}}%
                     =} 
\makeatother

\def\be{\begin{eqnarray}}
\def\ee{\end{eqnarray}}

\newcommand{\bea}{\begin{eqnarray}}
\newcommand{\eea}{\end{eqnarray}}
\def\ben{\begin{equation}}
\def\een{\end{equation}}

     \let\r=v

\def\be{\begin{equation}}
\def\ee{\end{equation}}
\def\ba{\begin{eqnarray}}
\def\ea{\end{eqnarray}}

\def\bal#1\eal{\begin{align}#1\end{align}}
\def\bs#1\es{\begin{split}#1\end{split}}

\allowdisplaybreaks  

\numberwithin{equation}{section}

\def\bx{\bar{X}}

\def\be{\begin{equation}}
\def\ee{\end{equation}}
\def\ba{\begin{eqnarray}}
\def\ea{\end{eqnarray}}
\def\bal#1\eal{\begin{align}#1\end{align}}

\def\r{\rightarrow}

\def\r{\right}

\def\lr{\leftrightarrow}

\def\br{\rbrace}

\def\ie{\begin{equation}\begin{aligned}}
\def\fe{\end{aligned}\end{equation}}

\usepackage{multirow}

\def \be {\begin{equation}}
\def \ee {\end{equation}}

\newcommand{\CL}{D}

\usepackage{framed}

\begin{document}

\onehalfspacing


\begin{center}

~
\vskip5mm


{\LARGE {Elastic stiffness of three-dimensional black holes \\[0.5em] and wormholes from Liouville line defects}
}



\vskip7mm
Jeevan Chandra$^1$ and Boris Post$^2$

\vskip5mm

{[1] \textit{Leinweber Institute for Theoretical Physics and Department of Physics, \\ University of California, Berkeley, USA
}
\\
{[2] \textit{Mathematical Institute, \\ University of Oxford, Oxford, United Kingdom}
}}

\vskip5mm

{\tt jcn1998@berkeley.edu, \tt boris.post@maths.ox.ac.uk}

\end{center}

\vspace{2mm}

\begin{abstract}

We study elastic deformations of thin-shell black holes and wormholes in AdS$_3$ gravity. These geometries are sourced by line defects in the dual conformal field theory, and their shape and mass distribution define elastic moduli of the gravitational saddle. We compute the quadratic response of the partition function to these deformations, defining stiffness kernels for both transverse shape fluctuations and inhomogeneous mass-density fluctuations. The computation of the stiffness kernels can be realized as a hyperbolic response to a conformal welding problem which reduces to the universal Schwarzian response in the heavy-shell limit. The stiffness kernels are two-point functions of defect-local operators in CFT: the displacement operator, which measures the response to shape deformations, and a mass-density operator, which measures the response to local changes in the shell density. We compute the spectrum of these operators in the semiclassical limit, in various black hole and wormhole backgrounds. The spectrum can be discrete or continuous depending on the existence of a non-compact direction transverse to the shell in the geometry. We also provide a Lorentzian interpretation for the stiffness kernels using linear response theory and compute the relaxation time scales towards the corresponding transient deformations in the dual holographic CFT. Lastly, we compute the effect of these elastic deformations on black hole microstate statistics and black hole entropy.

\end{abstract}

\pagebreak

\tableofcontents

\vspace{0.7in}

\section{Introduction and summary of results} \label{secIntro}

Black hole and wormhole solutions to the gravitational path integral sourced by extended matter in the CFT naturally come with
moduli associated with the support of the matter. In this
paper, the relevant extended object is a line defect placed on a closed curve in the
Euclidean boundary geometry. Besides the usual parameters specifying the strength,
mass, or tension of the defect, one can also vary the curve on which the defect is
inserted. The functions specifying this curve will be referred to as
\emph{elastic moduli}. Deformations of these functions change the shape of the
defect, and we will call them \emph{elastic deformations}. The response of the
black hole or wormhole partition functions to these moduli is measured by the
stiffness kernels defined below.

In this paper, we study this response for thin-shell black holes and wormholes in three-dimensional Einstein gravity with negative cosomological constant. The central observables are stiffness kernels, defined as the quadratic response of the gravitational partition function to shape and mass-density deformations of the shell. These kernels have three complementary interpretations: they measure the elastic rigidity of the gravitational saddle, they compute two-point functions of defect-local operators in the dual CFT, and they describe the response of a hyperbolic surface to changes in its gluing data. Their spectral decomposition also gives a quantitative model for impurity relaxation in holographic CFTs: a transient deformation of a localized non-critical impurity can relax by dissipating into the ambient elastic medium. In the examples studied below, this mechanism is controlled by whether the relevant defect spectrum is continuous, leading to Lorentzian relaxation, or discrete, leading instead to persistent finite-volume oscillations.

In the rest of this section, we will provide a detailed introduction to each of the key concepts utilized in the paper and also summarize the important findings of the paper.

\subsection{Elastic deformations of thin-shell black holes and wormholes}

Thin-shell black holes and wormholes in AdS$_3$ gravity \cite{Anous:2016kss, Chandra:2022fwi, Chandra:2024vhm, Sasieta:2022ksu, Bah:2022uyz} are sourced by a non-conformal line defect in the dual CFT \cite{Chandra:2024vhm}. A
useful way to think about this line defect is as the continuum limit of a dense
collection of heavy local operators placed along a curve. In the continuum limit,
the curve becomes the support of an extended defect, and the local operator weights
combine into an effective mass density along the curve. Since the defect introduces
a scale, it is not a conformal line defect. Its insertion therefore changes the
geometry in a way that depends not only on topological data, but also on the
detailed shape and density profile of the curve.

The simplest thin-shell saddles are obtained by placing the defect on highly
symmetric loci. For example, the single-sided thin-shell black hole considered in
\cite{Chandra:2022fwi,Chandra:2024vhm} can be constructed by inserting line defects on special
circles, such as latitudes of the sphere, and then filling in the resulting boundary
conditions with locally AdS geometries. On the gravity side, this construction
corresponds to gluing different locally AdS$_3$ regions across a massive shell. In
the spherically symmetric case, the shell trajectory is fixed by symmetry, and the
resulting on-shell action computes the semiclassical black-hole partition function,
or equivalently the norm of the CFT state prepared by the defect insertion. The
\emph{elastic moduli} that we study in this paper arise when we relax this symmetry
assumption and allow the supporting curve of the defect to wiggle.

Concretely, consider a black-hole saddle prepared by two defect insertions, one on
the upper half of the Euclidean preparation geometry and one on the lower half. In
the undeformed configuration, the two defects sit at fixed average locations. We
introduce a shape deformation by shifting the position of the defect by a function
$\xi(x)$ along the angular coordinate $x$. The function $\xi(x)$ is therefore an
elastic modulus for the black-hole saddle. We keep the average location fixed
by removing the zero mode,
\begin{equation}\label{eq:removezeromode}
    \frac{1}{L}\int_0^{L} 
    \dd x\, \xi(x) = 0 ,
\end{equation}
where $L$ is the circumference of the shell locus.
This condition ensures that we are not changing the modulus that controls the mean
separation or average position of the defects, but only their shape. In other
words, we separate genuine shape-changing elastic deformations from the
lower-dimensional moduli already present in the symmetric saddle. Pictorially, the elastic deformation of the black hole is described by changing the boundary conditions as depicted below,
 \begin{equation}
        \vcenter{\hbox{
\begin{overpic}[grid=false, scale=0.1]{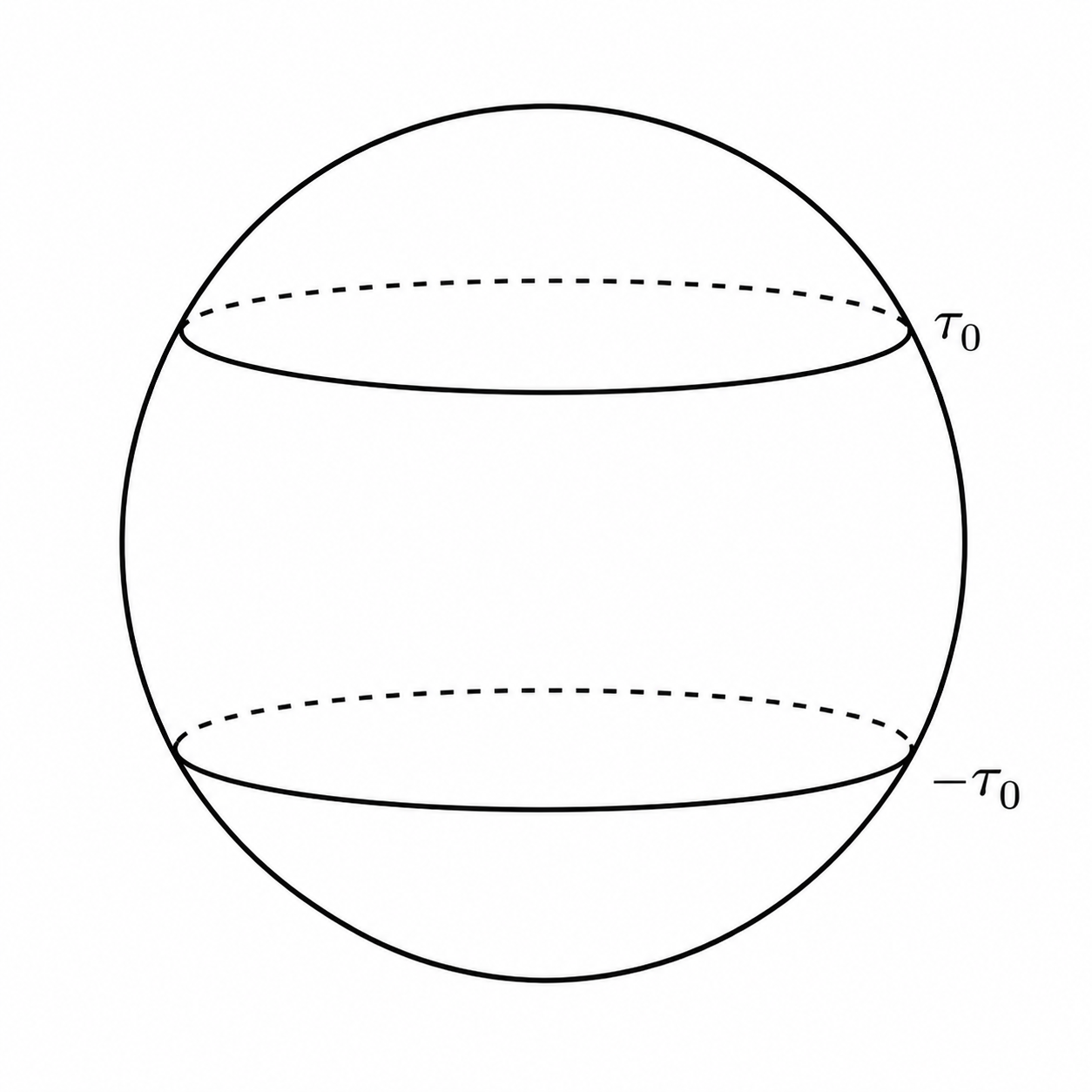}
\end{overpic}}} \qquad \xrightarrow{} \vcenter{\hbox{
\begin{overpic}[grid=false, scale=0.1]{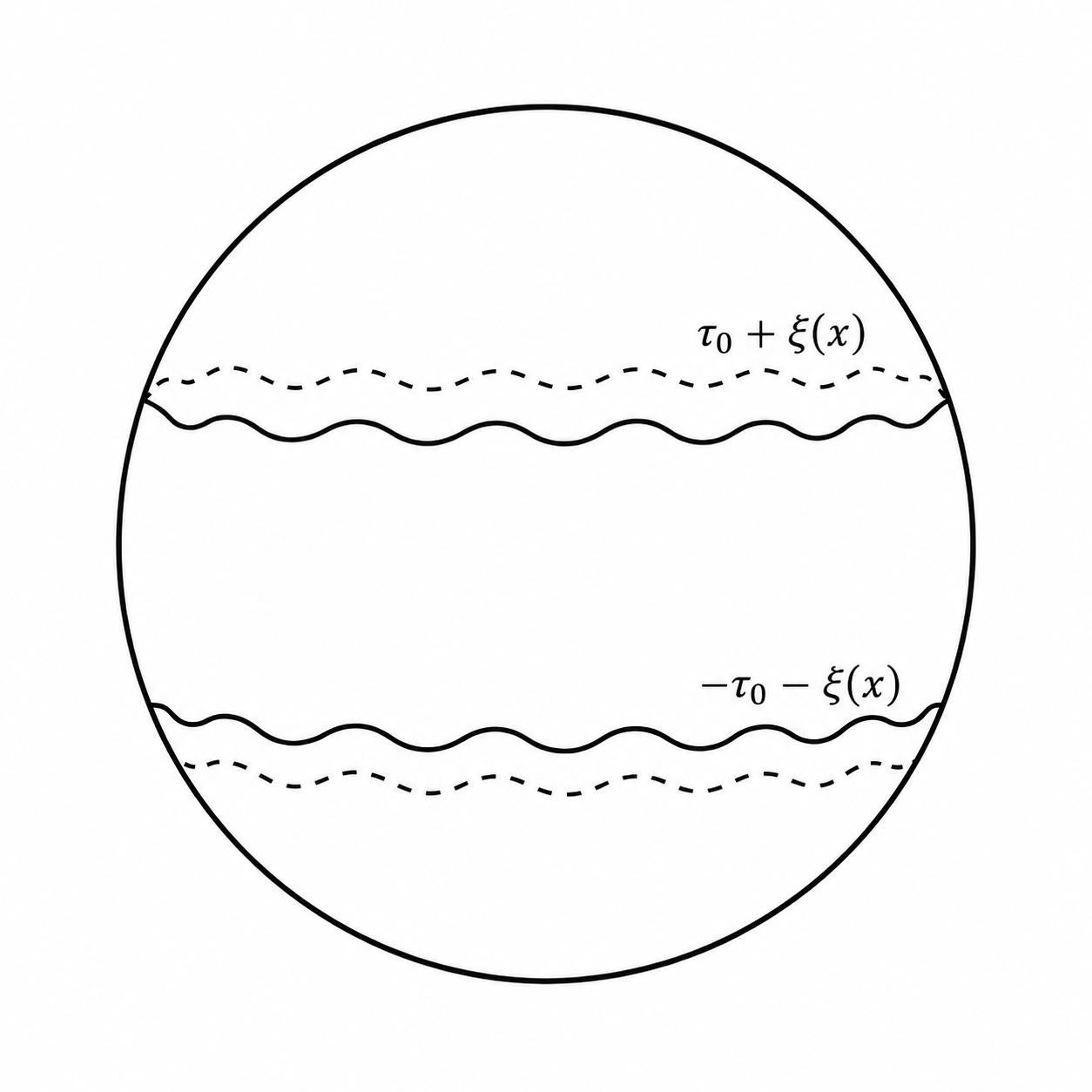}
\end{overpic}}}
    \end{equation}
For simplicity, we will mostly restrict to reflection-symmetric deformations. In
this sector, the deformation of the upper defect determines the deformation of the
lower defect. Thus the black-hole saddle is labeled by a single function $\xi(x)$
rather than by two independent functions. The black-hole partition function becomes
a functional of this elastic modulus,
\begin{equation}
    Z_{\rm BH}[\xi] \simeq e^{-I_{\rm BH}[\xi]} .
\end{equation}
The quadratic response to changing $\xi(x)$ defines the stiffness kernel for shape
deformations:
\begin{equation}
    \mathcal{K}^{\rm shape}_{\rm BH}(x,x')
    = L
    \left.
    \frac{\delta^2 \log Z_{\rm BH}[\xi]}
    {\delta \xi(x)\,\delta \xi(x')}
    \right|_{\xi=0}.
\end{equation}
Equivalently, in Fourier space one obtains coefficients
$\mathcal{K}^{\rm shape}_{{\rm BH},n}$ controlling the quadratic correction
\begin{equation}
    \log Z_{\rm BH}[\xi]
    =
    \log Z_{\rm BH}[0]
    +
    \frac12
    \sum_{n\in \mathbb{Z}}
    \mathcal{K}^{\rm shape}_{{\rm BH},n}\,
    \xi_n \xi_{-n}
    + \cdots .
\end{equation}
The omission of the $n=0$ mode implements the condition \eqref{eq:removezeromode} that the average location
of the shell is held fixed. The stiffness kernel therefore measures the response of
the semiclassical black-hole saddle to motion in the space of elastic moduli.

There is a second, closely related class of deformations. In the symmetric
constructions, the line defect is usually taken to arise from a uniform distribution
of local operator insertions along the curve. However, one can also make this
distribution non-uniform. In the continuum description, this corresponds to
replacing the constant mass density of the shell by a position-dependent density.
We call these \emph{mass deformations}. If $\mu(x)$ denotes the non-uniform part of
the mass density, then $\mu(x)$ is another modulus of the line defect. When the
total mass is held fixed, we remove its zero mode,
\begin{equation}
    \frac{1}{L}\int_0^L \dd x\, \mu(x) = 0 .
\end{equation}
The black-hole partition function then becomes a functional of this modulus, $Z_{\rm BH}[\mu]$.
The corresponding mass stiffness kernel is defined by
\begin{equation}
    \mathcal{K}^{\rm mass}_{\rm BH}(x,x')
    =L
    \left.
    \frac{\delta^2 \log Z_{\rm BH}[\mu]}
    {\delta \mu(x)\,\delta \mu(x')}
    \right|_{\mu=0}.
\end{equation}
Thus shape and mass deformations probe two complementary classes of moduli of the
same black-hole saddle: the first changes where the shell is placed, while the
second changes how the matter is distributed along the shell.

The same logic applies to thin-shell wormholes, which in the symmetric set-up were analyzed in \cite{Chandra:2024vhm, Sasieta:2022ksu}. A particularly simple example is
the sphere one-point wormhole. In that case, the undeformed configuration contains
an equatorial line defect. One can deform the equator into a wiggly curve, again
imposing that the zero mode of the deformation is removed so that the average
position of the defect is fixed. The wormhole partition function then becomes a
functional $Z_{\rm WH}[\xi]$, and the shape stiffness kernel is computed by a second functional derivative of the partition function just like in the black hole case.
Similarly, one can vary the mass density along the shell locus while keeping
the total mass fixed and quantify the effect of these mass inhomogeneities by computing the stiffness kernel by a second functional derivative of the wormhole partition function with respect to the mass deformation function.

For the wormhole examples studied below, there are in general deformations
associated with each asymptotic boundary. To simplify the analysis, we restrict to
the sector in which the deformations on the two boundaries are identical. This
symmetric sector is enough to exhibit the main physical point: thin-shell
wormholes, like thin-shell black holes, possess elastic moduli associated
with the shape and density profile of their extended matter source. The purpose of
the following sections is to compute the stiffness kernels associated with these
moduli and interpret their Lorentzian continuation in terms of a retarded response to a transient deformation.

\subsection{Computing the stiffness kernels using Liouville line defects}

The stiffness kernels defined above admit a useful interpretation both in the boundary CFT and
in the bulk theory. In the boundary language, we are asking how correlation functions of a
heavy line defect respond when the defect is deformed. In the bulk language, the same question
becomes a question about how a thin-shell black hole or wormhole responds when the shell is
displaced or when its mass density is made inhomogeneous. The basic bridge between these two
descriptions is the thin-shell line defect correspondence of \cite{Chandra:2024vhm}.
In that work, thin-shell black holes in $AdS_3$ are created by line defects $D_\Sigma$ in a
compact holographic CFT, while the universal large-$c$ defect data are computed by an
auxiliary Liouville line defect $L_\Sigma$. The three problems---line defects in compact CFTs,
line defects in Liouville CFT, and thin-shell solutions in $3d$ gravity---reduce semiclassically
to the same Liouville saddle.

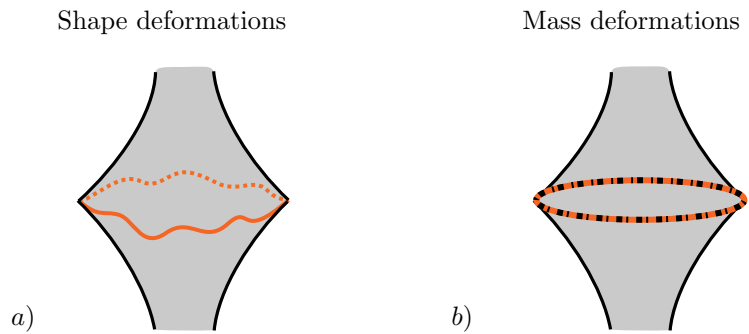
\begin{figure}
    \centering
\tikzset{every picture/.style={line width=1.2pt}} 
\begin{tikzpicture}[x=0.75pt,y=0.75pt,yscale=-1.3,xscale=1.3]
\draw  [draw opacity=0][fill={rgb, 255:red, 202; green, 202; blue, 202 }  ,fill opacity=1 ] (229.67,61.33) .. controls (230.11,59.22) and (232.78,59.33) .. (240.56,59.33) .. controls (248.33,59.33) and (250.75,58.5) .. (252.11,61) .. controls (253.47,63.5) and (252.83,73.25) .. (262.17,88.25) .. controls (271.5,103.25) and (280.22,110.53) .. (280.33,111.17) .. controls (280.44,111.81) and (271.93,118.69) .. (263.1,132.8) .. controls (254.27,146.91) and (253.42,160.79) .. (252,160.92) .. controls (250.58,161.04) and (243.62,161.02) .. (238.46,160.98) .. controls (233.29,160.94) and (228.58,160.79) .. (229,160.5) .. controls (229.42,160.21) and (227.83,149.87) .. (217.5,133.2) .. controls (207.17,116.53) and (199.56,110.92) .. (199.67,110.92) .. controls (199.78,110.92) and (212.33,99.58) .. (219.83,87.92) .. controls (227.33,76.25) and (229.22,63.44) .. (229.67,61.33) -- cycle ;
\draw [color={rgb, 255:red, 245; green, 101; blue, 35 }  ,draw opacity=1 ][line width=1.5]    (200,111) .. controls (208.25,117.88) and (209.5,114.63) .. (215.5,116.38) .. controls (221.5,118.13) and (224.21,130.04) .. (233.75,123.38) .. controls (243.29,116.71) and (248.43,128.14) .. (257.57,119.86) .. controls (266.71,111.57) and (262,127.38) .. (281,110.67) ;
\draw [color={rgb, 255:red, 245; green, 108; blue, 35 }  ,draw opacity=1 ][line width=1.5]  [dash pattern={on 1.5pt off 1.5pt on 1.5pt off 1.5pt}]  (200,111) .. controls (204.5,111.38) and (215.5,99.38) .. (223.75,103.38) .. controls (232,107.38) and (234.75,97.88) .. (244.25,100.38) .. controls (253.75,102.88) and (253.88,106.88) .. (264.25,105.13) .. controls (274.63,103.38) and (271.25,109.88) .. (281,110.67) ;
\draw    (229.67,61.33) .. controls (229.3,77.4) and (214.9,97.6) .. (200,111) ;
\draw    (252.11,61) .. controls (253.41,76.33) and (266.39,97) .. (281,110.67) ;
\draw    (229.33,160.58) .. controls (226.56,145) and (214.11,124.78) .. (199.67,110.92) ;
\draw    (252.33,161) .. controls (253.61,145.64) and (266.34,124.94) .. (280.67,111.25) ;
\draw  [draw opacity=0][fill={rgb, 255:red, 202; green, 202; blue, 202 }  ,fill opacity=1 ] (405.67,61.33) .. controls (406.11,59.22) and (408.33,59.11) .. (417,59.11) .. controls (425.67,59.11) and (426.75,58.75) .. (428.11,61) .. controls (429.47,63.25) and (428.83,73.25) .. (438.17,88.25) .. controls (447.5,103.25) and (456.22,110.53) .. (456.33,111.17) .. controls (456.44,111.81) and (447.93,118.69) .. (439.1,132.8) .. controls (430.27,146.91) and (429.42,160.79) .. (428,160.92) .. controls (426.58,161.04) and (419.62,161.02) .. (414.46,160.98) .. controls (409.29,160.94) and (404.58,160.79) .. (405,160.5) .. controls (405.42,160.21) and (403.83,149.87) .. (393.5,133.2) .. controls (383.17,116.53) and (375.56,110.92) .. (375.67,110.92) .. controls (375.78,110.92) and (388.33,99.58) .. (395.83,87.92) .. controls (403.33,76.25) and (405.22,63.44) .. (405.67,61.33) -- cycle ;
\draw    (405.67,61.33) .. controls (405.3,77.4) and (390.9,97.6) .. (376,111) ;
\draw    (428.11,61) .. controls (429.41,76.33) and (442.39,97) .. (457,110.67) ;
\draw    (405.33,160.58) .. controls (402.56,145) and (390.11,124.78) .. (375.67,110.92) ;
\draw    (428.33,161) .. controls (429.61,145.64) and (442.34,124.94) .. (456.67,111.25) ;
\draw  [color={rgb, 255:red, 245; green, 101; blue, 35 }  ,draw opacity=1 ][line width=2.25]  (376.67,110.67) .. controls (376.67,106.46) and (394.65,103.04) .. (416.83,103.04) .. controls (439.02,103.04) and (457,106.46) .. (457,110.67) .. controls (457,114.88) and (439.02,118.29) .. (416.83,118.29) .. controls (394.65,118.29) and (376.67,114.88) .. (376.67,110.67) -- cycle ;
\draw  [dash pattern={on 0.75pt off 1.5pt on 2.25pt off 3pt}][line width=2.25]  (376.67,110.67) .. controls (376.67,106.46) and (394.65,103.04) .. (416.83,103.04) .. controls (439.02,103.04) and (457,106.46) .. (457,110.67) .. controls (457,114.88) and (439.02,118.29) .. (416.83,118.29) .. controls (394.65,118.29) and (376.67,114.88) .. (376.67,110.67) -- cycle ;
\draw (190.5,36) node [anchor=north west][inner sep=0.75pt]   [align=left] {Shape deformations};
\draw (172.5,150) node [anchor=north west][inner sep=0.75pt]    {$a)$};
\draw (342.5,150) node [anchor=north west][inner sep=0.75pt]    {$b)$};
\draw (370,36) node [anchor=north west][inner sep=0.75pt]   [align=left] {Mass deformations};
\end{tikzpicture}
    \caption{The figure on the left is a sketch of a hyperbolic surface folded across a Liouville line defect placed along a deformed curve that is used to compute the stiffness kernels for shape deformations. The figure on the right depicts a position dependent coupling of the Liouville line defect that is used to compute the stiffness kernels for mass deformations. }
    \label{fig:placeholder}
\end{figure}

The Liouville line defect is defined by
\begin{equation}
L_\Sigma
=
\exp\left[
{m_0\over 2\pi b}
\int_\Sigma \dd \ell\,\phi
\right].
\end{equation}
In the semiclassical limit $b\to0$, with $\Phi=2b\varphi$ held fixed, the corresponding
classical Liouville action takes the schematic form
\begin{equation} \label{eq:Liouvactiondefect}
S_L[\Phi;\Sigma]
=
{1\over 4\pi}
\int_\Gamma \dd x\,\dd y
\left[
{1\over 4}(\partial\Phi)^2+e^\Phi
\right]
-
{m_0\over 4\pi}
\int_\Sigma \dd \ell\,\Phi
+\text{boundary terms}.
\end{equation}
The saddle describes a hyperbolic metric
\begin{equation}
\dd s^2=e^{\Phi(x,y)} (\dd x^2+\dd y^2),
\end{equation}
which is smooth away from $\Sigma$ but has a thin-shell junction condition across $\Sigma$. The full 3d geometry can be foliated by these hyperbolic slices, with the metric 
\begin{equation}
    \dd s^2 =\dd\rho^2+\cosh^2(\rho)e^{\Phi(x,y)}(\dd x^2+\dd y^2)\,.
\end{equation}
This metric applies to both the thin-shell wormholes that we will study and to the thin-shell black holes in hyperbolic slicing.
The Liouville field $\Phi$ is continuous across the shell, while its normal derivative jumps by an
amount fixed by the shell mass. In a local coordinate system where the shell is at fixed $y$,
this takes the form
\begin{equation}
\Phi_+=\Phi_-,
\qquad
\partial_y\Phi_+-\partial_y\Phi_-=-2m_0 .
\end{equation}
Thus the Liouville line defect is geometrically a thin shell: it glues two constant negative
curvature surfaces across a curve with a prescribed discontinuity in extrinsic curvature.

This is also why Liouville theory is the natural language for the CFT computation. The
large-$c$ Virasoro identity block is computed by the Zamolodchikov monodromy method \cite{ZamoRecursion}, and
the monodromy problem is equivalently a uniformization problem for the relevant Riemann
surface. Liouville theory performs precisely this uniformization. The same mechanism explains why Liouville theory appears in the semiclassical limit of pure AdS$_3$ gravity \cite{Verlinde:1989ua,Witten:1988hc,Brown:1986nw, Hartman:2013mia, Chandra:2022bqq, Collier:2023fwi}. In the present context, the
uniformizing metric is not smooth across the defect; rather, it is a hyperbolic metric with a
thin-shell junction condition. Therefore, the same Liouville saddle simultaneously computes
the semiclassical Virasoro block in the CFT and the on-shell action of the corresponding
$3d$ gravity saddle with a thin shell of matter. Schematically, the correspondence is summarized by the diagram,
\begin{equation}
\begin{tikzpicture}[scale=0.9]

\node[
    draw,
    minimum width=4.5cm,
    minimum height=1.7cm,
    align=center
] (upperleft) at (2.25,10.85)
{Thin shell black holes\\
and wormholes in $AdS_3$};

\node[
    draw,
    minimum width=4.5cm,
    minimum height=1.7cm,
    align=center
] (upperright) at (9.75,10.85)
{Statistics of line defects\\
$\CL_\Sigma$ in the dual CFT$_2$};

\node[
    draw,
    minimum width=5.9cm,
    minimum height=1.7cm,
    align=center
] (lower) at (5.95,7.85)
{Line defects $L_\Sigma$ in Liouville CFT};

\draw[thick,<->] (3.5,9.7) -- (4.8,8.9);
\draw[thick,<->] (8.4,9.7) -- (7.1,8.9);
\draw[thick,<->] (5,11) -- (7,11);

\end{tikzpicture}
\end{equation}
For example, for a black hole geometry by a pair of line defects, we may write
schematically\footnote{The holographic CFT correlator should be understood in a coarse-grained sense, as explained below.}
\begin{equation}
\langle D_\Sigma^\dagger D_\Sigma\rangle_{\rm CFT}
\;=\;
\left|\langle L_\Sigma\rangle_{\rm ZZ}\right|^2
\;=\;
Z_{\rm grav}^{\rm BH}.
\end{equation} 
The semiclassical relation between the central charge $c$, Liouville coupling $b$ and gravitational coupling $G_N$ is the usual
\begin{equation}
    c \sim \frac{6}{b^2} \sim \frac{3}{2G_N}.
\end{equation}
In this paper, we will work exclusively in the semiclassical limit $b\to 0$.

For thin-shell wormholes, the Liouville action computes the semiclassical contribution to the moments
of defect matrix elements. Black holes are naturally described by Liouville correlators with
ZZ boundaries, while wormholes are described by Liouville correlators on closed
surfaces, such as the sphere or torus. In this way, the same deformation problem can be viewed
either as a response problem for CFT defect correlators or as an elastic response problem for
thin-shell black holes and wormholes.

In the rest of the paper, we use this correspondence operationally. We first write the
undeformed Liouville saddle for the desired black hole or wormhole geometry, with undeformed shell locus $y_{\Sigma,0}$. We then perturb
the shell either by a transverse displacement,
\begin{equation}
y_\Sigma(x)=y_{\Sigma,0}+\epsilon\,\xi(x),
\end{equation}
or by a longitudinal mass-density deformation,
\begin{equation}
m(x)=m_0+\epsilon\,\mu(x),
\end{equation}
where $\epsilon$ is a small parameter.
The stiffness kernels are obtained by solving the linearized Liouville equation with the
corresponding linearized junction conditions and evaluating the on-shell action to quadratic
order. A useful feature of this procedure is that the quadratic action localizes to the shell:
for the stiffness kernels, the first-order Liouville field is enough.

A unifying way to organize the answer is through Dirichlet-to-Neumann eigenvalues. For each
Fourier mode $e^{inx}$, we solve the linearized Liouville equation in each region adjacent to
the shell, imposing the appropriate condition at the other end of the region: regularity at a
cap, smoothness around a torus cycle, a ZZ boundary condition, or a geodesic boundary
condition. If the solution is normalized to have unit value on the shell, the normal
derivative at the shell is the Dirichlet-to-Neumann eigenvalue, denoted schematically by
$\lambda_{\pm,n}$. More precisely, let $\widehat u_{\pm,n}$ denote the solution of the
linearized Liouville equation in the two regions $\mathcal{M}_{\pm}$
adjacent to the shell, with Fourier dependence $e^{inx}$, satisfying
the appropriate condition at the other end of the region and normalized
to unity on the shell locus $\Sigma$, $\left.\widehat u_{\pm,n}\right|_{\Sigma}=1$. In adapted coordinates for which
$\mathcal{M}_{-}$ lies below the shell and $\mathcal{M}_{+}$ lies above
it, the eigenvalues are given by
\begin{equation}
    \lambda_{\pm,n}
    =
    \left.\mp\,\partial_y\widehat u_{\pm,n}\right|_{\Sigma}.
\end{equation}
Thus $\lambda_{\pm,n}$ measures the normal response induced in each
region by prescribing a unit value of the linearized Liouville field
on the shell. 

In terms of these eigenvalues, the mass-deformation kernels universally take the form
\begin{equation}
\mathcal{K}^{\rm mass}_n
=
{1\over \lambda_{+,n}+\lambda_{-,n}},
\end{equation}
while the shape-deformation kernels take the schematic form
\begin{equation}
\mathcal{K}^{\rm shape}_n
=
-m_0
\left(
m_0\,{\lambda_{+,n}\lambda_{-,n}\over \lambda_{+,n}+\lambda_{-,n}}
+\text{local geometric terms on the shell}
\right).
\end{equation}
The first term in the shape kernel is the non-local response of the surrounding hyperbolic
geometry, while the local terms come from moving the shell and from the proper-length
correction to the shell action. 
This is the organizing principle behind the computations
in the subsequent sections.

\subsection{A displacement operator for the shell and its spectrum}

The shape-deformation kernels computed in this paper have a natural interpretation as two-point functions of a defect displacement operator \cite{Billo:2016cpy}. More generally, for a local line defect, the operator conjugate to normal
deformations of the defect embedding is the displacement operator, which we denote by $\mathcal{D}_\perp$. It appears
in the Ward identity for broken transverse translations,
\begin{equation}
    \partial_\mu T^{\mu i}(x,y)=\mathcal{D}^i_\perp(x)\,\delta_\Sigma(y),
\end{equation}
up to contact terms and possible defect-local improvement terms. For a conformal line defect, this operator is a protected defect primary, but for non-conformal defects like the shell defects considered here, its two-point
function is not fixed by symmetry.

For the shell defects considered here, the transverse displacement is the shape mode $\xi(x)$. If $Z[\xi]$ denotes the defect partition function, or equivalently the gravitational partition function of the deformed shell saddle, then a shape variation inserts the displacement operator:
\begin{equation}
    \delta_\xi \log Z[\xi]
    =\frac{1}{L}
    \int_0^L \dd x\,\xi(x)\,\langle \mathcal{D}_\perp(x)\rangle ,
\end{equation}
where $L$ is the circumference of the shell locus.
Expanding to quadratic order gives
\begin{equation}
    \log Z[\xi]
    =
    \log Z[0]
    +
    {1\over 2L}
    \int \dd x\,\dd x'\,
    \xi(x)\,
    \mathcal{K}^{\rm shape}(x,x')\,
    \xi(x')
    +\cdots .
\end{equation}
Thus, up to contact terms,
\begin{equation}
    \mathcal{K}^{\rm shape}(x,x')
    =
    \langle \mathcal{D}_\perp(x)\mathcal{D}_\perp(x')\rangle_{\rm sep}
    +
    \text{contact terms}.
\end{equation}
The contact terms depend on local counterterms and on the choice of conformal frame for the non-conformal shell defect. The separated-point part is the part that admits a spectral interpretation. Therefore the stiffness kernel is not only a measure of mechanical rigidity: it is the Euclidean two-point function that quantifies how defect observables respond to changing the shape of the line on which the defect is supported.

For the sphere 1-point wormhole, the shell defect sits on an equatorial circle in the background Liouville saddle. Setting the radius of the shell locus to $1$, the separated-point kernel is reflection-positive and therefore has the spectral representation
\begin{equation}
    \mathcal{K}^{\rm shape}_{\rm WH}(x)
    =
    \int_0^\infty \dd \omega\,
    \rho_D(\omega)\,
    {\cosh\!\left[\omega(\pi -x)\right]\over \sinh(\pi \omega)} .
\end{equation}
In section \ref{sec:2}, we compute the spectral density for the sphere 1-point wormhole, finding
\begin{equation}
    \rho_D(\omega)
    =
    {c m_0^2\over 6\pi}\,
    {\omega(\omega^2+1)\over \omega^2+{m_0^2\over 4}} .
    \label{eq:intro-sphere-one-point-displacement-density}
\end{equation}
The spectrum is manifestly positive. It is also continuous. This is explained geometrically by the fact that the complete set of defect states is inserted on the cycle dual to the shell locus. For the sphere 1-point wormhole this dual slice is non-compact, and hence the spectral parameter $\omega$ is continuous. The high-energy growth of $\rho_D(\omega)$ encodes the short-distance OPE singularity of the displacement two-point function, while the finite-$\omega$ structure contains the nonlocal response of the hyperbolic geometry glued across the shell.

The torus 1-point wormhole gives a useful contrast, which will be studied in section \ref{sec:torus1point}. The shell again supports a displacement operator, but the dual cycle on which the defect Hilbert space is quantized now has finite length $\beta$. Consequently, the spectrum is discrete. The non-contact part of the shape kernel has poles at real frequencies $\omega_{j,D}$ solving
\begin{equation}
    m_0
    =
    2\omega_{j,D}
    \tan\!\left({\beta\omega_{j,D}\over 2}\right),
    \label{eq:intro-torus-displacement-pole-condition}
\end{equation}
with one solution in each interval
\begin{equation}
    \omega_{j,D}
    \in
    \left(
    {2\pi j\over \beta},
    {(2j+1)\pi\over \beta}
    \right),
    \qquad
    j=0,1,2,\ldots 
\end{equation}
Hence the physical spectral density can be expressed by the following discrete sum
\begin{equation}
    \rho_D(\omega;\beta)
    =
    \sum_{j=1}^{\infty}
    \rho_{j,D}\,\delta(\omega-\omega_{j,D}),
\end{equation}
with the spectral coefficients given by
\begin{equation}
    \rho_{j,D}
    =
    {c\, m_0^2(\omega_{j,D}^2-r_H^2)\over
    6\left[
    \tan\!\left({\beta\omega_{j,D}\over 2}\right)
    +
    {\beta\omega_{j,D}\over 2}
    \sec^2\!\left({\beta\omega_{j,D}\over 2}\right)
    \right]} .
    \label{eq:intro-torus-displacement-spectral-weights}
\end{equation}
Here $r_H$ is fixed by the background equation
\begin{equation}
    m_0
    =
    2r_H
    \tan\!\left({r_H\beta\over 2}\right).
\end{equation}
The weights $\rho_{j,D}$ are positive. At high temperature the levels are sparse, with uniform thermal spacing $\Delta \omega \sim \frac{2\pi}{\beta}$,
while at low temperature the poles become dense and condense into the branch cut of the sphere one-point answer \eqref{eq:intro-sphere-one-point-displacement-density}. Thus the sphere and torus 1-point wormholes provide two complementary realizations of the same displacement spectrum: a continuous spectrum for a non-compact dual cycle and a discrete spectrum for a compact dual cycle.

\begin{figure}
    \centering
    \includegraphics[width=0.75\linewidth]{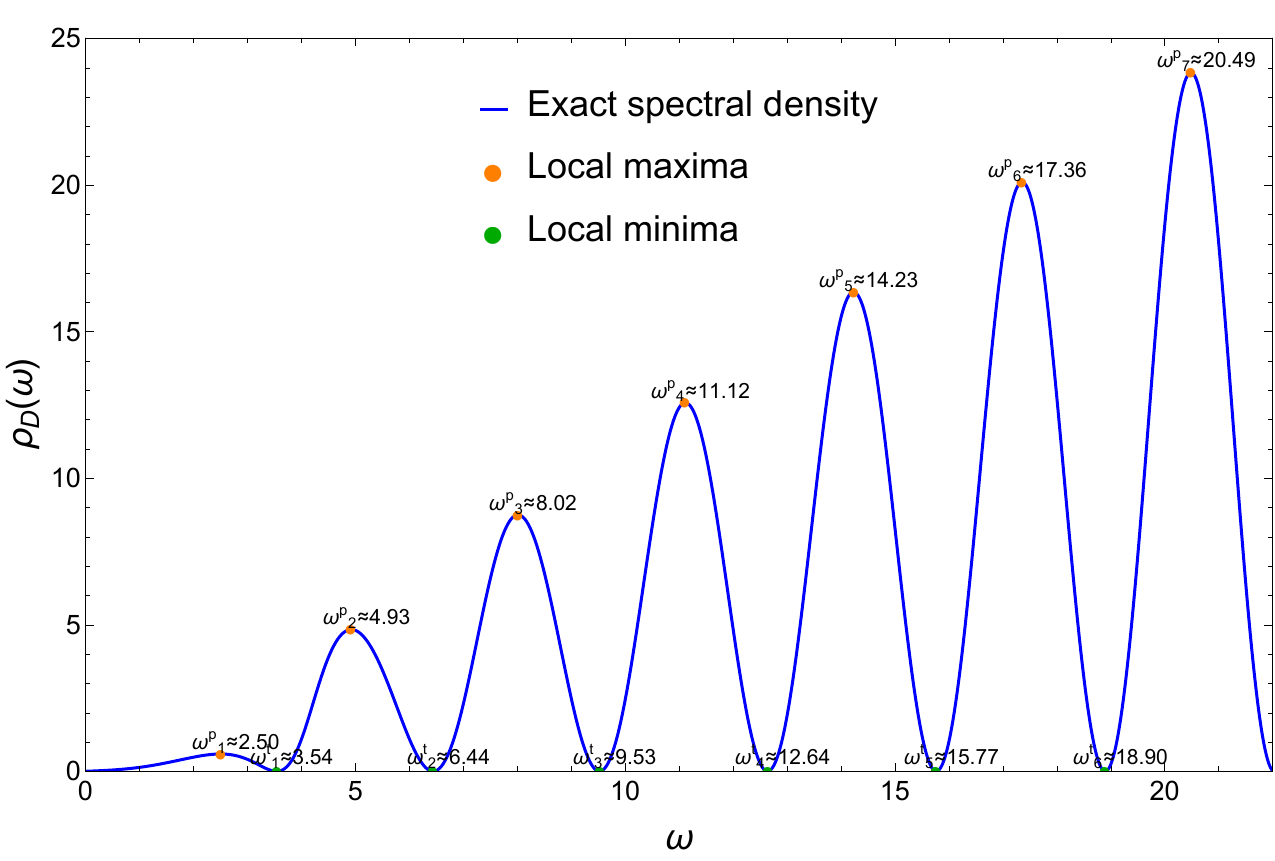}
    \caption{A plot of the spectral density of the displacement operator for the one-sided black hole drawn for the horizon radius $r_H=2$ and shell separation parameter $\tau_0=1$. The locations of the local maxima and minima have been marked.}
    \label{fig:shapespectrumBH}
\end{figure}

The thin-shell black holes studied in sections \ref{sec:one-sided} and \ref{sec:PETS} follow the same organizing principle. For the one-sided black hole, the Liouville saddle is constructed by placing the shell defect in a cylinder geometry with a ZZ boundary. The defect Hilbert space is again defined on a non-compact slice, so the displacement spectrum is continuous. The corresponding spectral density is written explicitly in \eqref{eq:one-sided-BH-displacement-spectral-density}. Its detailed functional form depends on the Dirichlet-to-Neumann data of the two regions adjacent to the shell. Conceptually, this spectral density contains sharp information about the finite geometric region between the shell and the ZZ boundary. In particular, it has alternating local maxima and minima as plotted in Fig.~\ref{fig:shapespectrumBH}. These features arise, respectively, from constructive and destructive interference of the linearized Liouville modes in the cavity between the shell and the ZZ boundary. Thus, the one-sided black hole has a continuous displacement spectrum and its spectral density retains a detailed imprint of the intermediate black hole region.

Finally, the symmetric PETS black hole gives the finite-temperature two-sided black hole example. The relevant Euclidean geometry is an annulus with ZZ boundaries, and the shell lies at the midpoint in the symmetric configuration. The displacement spectrum is discrete, as in the torus wormhole, because the relevant dual cycle has a finite length. However, the pole equation differs from the torus wormhole because the relation between $r_H$, $m_0$, and $\beta$ is different in the black-hole saddle. The corresponding spectral density is written in \eqref{eq:PETS-displacement-spectral-density}; it is a sum of delta functions with positive spectral weights. Thus, the black-hole examples mirror the wormhole examples. Continuous spectra arise when the dual slice supporting the defect Hilbert space is non-compact, while compact dual cycles produce discrete spectra. The one-sided black hole realizes the continuous case with additional resonance structure due to the ZZ-boundary cavity, whereas the symmetric PETS black hole realizes the finite-temperature discrete case.

\subsection{A mass-density operator for the shell and its spectrum}

The mass-deformation kernels admit an interpretation closely parallel to the displacement kernels discussed above, but now the relevant defect-local operator is conjugate to changes in the strength of the line defect rather than to transverse motions of its support. We define a mass-density operator $\mathcal{M}(x)$ on the shell by saying that a local change of the shell mass,
\begin{equation}
m(x)
=
m_0+\epsilon\,\mu(x),
\end{equation}
inserts $\mathcal{M}(x)$ in the defect correlator. Equivalently, if $Z[\mu]$ denotes the partition function in the presence of the non-uniform mass profile, then
\begin{equation}
\delta_\mu \log Z[\mu]
=
\frac{1}{L}\int_0^L \dd x\,\mu(x)\,\langle \mathcal{M}(x)\rangle .
\end{equation}
Expanding to quadratic order gives
\begin{equation}
\log Z[\mu]
=
\log Z[0]
+
{1\over 2L}
\int \dd x\,\dd x'\,
\mu(x)\,
\mathcal{K}^{\rm mass}(x,x')\,
\mu(x')
+\cdots ,
\end{equation}
and therefore, up to possible contact terms,
\begin{equation}
\mathcal{K}^{\rm mass}(x,x')
=
\langle \mathcal{M}(x)\mathcal{M}(x')\rangle_{\rm sep}
+
\text{contact terms}.
\end{equation}
The zero mode of $\mu(x)$ is removed so that we keep the total mass of the shell fixed. Since $m_0$ is the coupling constant of the Liouville line defect, mass deformations correspond to making this defect coupling position dependent along the line. This perspective is not special to the thin-shell defects studied here: for a general line defect with a local coupling $g$, one can similarly ask for the defect-local operator conjugate to an inhomogeneous coupling profile $g(x)$. In the present paper, $\mathcal{M}$ is the operator that measures the response to such local changes in the shell mass density. 
In the defect-RG literature, the operator conjugate to a local change of the defect coupling is usually described as the defect stress tensor $T_D$ \cite{Cuomo:2021rkm}. Upon promoting the coupling to a dilaton, $m(x)=m_0e^{\sigma(x)}$, it appears together with the displacement operator in the modified Ward identity\footnote{We thank Zohar Komargodski for pointing out that the mass deformation function $\mu(x)$ in our setup is the dilaton/spurion that is well studied in the context of defect-RG flows \cite{Cuomo:2021rkm}.}
\begin{equation}
\nabla_\mu T_{\rm bulk}^{\mu\nu}
=
-\delta_\Sigma\,\dot X^\nu
\bigl(\partial_s T_D-(\partial_s\sigma)T_D\bigr)
-\delta_\Sigma\,n^\nu \mathcal{D}_\perp .
\end{equation}
Here $\delta_\Sigma$ is a delta function supported on the line defect,
and $X^\mu(s)$ is the embedding of the line into the 2d surface.
The mass kernel is therefore the 2-point function of the defect stress tensor up to contact terms,
\begin{equation}
    \langle T_D(x) T_D(x') \rangle =m_0^2\, \langle \mathcal{M}(x)\mathcal{M}(x')\rangle + \text{contact terms}.
\end{equation}

For the sphere $1$-point wormhole, the spectrum of $\mathcal{M}$ is continuous for the same geometric reason as the displacement spectrum: the complete set of defect states is inserted on the non-compact cycle dual to the shell circle. After reinstating the radius $R$ of the shell locus (i.e. $L = 2\pi R$), the spectral density computed in section \ref{sec:2} is found to be
\begin{equation}
\rho_{M}(\omega;R)
=
{8c\,\omega(\omega^2+R^{-2})\over
3\pi\left[\left(4\omega^2-m_0^2+4R^{-2}\right)^2
+4m_0^2\omega^2\right]} .
\label{eq:intro-sphere-one-point-mass-density}
\end{equation}
Unlike the displacement spectral density, this expression has a softer high-energy behavior. This is the spectral counterpart of the universal logarithmic short-distance singularity of the mass-density two-point function. Thus, in the sphere $1$-point wormhole, both $\mathcal{D}_\perp$ and $\mathcal{M}$ have continuous spectra, but they probe different components of the defect response: $\mathcal{D}_\perp$ measures transverse motion of the shell, while $\mathcal{M}$ measures local variations of the defect coupling.

\begin{figure}
    \centering
    \includegraphics[width=0.75\linewidth]{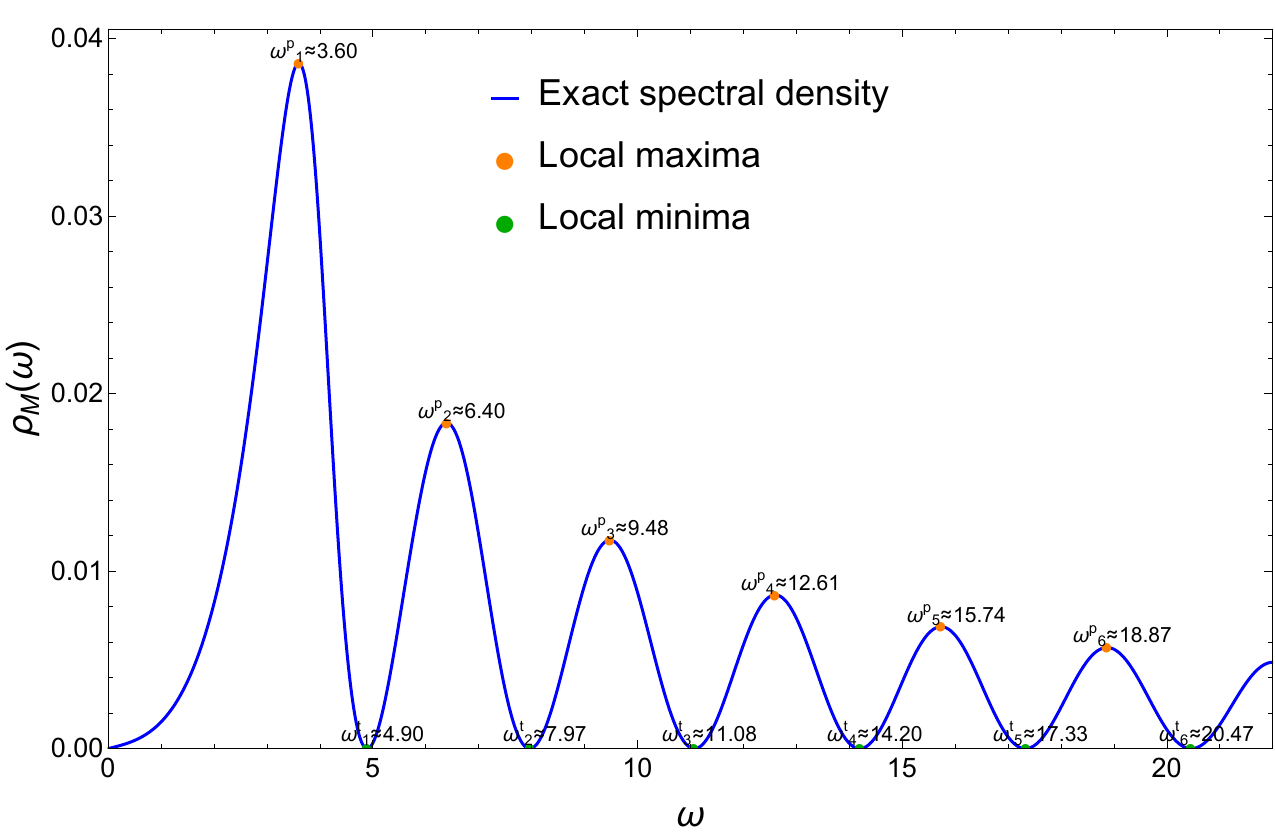}
    \caption{A plot of the spectral density of the mass-density operator for the one-sided black hole drawn for the horizon radius $r_H=2$ and shell separation parameter $\tau_0=1$. The locations of the local maxima and minima have been marked.}
    \label{fig:massspectrumBH}
\end{figure}

For the torus $1$-point wormhole, the dual cycle has finite length $\beta$, so the mass-density spectrum is discrete, as shown in section \ref{sec:torus1point}. The spectral poles $\omega_{j,M}$ are determined by
\begin{equation}
m_0\omega_{j,M}
+
2\left(
r_H^2+{m_0^2\over 4}-\omega_{j,M}^2
\right)
\tan\left({\beta\omega_{j,M}\over 2}\right)
=
0 .
\label{eq:intro-torus-mass-pole-condition}
\end{equation}
This should be compared with the equation for the displacement pole in \eqref{eq:intro-torus-displacement-pole-condition}. The two equations are different, so the displacement operator and the mass-density operator are in general supported on distinct discrete spectra. The mass-density operator spectral density takes the form
\begin{equation}
\rho_{M}(\omega;\beta)
=
\sum_{j=1}^{\infty}
\rho_{j,M}
\delta(\omega-\omega_{j,M}),
\end{equation}
with
\begin{equation}
\rho_{j,M}
=\frac{c}{6}
{
2\omega_{j,M}
+
m_0
\tan\left({\beta\omega_{j,M}\over 2}\right)
\over
4\omega_{j,M}
\tan\left({\beta\omega_{j,M}\over 2}\right)
+
\beta\left(
\omega_{j,M}^2-r_H^2-{m_0^2\over 4}
\right)
\sec^2\left({\beta\omega_{j,M}\over 2}\right)
-m_0} .
\label{eq:intro-torus-mass-spectral-weights}
\end{equation}
Here $r_H$ is fixed by the torus background equation
\begin{equation}
m_0
=
2r_H
\tan\left({r_H\beta\over 2}\right).
\end{equation}
The weights $\rho_{j,M}$ are positive. As in the displacement case, the spectral lines become denser at low temperature and condense into the continuous spectrum of the sphere $1$-point wormhole \eqref{eq:intro-sphere-one-point-mass-density}. The important new point is that the two natural defect-local operators, $\mathcal{D}_\perp$ and $\mathcal{M}$, need not share the same microscopic spectral support even though their Euclidean kernels are computed from the same linearized Liouville problem.

The black-hole examples follow the same pattern. For the one-sided black hole of section \ref{sec:one-sided}, the relevant defect Hilbert space is defined on a non-compact slice, and the mass-density operator therefore has a continuous spectrum. Its spectral density is obtained from the mass-deformation kernel built out of the same Dirichlet-to-Neumann data that controlled the shape response, but with the simpler universal structure characteristic of coupling deformations. At high frequencies this kernel again has the universal behavior that produces a logarithmic short-distance singularity in the two-point function of  $\mathcal{M}$. Just as for shape deformations, the spectral density has alternating local maxima and minima as plotted in Figure \ref{fig:massspectrumBH}. The locations of the minima for the mass deformations case agree approximately with the locations of the maxima for shape deformations case and vice versa as is evident by comparing the plots \ref{fig:massspectrumBH} and \ref{fig:shapespectrumBH}. 

For the symmetric PETS black hole studied in section \ref{sec:PETS}, the Euclidean geometry is an annulus with ZZ boundaries, so the relevant dual cycle is compact. The mass-density spectrum is therefore discrete, with spectral lines determined by the PETS mass-deformation kernel. Thus, just as for shape deformations, the one-sided black hole realizes the continuous-spectrum case, while the symmetric PETS black hole realizes the finite-temperature discrete-spectrum case.

\subsection{Lorentzian interpretation using linear response theory}\label{sec:linearresponse}

The Euclidean stiffness kernels computed above have a direct Lorentzian
interpretation in terms of linear response.  The shape modulus $\xi$ is the
source for the transverse displacement operator $\mathcal{D}_\perp$, while the
mass-density deformation $\mu$ is the source for a local defect operator
$\mathcal{M}$ that measures the response to changes in the defect coupling.
With the source convention
\begin{equation}
    H_{\rm pert}(t)
    =
    -\delta \xi(t)\mathcal{D}_\perp(t)
    -\delta\mu(t)\mathcal{M}(t),
\end{equation}
the corresponding causal response functions are
\begin{align}
    \delta\langle \mathcal{D}_\perp\rangle(t)
    &=
    \int_{-\infty}^{\infty} \dd t'\,
    \mathcal{K}^{\rm shape}_{\rm ret}(t-t')\,\delta\xi(t'),
    \\
    \delta\langle \mathcal{M}\rangle(t)
    &=
    \int_{-\infty}^{\infty} \dd t'\,
    \mathcal{K}^{\rm mass}_{\rm ret}(t-t')\,\delta\mu(t'),
\end{align}
where
\begin{align}
     \mathcal{K}^{\rm shape}_{\rm ret}(t)
    =
    -i\Theta(t)\,
    \langle [\mathcal{D}_\perp(t),\mathcal{D}_\perp(0)]\rangle,\qquad
    \mathcal{K}^{\rm mass}_{\rm ret}(t)
    =
    -i\Theta(t)\,
    \langle [\mathcal{M}(t),\mathcal{M}(0)]\rangle .
\end{align}
Equivalently, in frequency space,
\begin{equation}
    \delta\langle \mathcal{D}_\perp\rangle(\omega)
    =
    \mathcal{K}^{\rm shape}_{\rm ret}(\omega)\,\delta\xi(\omega),
    \qquad
    \delta\langle \mathcal{M}\rangle(\omega)
    =
    \mathcal{K}^{\rm mass}_{\rm ret}(\omega)\,\delta\mu(\omega).
\end{equation}
These formulae are the shell-defect analogue of the standard Kubo response
formula, which expresses the response to a weak external source in terms of an
equilibrium commutator of the operator conjugate to that source
\cite{Green1954,Kubo:1957mj}.  In the Euclidean calculation, the stiffness kernels
$\mathcal{K}^{\rm shape}(i\omega_n)$ and
$\mathcal{K}^{\rm mass}(i\omega_n)$ are thermal Matsubara correlators \cite{Matsubara1955, Martin:1959jp} of
$\mathcal{D}_\perp$ and $\mathcal{M}$, respectively.  Their retarded counterparts are
obtained by the usual analytic continuation
\begin{equation}
   \mathcal{K}^{\rm shape}_{\rm ret}(\omega)
    =
    \mathcal{K}^{\rm shape}(i\omega_n\rightarrow \omega+i0^+),
    \qquad
    \mathcal{K}^{\rm mass}_{\rm ret}(\omega)
    =
    \mathcal{K}^{\rm mass}(i\omega_n\rightarrow \omega+i0^+),
\end{equation}
up to local contact terms.  The associated spectral density,
\begin{equation}
    \rho_{D}(\omega)
    =
    \frac{1}{\pi}\operatorname{Im} \mathcal{K}^{\rm shape}_{\rm ret}(\omega),
    \qquad
     \rho_{M}(\omega)
    =
    \frac{1}{\pi}\operatorname{Im} \mathcal{K}^{\rm mass}_{\rm ret}(\omega)\,,
\end{equation}
controls both the Euclidean spectral representation and the Lorentzian
relaxation.  In particular,
\begin{equation}
    \mathcal{K}^{\rm shape}_{\rm ret}(t)
    =
    -2\Theta(t)\int_0^\infty \dd \omega\,
    \rho_{D}(\omega)\sin(\omega t), \qquad  \mathcal{K}^{\rm mass}_{\rm ret}(t)
    =
    -2\Theta(t)\int_0^\infty \dd \omega\,
    \rho_{M}(\omega)\sin(\omega t)\,.
\end{equation}
Hence, the late-time behavior is determined by the low-lying singularities of the
retarded correlator in the complex frequency plane.  Continuous spectra, or
equivalently branch cuts and poles away from the real axis, lead to dissipative
relaxation, while a purely discrete real spectrum gives oscillatory response
without late-time decay.  This is the sense in which the relaxation times
extracted below are Kubo relaxation times for the defect operators conjugate to
the elastic shape and mass-density moduli of the shell. This viewpoint is consistent with the correlation-function approach to
relaxation and transport: the slow relaxation processes that appear in the
response to an external perturbation are encoded in the analytic structure of
equilibrium response functions, as in the classic hydrodynamic analysis of
Kadanoff and Martin~\cite{KadanoffMartin1963}. The computation of the relaxation time from the exponential decay of the retarded correlator of defect-local operators presented in this paper is exactly analogous to the computation of quasinormal modes from retarded thermal correlators in CFTs (see for example \cite{Birmingham:2002ph}).

\begin{figure}
    \centering
    \includegraphics[width=0.7\linewidth]{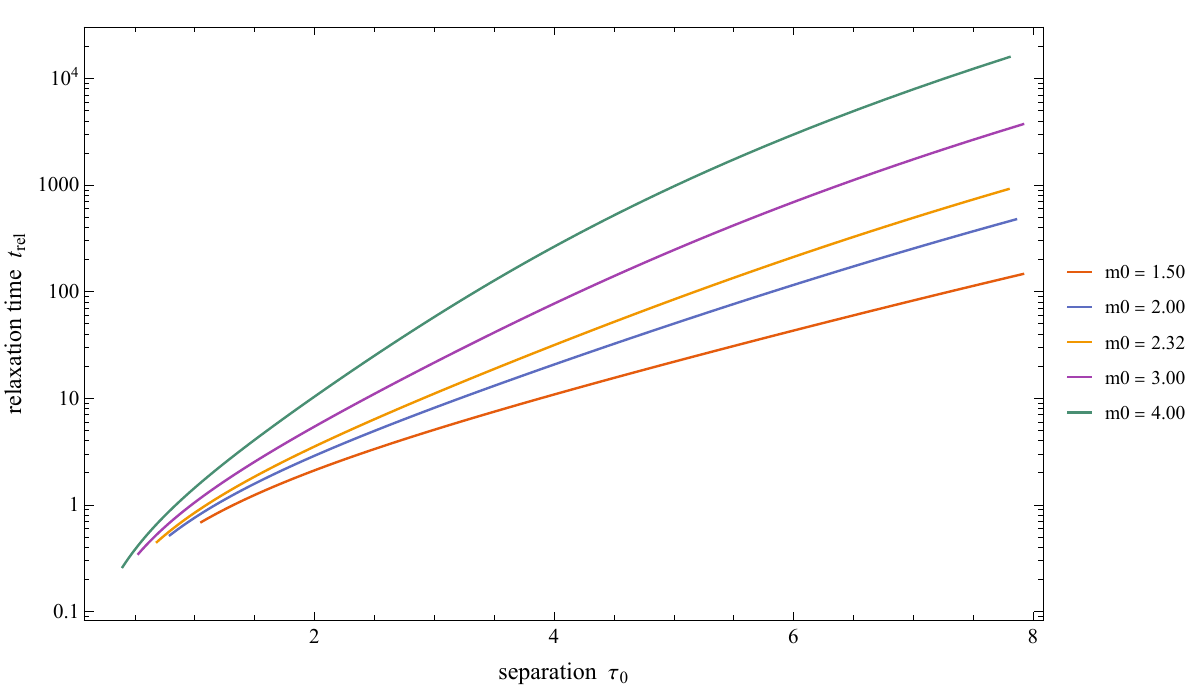}
    \caption{A plot of the relaxation time $t_{\rm rel}$ against the separation $\tau_0$ between the pair of defects in the holographic CFT sourcing the one-sided black hole, for various values of the mass parameter/ defect coupling. The slope is positive for each of the curves, thereby illustrating that the relaxation time increases monotonically with the defect separation. Since the relaxation times are identical for the shape or mass deformations, the same plot applies in either case.}
    \label{fig:relaxation_time}
\end{figure}

Applying these ideas from linear response theory to the analytically-continued stiffness kernels for the sphere 1-point wormhole, we derive the following relaxation times:
\begin{equation}
    \begin{split}
        & \delta \langle \mathcal{D}_\perp\rangle (t) \sim e^{-\frac{m_0 t}{2}} \implies  t_{\rm rel}=\frac{2}{m_0},\\
        & \delta \langle \mathcal{M}\rangle (t) \sim e^{-\frac{m_0 t}{4}}\left[\cos\left(\tfrac{\sqrt{3}}{4}m_0t\right)+\frac{1}{\sqrt{3}}\sin\left(\tfrac{\sqrt{3}}{4}m_0t\right)\right] \implies  t_{\rm rel}=\frac{4}{m_0}\,.
    \end{split}
\end{equation}
It is important to note that the above relaxation scales do not correspond to a conventional mass gap since the spectrum of the displacement or mass-density operators is continuous all the way down to zero energy. Since relaxation occurs due to a complex pole in the spectral density,
\begin{equation} \label{eq:defectQNM}
    \omega_*^{\rm shape}=-\frac{im_0}{2}, \qquad \omega_*^{\rm mass}=-\frac{im_0}{4}\pm \frac{\sqrt{3}m_0}{4}\,,
\end{equation}
we attribute it to a resonance between the shell deformation modes and the continuum Liouville modes in the bulk, with $m_0$ controlling the width of the resonance. Due to the close analogy with quasinormal modes, we could refer to (\ref{eq:defectQNM}) as \textit{defect quasinormal mode frequencies}.

Geometrically, the relevant Lorentzian continuation of the sphere 1-point wormhole describes a traversable wormhole sourced by a timelike membrane obtained by analytically continuing the shell worldvolume \cite{Wang:2025eow}. The responses computed above are toward a transient transverse deformation of the membrane and to a transient change in the coupling $m_0$ of the membrane $m_0\to m_0+\epsilon \mu(t)$. The response to a transverse deformation is exponentially decaying with a relaxation time of $t_{\rm rel}=\frac{2}{m_0}$ and is analogous to an overdamped harmonic oscillator, while the response to a change in coupling is oscillatory with a decaying envelope that determines the relaxation time to be $t_{\rm rel}=\frac{4}{m_0}$, and is analogous to an underdamped harmonic oscillator. 

Since the stiffness kernels are computed using the Liouville line defect, it is natural to interpret these responses also in Liouville CFT. The Euclidean line defect analytically continues to the worldline of a localized impurity. So, the above quantities are used to compute the response of Liouville CFT with a localized impurity that is away from criticality\footnote{The shell-defect sources a relevant perturbation that is localized on a line. So, the corresponding impurity is non-critical and hence exhibits slow relaxation. An impurity at the critical point where it is conformal relaxes instantaneously.} to transient deformations of the impurity. As noted above, relaxation occurs only if the spectrum of the corresponding defect operator is continuous which is true only at infinite spatial volume due to dephasing of the spectral correlations. Since the spectra of the shell-defect operators are discrete for the torus one-point wormhole, its Lorentzian continuation exhibits a persistent, generically quasiperiodic oscillatory response rather than late-time relaxation, as expected for finite-volume linear response governed by real normal-mode poles \cite{Birmingham:2002ph}. This is also transparent in the Liouville impurity picture since the impurity is now placed at finite ambient volume so the deformations do not dissipate away at late times.

The linear response analysis also generalizes to the black hole geometries. The displacement and mass-density operators have continuous spectra for the one-sided black hole. Hence, the one-sided black hole exhibits relaxation. In the dual holographic CFT, since the one-sided black hole is dual to the two-point function of the line defect, the Lorentzian continuation gives a pair of localized impurities at a spatial separation set by the Euclidean evolution time $\tau_0$ of the excited state dual to the black hole. In this paper, we have analyzed the symmetric mode of the deformations where both the impurities are subject to the same transient deformation. Using this setup, we have therefore analyzed the correlated relaxation of a pair of non-critical impurities in a large-$c$ holographic CFT. Interestingly, it turns out that the relaxation time for the two types of deformations is the same and is given by the lowest zero of the sum of the Dirichlet-to-Neumann factors $\lambda_++\lambda_-=0$ that enter the expression for the stiffness kernels, see Eq.~\eqref{eq:reltime}. This is to be contrasted with the sphere 1-point wormhole case where the traversable wormhole has different relaxation times for the two types of deformations. In addition, the relaxation time increases monotonically as the separation between the impurities is increased, as illustrated by the plot in Fig.~\ref{fig:relaxation_time}. This means the system relaxes quicker when there are stronger correlations between the pair of impurities. The two-sided PETS black hole, on the other hand, has a discrete spectrum for the defect operators and hence does not exhibit relaxation. Equivalently, since the two impurities are now placed at finite ambient volume, there is no relaxation in the dual CFT.

\subsection{Imprint on black hole microstate statistics and black hole entropy}

\paragraph{Black hole microstate statistics.}
Semiclassical gravity is intrinsically a coarse-grained description: it does not resolve the
erratic microscopic data of individual black hole states, but it can nevertheless determine
smooth statistical properties of this data.  This viewpoint is closely related to the eigenstate
thermalization hypothesis (ETH), according to which the matrix elements of a sufficiently simple
operator in a chaotic high-energy subsector take the schematic form \cite{Deutsch1991ETH,Srednicki1994ETH}
\begin{equation}
    \langle E_m|\mathcal{O}|E_n\rangle
    =
    \overline{\mathcal{O}}(E)\,\delta_{mn}
    +e^{-S(E)/2}f_{\mathcal{O}}(E,\omega)R_{mn},
    \qquad
    E=\tfrac{E_m+E_n}{2},
    \quad
    \omega=E_m-E_n .
    \label{eq:ETH_introduction}
\end{equation}
Here the first term is smooth, while $R_{mn}$ contains the erratic microscopic information and
has approximately Gaussian random statistics.
Euclidean wormholes are naturally sensitive to precisely these statistical quantities: a
single-boundary saddle determines a smooth variance, while connected multi-boundary
wormholes compute correlations and higher moments of the microscopic data
\cite{Belin:2020hea,Belin:2021ryy,Chandra:2022bqq,Chandra:2024bqz,Chandra:2022fwi, Chandra:2023rhx, Sasieta:2022ksu, deBoer:2023vsm, Collier:2024mgv, deBoer:2024znb, Saad:2019pqd,Chandra:2025fef,Jafferis:2024random, Belin:2026pko}.
For thin-shell black holes, these ideas can be made particularly
explicit because the black-hole-creating operator is a line defect, whose high-energy matrix
elements are governed by a Liouville line defect
\cite{Chandra:2024vhm}.

The two-sided thin-shell black hole studied in section \ref{sec:PETS} prepares a partially entangled thermal state (PETS) \cite{Goel:2018ubv,Sasieta:2022ksu}, and its norm is a thermal two-point function of the line defect $D_\Sigma$ in the holographic 2d CFT sourcing the shell. Semiclassically, it is computed by the on-shell gravitational action, from which we can extract the variance $\overline{|\langle h'|D_\Sigma|h\rangle|^2}$ of the line-defect matrix elements after performing a Legendre transform. The variance, in turn, determines the ETH envelope function $f_{\mathcal{O}}$. In section \ref{sec:PETS_ETH}, we compute how this ETH envelope changes when the shell is made non-spherical due to shape deformations $\xi(x)$, with the result
\begin{align}
 \frac{\overline{\big|\!\mel{ h'}{D_{\Sigma}[\xi]}{h}\!\big|^2}}
               {\overline{\big|\!\mel{ h'}{D_{\Sigma}[0]}{h}\!\big|^2}}
 &= \exp(
 \frac{c\epsilon^2}{6}
 \sum_{n\neq0}
 \left[
 m_0\!\left(r_0^2+n^2\log r_0\right)
 -m_0^2\frac{\lambda_{+,n}\lambda_{-,n}}
                 {\lambda_{+,n}+\lambda_{-,n}}
 \right]\xi_n\xi_{-n}
 +O(\epsilon^3)).
\label{eq:introduction_microstate_variance}
\end{align}
Here $D_\Sigma[\xi]$ is the operator creating the shell whose locus is deformed by $\xi(x)$, and we have compared its variance to the variance of the undeformed shell operator.

Similarly, we also compute how the variance changes to leading order due to mass deformations. In terms of the Fourier modes $\mu_n$, the quadratic correction is found to be
\begin{equation}
    \frac{\overline{|\langle h'|D_{\Sigma}[\mu]|h\rangle|^2}}
               {\overline{|\langle h'|D_\Sigma[0]|h\rangle|^2}}
 =
\exp( \frac{c\epsilon^2}{6}
 \sum_{n\neq0}
 \frac{\mu_n\mu_{-n}}{\lambda_{+,n}+\lambda_{-,n}}
 +O(\epsilon^3)).
\end{equation}
Thus the response of the shape variance contains competing local and non-local geometric contributions, whereas the mass-deformation answer is universally controlled by the inverse sum of the two Dirichlet-to-Neumann eigenvalues.  Since $\lambda_{\pm,n}>0$, every non-zero mass-density mode enhances the variance of the corresponding black hole microstate matrix elements.

\begin{figure}
    \centering

    \begin{subfigure}[t]{0.48\textwidth}
        \centering
        \includegraphics[width=\linewidth]{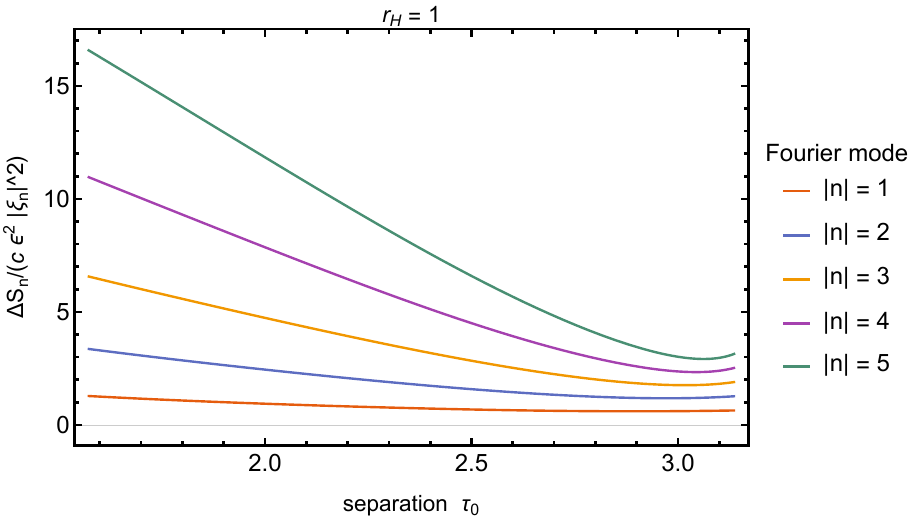}
    \end{subfigure}
    \hfill
    \begin{subfigure}[t]{0.48\textwidth}
        \centering
        \includegraphics[width=\linewidth]{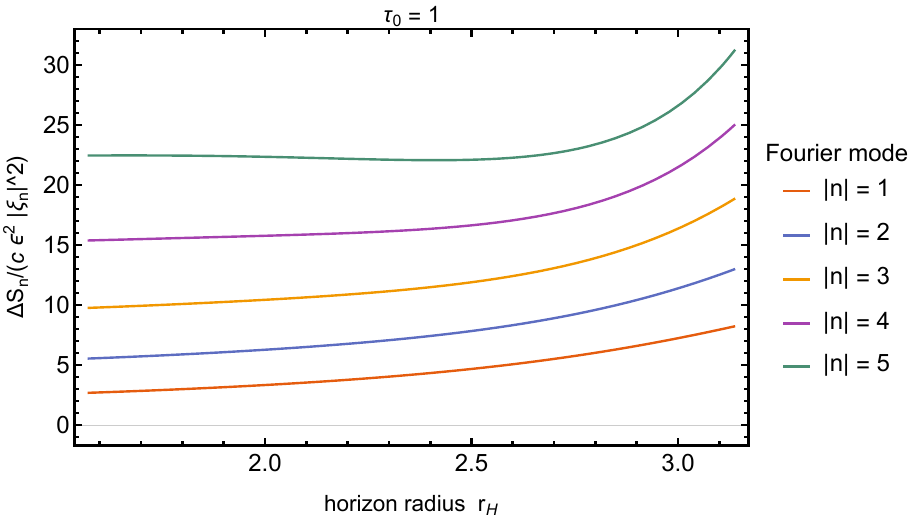}
    \end{subfigure}
    \caption{Plot of the mode corrections to the apparent-horizon entropy $\Delta S_n$, as a function of the separation $\tau_0$ at fixed $r_H$ (left) and as a function of the horizon radius $r_H$ at fixed $\tau_0$ (right). Here \(\Delta S_n\) denotes the contribution of the real \(\pm n\) mode pair, so
\(\Delta S_n/(c\epsilon^2|\xi_n|^2)=\frac{\pi r_H}{3} Q_n^{\rm shape}\) for \(n>0\). All plots are made in the range $\frac{\pi}{2}<r_H\tau_0 <\pi$ where the shell goes behind the horizon on the time-symmetric slice. The plot shows the first 5 non-zero modes.\vspace{2mm}}
    \label{fig:AHentropy}
\end{figure}
\vspace{2cm}
\begin{figure}
    \centering

    \begin{subfigure}[t]{0.48\textwidth}
        \centering
        \includegraphics[width=\linewidth]{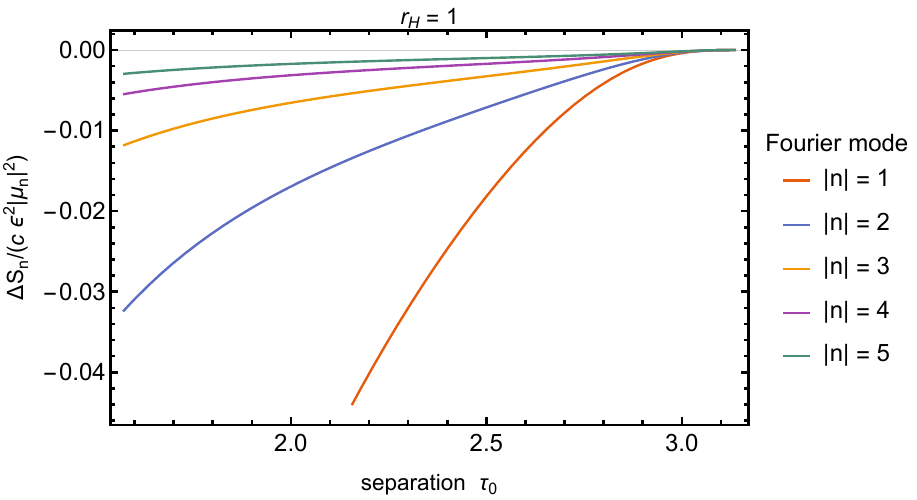}
    \end{subfigure}
    \hfill
    \begin{subfigure}[t]{0.48\textwidth}
        \centering
        \includegraphics[width=\linewidth]{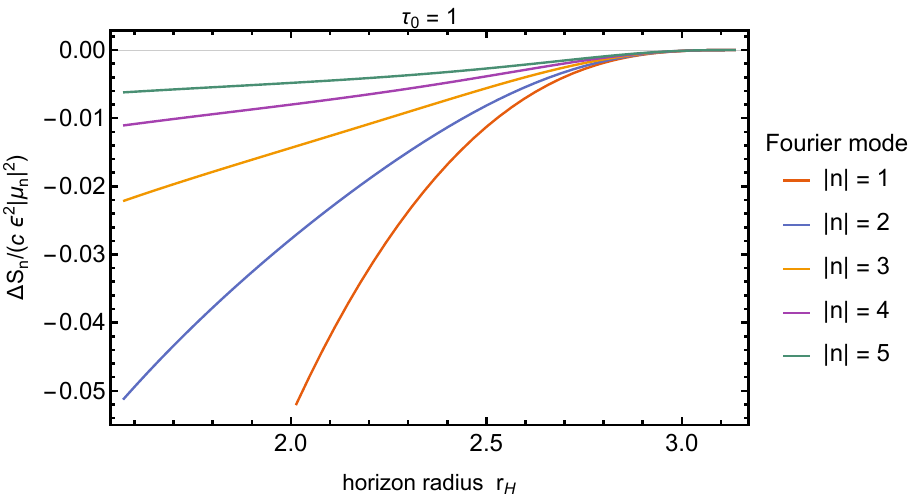}
    \end{subfigure}
    \caption{Mode corrections to the apparent-horizon entropy $\Delta S_n$ due to mass deformations, as a function of the separation $\tau_0$ at fixed $r_H$ (left) and as a function of the horizon radius $r_H$ at fixed $\tau_0$ (right), in the range $\frac{\pi}{2}<r_H\tau_0 <\pi$ where the shell goes behind the horizon on the time-symmetric slice. Here \(\Delta S_n\) denotes the contribution of the real \(\pm n\) mode pair, so
$
\Delta S_n/(c\epsilon^2|\mu_n|^2)=\frac{\pi r_H}{3} Q_n^{\rm mass}
$
for \(n>0\). The plot shows the first 5 non-zero modes.}
    \label{fig:AHentropy_mass}
\end{figure}

\begin{figure}
    \centering
    \includegraphics[width=0.7\linewidth]{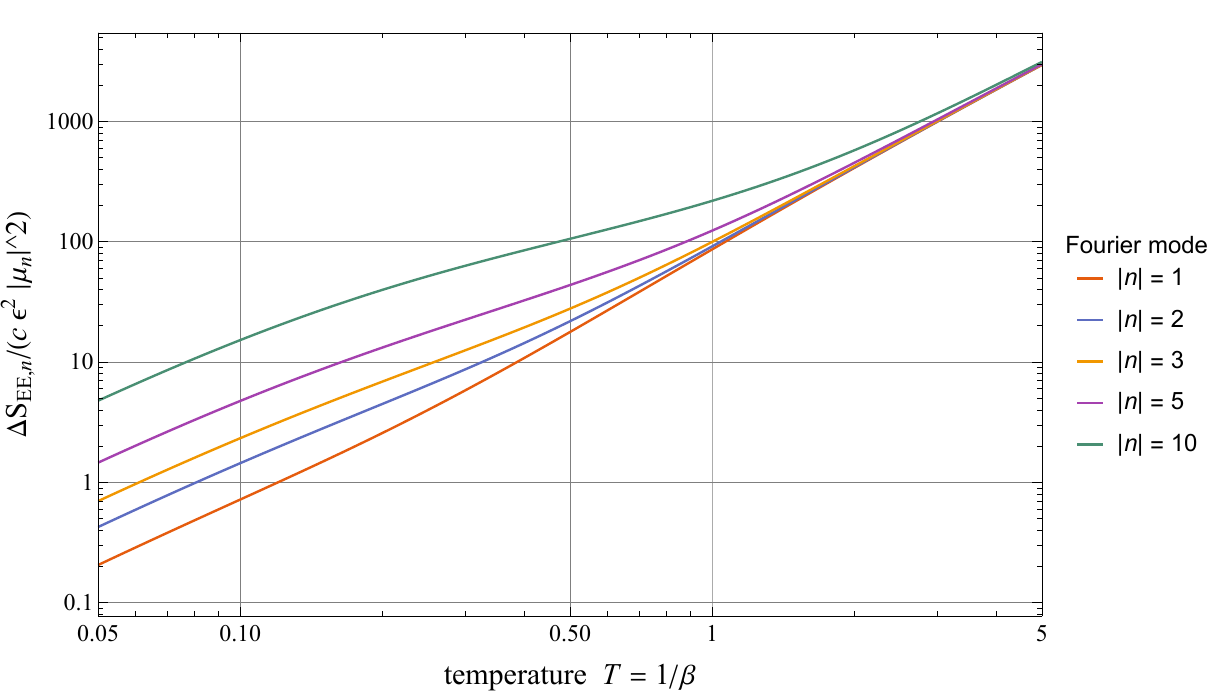}
    \caption{Correction to PETS entanglement entropy induced by shape deformation modes $\xi_n$, plotted on a log-log scale as a function of temperature, for several values of $n$. Here \(\Delta S_{EE,n}\) denotes the contribution from the Fourier pair $\pm n$, so that
\(\Delta S_{EE,n}/(c\epsilon^2|\xi_n|^2)=\pi r_H \mathcal{Q}^{\rm shape}_n/3\).}
    \label{fig:PETSentanglement}
\end{figure}

\begin{figure}
    \centering
    \includegraphics[width=0.7\linewidth]{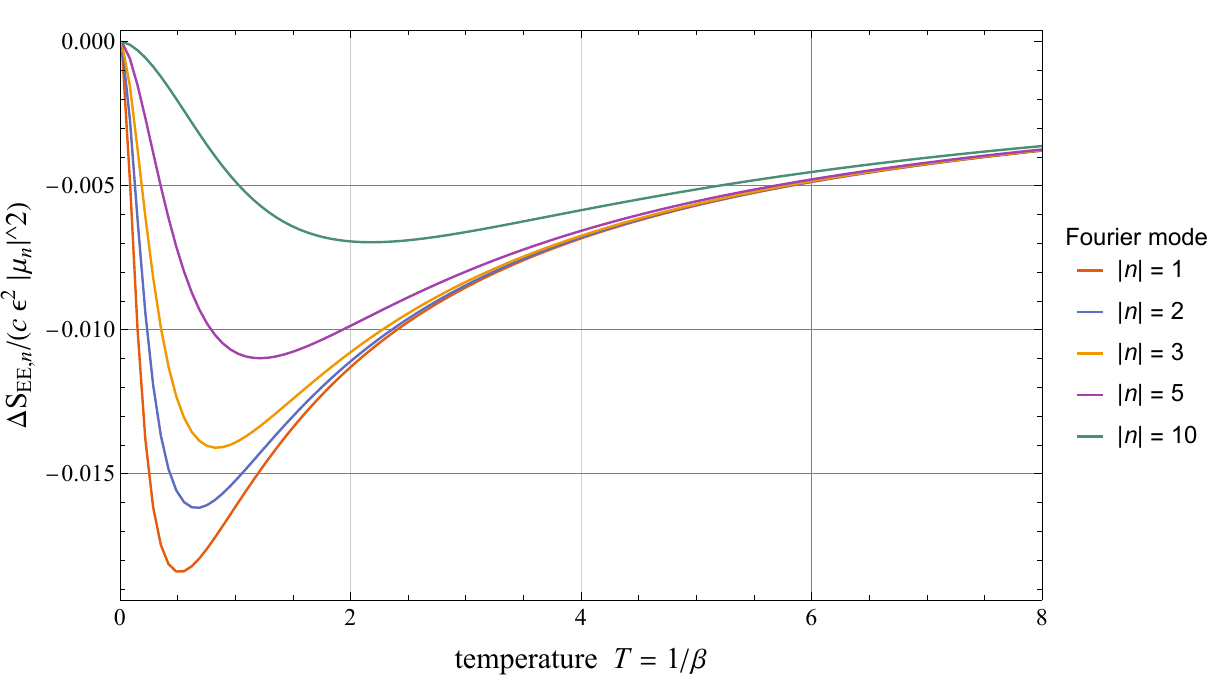}
    \caption{Correction to PETS entanglement entropy induced by mass deformation modes $\mu_n$, plotted as a function of temperature, for several values of $n$. Here \(\Delta S_{EE,n}\) denotes the contribution from the Fourier pair $\pm n$, so that
\(\Delta S_{EE,n}/(c\epsilon^2|\mu_n|^2)=\pi r_H \mathcal{Q}^{\rm mass}_n/3\).}
    \label{fig:PETSentanglement2}
\end{figure}

\paragraph{Black hole entropy.}

The black hole entropy is the most basic example of microscopic information encoded in a smooth
geometrical observable: the Bekenstein-Hawking formula assigns an entropy equal to one-quarter of the horizon area \cite{Bekenstein1973Entropy,Hawking1975Radiation}.  In holography, closely related extremal-area prescriptions compute the entanglement entropy of boundary states \cite{RyuTakayanagi2006,HubenyRangamaniTakayanagi2007}.  For a one-sided black hole prepared as a pure state in the dual CFT, the area of the time-symmetric apparent horizon instead computes a natural coarse-grained entropy obtained by discarding microscopic phase correlations \cite{Chandra:2022fwi}.  For a two-sided partially entangled thermal state,
the minimal surface computes the coarse-grained entanglement between the two CFT copies
\cite{Goel:2018ubv, Chandra:2023rhx}.  More generally, wormhole overlaps between semiclassical states provide
a mechanism by which the dimension of the resulting black hole Hilbert space is reduced to
$e^{S_{\mathrm{BH}}}$
\cite{Penington:2019kki,Balasubramanian:2022gmo,Balasubramanian:2025zey}.

In sections \ref{sec:one-sided} and \ref{sec:PETS},
we compute how these area-law entropies respond to elastic deformations of
the shell.  Both the one-sided apparent-horizon entropy and, at the symmetric point
$\tau_0=\beta/4$, the two-sided PETS entanglement entropy take the mode-space form
\begin{equation}
    S[\alpha]
    =
    \frac{2\pi r_H}{4G_N}
    +\epsilon^2\frac{2\pi r_H}{4G_N}
      \sum_{n>0}\mathcal{Q}_n^{\alpha}|\alpha_n|^2
    +O(\epsilon^3),
    \qquad
    \alpha=\xi,\mu ,
    \label{eq:introduction_entropy_correction}
\end{equation}
where the sum over $n>0$ denotes a real pair of Fourier modes $\pm n$.
For the one-sided black hole with the shell going behind the horizon on the time-symmetric slice, the mode-by-mode evaluation gives $\mathcal{Q}_n^{\rm shape}>0$ and $\mathcal{Q}_n^{\rm mass}<0$: shape fluctuations increase the apparent-horizon entropy, while redistributing the shell mass at fixed total mass decreases it. This is illustrated by the plots in figures \ref{fig:AHentropy} and \ref{fig:AHentropy_mass} respectively. The symmetric PETS black hole displays the same qualitative distinction: shape deformations give a positive correction to the entanglement entropy as illustrated by the plot in figure \ref{fig:PETSentanglement}, whereas mass-density deformations give a negative correction, as shown in figure \ref{fig:PETSentanglement2}. 
Thus, shape and mass deformations of the shell have opposite effects on the amount of coarse-grained information hidden behind the horizon.

\section{The sphere 1-point wormhole}\label{sec:2}

\begin{figure}
    \centering
    \includegraphics[width=0.5\linewidth]{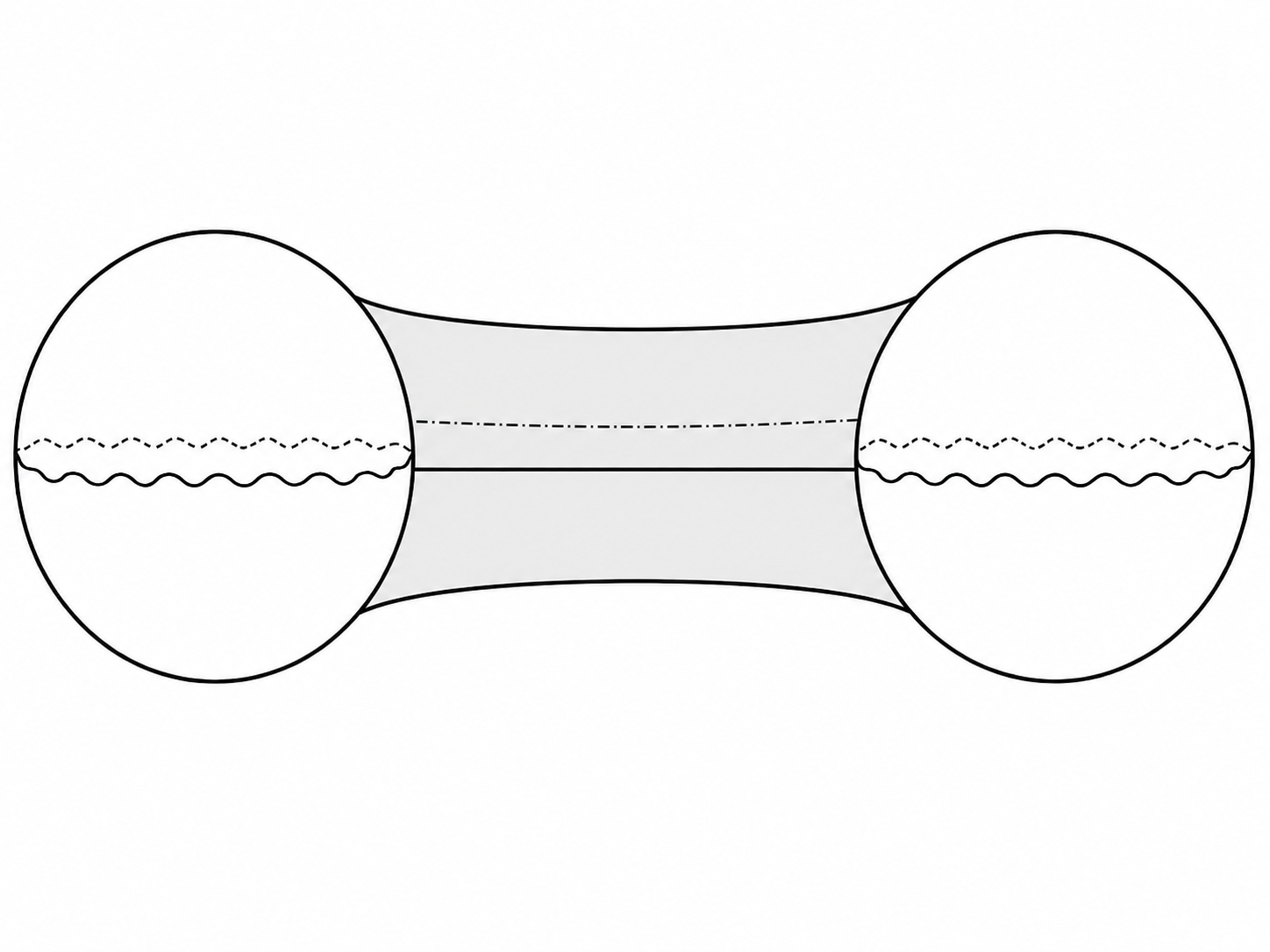}
    \caption{The figure shows the sphere 1-point wormhole used to compute the stiffness kernel for shape deformations of the shell. Although for illustration, we have drawn the boundaries in this figure in the sphere conformal frame, the calculations of the stiffness kernels in this paper are mostly in the cylinder frame.}
    \label{fig:1pointWH}
\end{figure}

The first example that we study is the sphere 1-point wormhole, which is the simplest setup where the stiffness kernels and their properties can be analyzed explicitly. In hyperbolic slicing coordinates, the metric on the wormhole is given by
\begin{equation} \label{eq:MMmetric}
    \dd s^2=\dd\rho^2+\cosh^2(\rho)e^{\Phi(x,y)}(\dd x^2+\dd y^2)\,.
\end{equation}
Here $\rho \in \mathbb{R}$ is the radial coordinate along the wormhole and $\Phi(x,y)$ is the Liouville field governing the hyperbolic metric on each transverse slice of the wormhole. The background saddle with a cylindrical shell extending between the equators of the two boundary spheres was constructed in \cite{Chandra:2024vhm}, with the Liouville field given by
\begin{equation}
    \Phi_0(y)=-2\log (\sinh(|y|+A_0)), \qquad A_0=\frac{1}{2}\log\left(\frac{m_0+2}{m_0-2}\right).
\end{equation}
We are working in the cylinder conformal frame, with $x\sim x+2\pi$ and $y \in \mathbb{R}$. Here, $m_0$ is the mass of the shell and the radius of the cylinder has been set to $1$. In these coordinates, the shell is placed along the circle at $y=0$. Given this spherically symmetric solution $\Phi_0(y)$, our goal is to compute to the response of the gravitational partition functions to deformations of $\Phi(x,y)$ that break spherical symmetry. The set-up is illustrated in figure \ref{fig:1pointWH}.

To compute the stiffness kernels, the general plan is to analyze the effect of a small deformation of the shell by solving the linearized Liouville equation, as well as the linearized junction conditions, and then compute the quadratic on-shell action associated with these deformations. When we deform the boundary conditions for the shell symmetrically on the two boundaries of the wormhole, the metric of the form (\ref{eq:MMmetric}) still solves the Einstein's equations but now the perturbed Liouville field $\Phi(x,y)$ is no longer axisymmetrical. More generally, one could deform the shell on the two asymptotic boundaries by different functions, in which case the metric (\ref{eq:MMmetric}) would no longer be the right ansatz. Instead, one needs to use an ``almost-Fuchsian" metric ansatz like the one used in \cite{Krasnov:2005dm, Chandra:2022bqq} where the deformation of the shell now smoothly interpolates between the two boundary values through the bulk. Leaving the analysis of these more general wormholes to future work, we give some preliminary comments on the set-up in the discussion.

Let us now set up the analysis of the linearized Liouville equations in order to compute the stiffness kernels.
We parametrize the linearized Liouville solution around the background $\Phi_0$ as
\begin{equation}
    \Phi_\pm(x,y) = \Phi_0(y)+\epsilon\hspace{0.2mm} \varphi_\pm(x,y),
\end{equation}
where $\Phi_0$ is the background solution and $\epsilon$ is a deformation parameter. Here, the subscripts $\pm$ denote the Liouville solution above and below the deformed locus which we write as $\Sigma_\epsilon: y=\epsilon \xi(x)$. The linearized Liouville equation takes the form,
\begin{equation}
    (\partial_x^2+\partial_y^2)\varphi_\pm=2e^{\Phi_0(y)}\varphi_\pm.
\end{equation}
Using the background solution, we get the following PDEs for the fluctuation fields,
\begin{equation}
    \begin{split}
        &  (\partial_x^2+\partial_y^2-2\csch^2(y+A_0))\varphi_+(y)=0 ,\qquad y>0,\\
        &  (\partial_x^2+\partial_y^2-2\csch^2(A_0-y))\varphi_-(y)=0 ,\qquad y<0.
    \end{split}
\end{equation}
We can readily solve the above ODEs. For the $y>0$ ODE, it is convenient to introduce a shifted coordinate $u=A_0+y$. Going to Fourier space, the ODE then takes the form
\begin{equation}
    f''_n(u)-(n^2+2\csch^2(u))f_n(u)=0.
\end{equation}
A solution of this ODE which decays for $n\neq 0$ as $y\to \infty$ is 
\begin{equation}
    f_n(u)=e^{-|n|u}(|n|+\coth(u)).
\end{equation}
Hence, the general solution which decays at infinity for the linearized Liouville field is given by
\begin{equation}
\begin{split}
    & \varphi_+(x,y)=\sum_{n\in \mathbb{Z}}a_n e^{inx}e^{-|n|y}\left(|n|+\coth(A_0+y)\right), \qquad y>0,\\
    & \varphi_-(x,y)=\sum_{n\in \mathbb{Z}}b_n e^{inx}e^{|n|y}\left(|n|+\coth(A_0-y)\right), \qquad y<0.
  \end{split}  
\end{equation}
We can solve for the mode coefficients $a_n$ and $b_n$ using the linearized junction conditions. So far, the above analysis applied to both the shape and mass deformations; but they differ in the junction conditions, to which we turn next.

\subsection{Stiffness towards shape deformations}

To study small shape deformations, we parametrize the shell locus by $y=\epsilon \xi(x)$, where $\epsilon$ is a small parameter and $\xi(x)$ is the function governing the size of the deformation at each point on the background locus. The continuity condition at the shell junction is 
\begin{equation}
    \Phi_+(x,\epsilon \xi(x))=\Phi_-(x,\epsilon \xi(x)).
\end{equation}
Expanding to linear order in $\epsilon$, we find
\begin{equation} \label{eq:sphere1pointLmodes}
    \varphi_+(x,0)-\varphi_-(x,0) =\xi(x)(\Phi_{0,-}'(0)-\Phi_{0,+}'(0))  = 2m_0 \xi(x).
\end{equation}
Now we turn to the jump in the normal derivative. We will show that the jump is unaffected at linear order by the shape deformation. Firstly, the unit-normalized normal vector to the shell locus is given by
\begin{equation}
    \Hat{n}=\frac{1}{\sqrt{1+\epsilon^2\xi^2(x)}}(-\epsilon \xi'(x),1)=(-\epsilon \xi'(x),1)+O(\epsilon^2).
\end{equation}
The exact jump condition is given by
\begin{equation}
    \Hat{n}\cdot\nabla \Phi_+|_{\Sigma_\epsilon}-\Hat{n}\cdot\nabla \Phi_-|_{\Sigma_\epsilon}=-2m_0.
\end{equation}
Expanding to linear order in $\epsilon$, we get
\begin{equation}
    \partial_y \varphi_+(x,0)-\partial_y \varphi_-(x,0)+\xi(x)(\Phi''_{0,+}(0)-\Phi''_{0,-}(0))=0.
\end{equation}
For the above background solution, the difference in second derivatives vanishes, hence the linearized jump condition on the shell locus takes the form
\begin{equation}\label{eq:linjump}
    \partial_y\varphi_+(x,0)=\partial_y \varphi_-(x,0).
\end{equation}

Having set up the linearized equations, we will now solve them by Fourier decomposition. The deformed locus has a mode expansion,
\begin{equation}
    \xi(x)=\sum_{n \in \mathbb{Z}}\xi_n e^{inx}, \qquad \xi_{-n}=\xi^*_n.
\end{equation}
The latter condition ensures reality. We set the rigid translation mode to zero, $\xi_0 = 0$. Solving for the mode coefficients $a_n$ and $b_n$ in the linearized Liouville fields (\ref{eq:sphere1pointLmodes}) using the linearized junction conditions, we obtain the linearized solution in closed form,
\begin{equation}\label{eq:linsolwh}
    \varphi_\pm(x,y)=\pm 2m_0\sum_{n \in \mathbb{Z}} \frac{e^{in x}e^{-|n||y|}(|n|+\coth(A_0+|y|))}{2|n|+m_0}\xi_n\,.
\end{equation}

\paragraph{Wiggle stress energy and zero modes.}

Before we use the above linearized solution to compute the action for the shape deformations, we would like to analyze the stress energy associated with the shape deformation wiggles parametrized by $\xi$. Using the convention
\begin{equation}
    T_{zz}=\frac{1}{8\pi}\left[(\partial_z \Phi)^2-2\partial_z^2 \Phi\right]
\end{equation}
for the holomorphic stress tensor, we can compute its fluctuations just above and just below the shell locus, by substituting the linearized solutions \eqref{eq:linsolwh}:
\begin{equation}
\begin{split}
     \delta T_{zz}(z)=\frac{m_0}{2\pi}\sum_{n>0}\frac{n(n^2-1)}{m_0+2n}\xi_n e^{inz},\qquad & y>0\\
     \delta T_{zz}(z)=-\frac{m_0}{2\pi}\sum_{n<0}\frac{|n|(n^2-1)}{m_0-2n}\xi_n e^{inz}, \qquad & y<0\,. 
    \end{split}
\end{equation}
Similarly, the anti-holomorphic components of the wiggle stress tensor give
\begin{equation}
\begin{split}
     \delta T_{\overline{z}\overline{z}}(\overline{z})=\frac{m_0}{2\pi}\sum_{n<0}\frac{|n|(n^2-1)}{m_0-2n}\xi_n e^{in\overline{z}},\qquad & y>0\\
     \delta T_{\overline{z}\overline{z}}(\overline{z})=-\frac{m_0}{2\pi}\sum_{n>0}\frac{n(n^2-1)}{m_0+2n}\xi_n e^{in\overline{z}}, \qquad & y<0\,. 
    \end{split}
\end{equation}
Notice that the stress tensors only have support for $n>0$ or $n<0$ on the two halves of the cylinder. This is expected based on the decay of the wiggle modes of the shell into the two halves of the cylinder.
So, the jump in the wiggle stress tensors across the shell is given by
\begin{equation}
\begin{split}
    \delta(T^+_{zz}- T^-_{zz})(x)= \delta( T^+_{\overline{z}\overline{z}}-T^-_{\overline{z}\overline{z}})(x)=\frac{m_0}{2\pi}\sum_{n \in \mathbb{Z}}\frac{|n|(n^2-1)}{m_0+2|n|}\xi_n e^{inx} \,.
 \end{split}
\end{equation}
Note that the jump is even under chirality reversal $n\to -n$. Also note that the jump in the holomorphic and antiholomorphic components is equal, as there is no net longitudinal momentum flux along the shell. This is expected since the deformation that we introduced is transverse to the shell.

Interestingly, the above jump takes a universal form in the heavy shell limit $m_0 \to \infty$,
\begin{equation}
    \delta(T^+_{zz,n}- T^-_{zz,n})=\delta( T^+_{\overline{z}\overline{z},n}-T^-_{\overline{z}\overline{z},n})=\frac{1}{2\pi}|n|(n^2-1)\xi_n\,.
\end{equation}
In \ref{sec:conformalweldingcylinder}, we interpret the above jump in terms of a linearized Schwarzian response \cite{Maldacena:2016upp,Stanford:2017thb} to a conformal welding problem.
We see that the jump in the holomorphic or anti-holomorphic components of the stress tensor has three zero modes $n=0,\pm 1$. We expect three zero modes because the stabilizer subgroup of the equatorial circle on the sphere is a PSL$(2,\mathbb{R})$ subgroup of the global conformal group PSL$(2,\mathbb{C})$. The action of the conformal group on the unit circle can be derived by restricting an infinitesimal Mobius transformation $v(z)=\alpha+\beta z+ \gamma z^2$ to the circle $z=e^{i\theta}$, giving $\eta(\theta)=\text{Re}(e^{-i\theta}v(e^{i\theta}))$. So the three harmonics $\eta(\theta)=\{1,\cos(\theta),\sin(\theta)\}$ correspond to the action of infinitesimal conformal transformations on the shell.

\paragraph{The stiffness kernel.}

We now compute the quadratic kernel for the shape deformations by evaluating the on-shell action for the linearized solution that we obtained above. In Appendix \ref{sec:appA}, we derive the quadratic effective action for shape deformations of the shell, finding
\begin{align}
    S_2[\xi]
    =
    -\frac{m_0}{8\pi}
    \int_0^{2\pi}\dd x\,\left(\xi(x) \partial_y\varphi(x,0)
    +\Phi''_0(0)\xi(x)^2+\Phi_0(0)\xi'(x)^2
    \right).
\end{align}
Here $\partial_y\varphi$ is the normal derivative of the linearized solution $\varphi_\pm$ (we can use either $\pm$ sign, since they agree on the shell locus by \eqref{eq:linjump}).
Using the linearized solution that we found in \eqref{eq:linsolwh} and converting to Fourier space, we find
the quadratic action is
\begin{equation}
    S_2[\xi]
    =-
    \frac{1}{2}\sum_{n\in\mathbb Z}
    \mathcal{K}^{{\rm shape}}_n
    \xi_n\xi_{-n},
\end{equation}
\begin{equation} \label{eq:cylkernelsphere1point}
    \mathcal{K}^{{\rm shape}}_n
    =
    -\frac{m_0 |n|(m_0|n|+2)}{(m_0+2|n|)}
    +
    \frac{m_0 n^2}{2}
    \log\left(\frac{m_0^2-4}{4}\right).
\end{equation}
We observe that the kernel has only one zero mode $n=0$ corresponding to rigid translations along the cylinder. The other two PSL($2,\mathbb{R}$) modes $n=\pm 1$ are not zero modes of the kernel because they are not isometries of the flat cylinder, rather they are conformal isometries. Since the line defect dual to the shell is not conformal, its dynamics are not expected to be invariant under conformal isometries, hence the kernel exhibits only one zero mode. 

Another consequence of the non-conformality of the shell defect is that the stiffness kernel is frame dependent. The expression derived above is the stiffness kernel in the flat cylinder frame. Two other natural choices of the conformal frame are the round sphere and the flat plane. The kernel evaluated in these frames is given by\footnote{For the flat-plane frame, we use the standard support-function parametrization of a convex plane curve \cite[Sec.~1.7]{Schneider:2014convex}. If \(n(\alpha)=(\cos\alpha,\sin\alpha)\) and \(t(\alpha)=(-\sin\alpha,\cos\alpha)\) are respectively the normal and tangential unit vectors, a curve with support function \(p(\alpha)\) is reconstructed as \(X(\alpha)=p(\alpha)n(\alpha)+p'(\alpha)t(\alpha)\).  For deformations of the unit circle we write \(p(\alpha)=1+\epsilon u(\alpha)\), with \(u(\alpha)=\sum_{n\in\mathbb Z}u_n e^{in\alpha}\), and use the Fourier modes \(u_n\) to compute the kernel. The advantage of using the support-function parametrization to compute the kernel is that the two translation zero modes are transparent and realized as $n=\pm 1$ modes of $u(\alpha)$.}
\begin{equation} 
    \mathcal{K}_n^{\rm shape}=\begin{cases}
         -\frac{m_0 |n|(m_0|n|+2)}{(m_0+2|n|)}
    +
    \frac{m_0 n^2}{2}
    \log\left(\frac{m_0^2-4}{4}\right), \qquad & \text{flat cylinder},\\
        -m_0 (n^2-1)\left(\frac{m_0}{(m_0+2|n|)}-\frac{1}{2}\log\left(\frac{m_0^2-4}{4}\right)\right), \qquad & \text{round sphere},\\
        -\frac{m_0(n^2-1)P_{|n|}(m_0)}{2(m_0+2|n|)(4n^2+m_0^2-4+2|n|m_0)}, \qquad & \text{flat plane}\,.
    \end{cases}
\end{equation}
where the polynomial appearing in the plane kernel is given by
\begin{equation}
    P_k(x)=x^4-4x^2+8x+4k(x^3-x^2-2x+8)+8k^2x(x-1)+8k^3(x-2)\,.
\end{equation}
The zero modes appearing in the kernels have a clear explanation in terms of isometries. On the round sphere, the shell is placed along the equator and deformed away from it. There are two zero modes $n=\pm 1$ corresponding to the two rotational isometries. Note that unlike the kernel in the cylinder frame, $n=0$ is not a zero mode because translating the shell to a different latitude is not an isometry of the round metric on the sphere. On the flat plane, the shell is placed along the unit circle and deformed away from it. The two zero modes $n=\pm 1$ correspond to translations in the $x$ and $y$ directions.

The stiffness kernel in position space measures correlations between the shape deformations at two points on the shell. After reinstating the normalization factor of $\frac{c}{3}$ (where $c$ is the Brown-Henneaux central charge $c=\frac{3}{2G_N}$) which arises from the relation between the gravitational on-shell action of the wormhole and the Liouville action (\ref{eq:Liouvactiondefect}), $I_{\rm WH}=\frac{c}{3}S_L$, the stiffness kernel in position space is given in terms of the Fourier components computed above by
\begin{equation}
    \mathcal{K}^{\rm shape}_{\rm WH}(x,x')=\frac{c}{3}\frac{1}{2\pi}\sum_{n \in \mathbb{Z}}\mathcal{K}^{\rm shape}_n e^{in (x-x')}\,.
\end{equation}
The short-distance behaviour in position space has a universal non-local piece which is the same in the three frames,
\begin{equation} \label{eq:univsepsing}
    \mathcal{K}^{\rm shape}_{\rm WH}(x,x') \sim \frac{c m_0^2}{6\pi(x-x')^2}+\text{contact terms}, \qquad |x-x'|\to 0\,.
\end{equation}
so the short-distance singularity is governed by both the central charge $c$ and the shell mass $m_0$.

\subsection{Relation to conformal welding and the vacuum Virasoro coadjoint orbit} \label{sec:conformalweldingcylinder}

In this subsection, we comment on the relation between our analysis of shape deformations of the thin shell and the mathematical problem of conformal welding of two disks across a Jordan curve on the sphere \cite{Bers:1960Simultaneous, Bers:1960BoundedDomains}. Namely, the shape deformation of the shell locus $\xi(x)$ determines an infinitesimal conformal welding datum. The linearized Liouville solution is the unique hyperbolic metric response to the welding datum subject to the imposed junction conditions.

\begin{figure}
    \centering
    \includegraphics[width=0.8\linewidth]{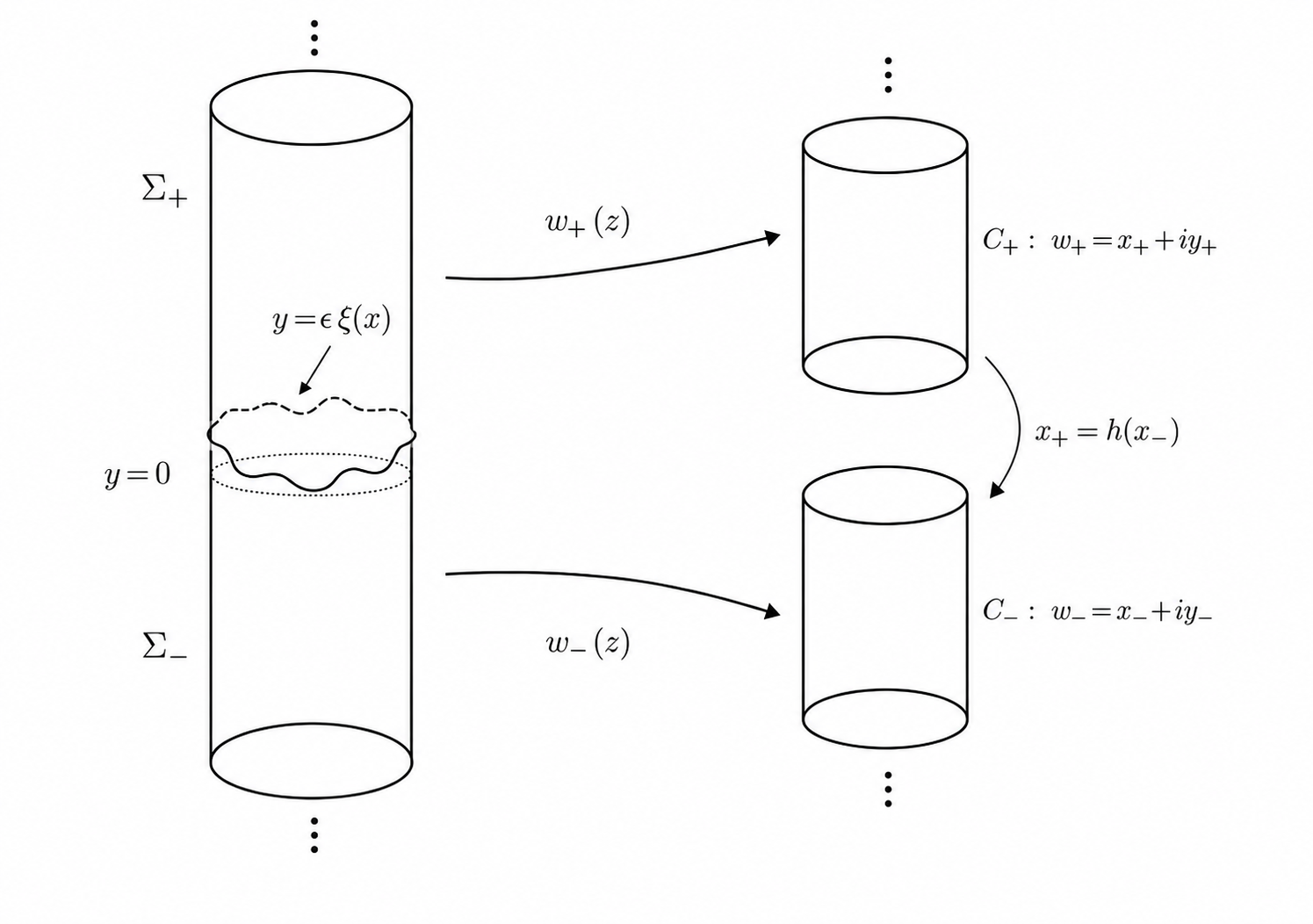}
    \caption{This figure describes the conformal welding across the deformed shell locus $y=\epsilon \xi(x)$ in the cylinder frame. The conformal maps $w_\pm :\Sigma_\pm \to C_\pm$ from the two halves of the cylinder are constructed such that they solve the welding condition $h(x_-)=(w_+\circ w_-^{-1})(x_-)$ given the welding diffeomorphism $h$, which is expressible in terms of the deformed locus by a Hilbert transform.}
    \label{fig:confweldcyl}
\end{figure}

To make this connection more precise, it is convenient to go to the ``welding-gauge'' where the two semi-infinite cylinders, $(x_+,y_+)$ and $(x_-,y_-)$ are glued across their boundary circles $y_+=y_-=0$ with a diffeomorphism $h: S^1 \to S^1$ relating the coordinates on the circles, $x_+=h(x_-)$. This is the standard welding datum in the Bers description of Teichm\"uller space \cite{Bers:1960Simultaneous, Bers:1960BoundedDomains}. We take $h \in \text{Diff}^+(S^1)$ so it satisfies the winding condition $h(x+2\pi)=h(x)+2\pi$ and is orientation preserving, $h'(x)>0$. Going to the welding gauge amounts to straightening the deformed locus $y=\epsilon \xi(x)$ to $y_+=y_-=0$, as illustrated by figure \ref{fig:confweldcyl}. To this end, define the conformal maps,
\begin{equation}
    w_\pm =z + \epsilon \alpha_\pm (z)+O(\epsilon^2), \qquad w_\pm =x_\pm +iy_\pm\,.
\end{equation}
Here, $\epsilon\alpha_\pm(z)$ are the infinitesimal conformal maps that solve the welding problem. These maps are subject to $\text{Im}(w_\pm) |_{z=x+i\epsilon \xi(x)}=0$, which implies 
\begin{equation}
\begin{split}
         \alpha_+(z)= -i\xi_0-2i\sum_{n>0} \xi_n e^{inz}, \qquad & y>0,\\
         \alpha_-(z)= -i\xi_0-2i\sum_{n<0} \xi_n e^{inz}, \qquad & y<0\,. 
\end{split}
\end{equation}
The chiral splitting of the above welding maps is natural since the modes $\xi_n$ of positive frequency $(n>0)$ decay into the top cylinder while the modes of negative frequency $(n<0)$ decay into the bottom cylinder.
Restricting the above conformal maps to the boundaries of the semi-infinite cylinders, we have
\begin{equation}
    x_+=x_-+2\epsilon H\xi(x_-)+O(\epsilon^2), \qquad H e^{inx}=-i\text{sgn}(n)e^{inx}\,,
\end{equation}
where $H$ is the Hilbert transform on the circle.
We can thus relate the shape deformation function $\xi(x)$ to the welding diffeomorphism $h$,
\begin{equation}
    h_\epsilon(x)=x+\epsilon \psi(x)+O(\epsilon^2), \qquad \psi(x)=2H \xi(x)\,.
\end{equation}
Therefore, we have solved the conformal welding problem associated with infinitesimal shape deformations of the shell locus. 

The linearized Liouville solutions can now be viewed as hyperbolic metric responses to the conformal welding, and can be expressed in terms of the welding diffeomorphism $\psi(x)$ as\footnote{Since the Hilbert transform $H$ annihilates the zero mode, this expression as written is applicable only for the non-zero modes of the linearized Liouville field. But since we set the zero mode $\xi_0$ which corresponds to rigid translations along the cylinder to zero in our setup, the expression is correct as written.}
\begin{equation}
    \varphi_\pm (x,y)=\mp m_0\sum_{n \in \mathbb{Z}}e^{inx}e^{-|n||y|}\frac{|n|+\coth(A_0+|y|)}{2|n|+m_0}(H\psi)_n\,.
\end{equation}
The jump in the holomorphic stress tensor can also be expressed in terms of $\psi(x)$ as 
\begin{equation} \label{eq:stressjumphypresp}
    \delta(T^+_{zz,n}- T^-_{zz,n})=-\frac{m_0}{4\pi(m_0+2|n|)}\left[(\partial_x^3+\partial_x)\psi(x)\right]_n \,.
\end{equation}
Notice that $\partial_x^3+\partial_x$ is the Schwarzian operator. In the heavy shell limit, $m_0\to \infty$, the above response reduces to the universal Schwarzian response to conformal welding,
\begin{equation}
     \delta(T^+_{zz}- T^-_{zz})(x) \longrightarrow -\frac{1}{4\pi}(\partial_x^3+\partial_x)\psi(x), \qquad m_0\to \infty \,.
\end{equation}
The same Schwarzian structure is visible directly in the stiffness kernel once local contact terms are separated. Indeed, the cylinder-frame kernel in (\ref{eq:cylkernelsphere1point}) can be rearranged exactly as
\begin{equation}
    \mathcal{K}^{\rm shape}_n
=
m_0 n^2
\left[
\frac12\log\left(\frac{m_0^2-4}{4}\right)-1
\right]
+
\frac{2m_0}{m_0+2|n|}
|n|(n^2-1).
\end{equation}
The first term is local in position space, so the separated-point part of the stiffness kernel is
\begin{equation}
\mathcal{K}^{\rm shape}_{n,\mathrm{sep}}
=
\frac{2m_0}{m_0+2|n|}
|n|(n^2-1).
\end{equation}
Using \(\psi=2H\xi\), we obtain
\begin{equation} \label{eq:shapestiffhypresp}
\bigl(\mathcal{K}^{\rm shape}_{\mathrm{sep}}\xi\bigr)_n
=
-\frac{m_0}{m_0+2|n|}
\bigl[(\partial_x^3+\partial_x)\psi\bigr]_n.
\end{equation}
Comparing with (\ref{eq:stressjumphypresp}), we see that
\begin{equation}
\bigl(\mathcal{K}^{\rm shape}_{\mathrm{sep}}\xi\bigr)_n
=
4\pi\,\delta(T^+_{zz,n}-T^-_{zz,n}),
\end{equation}
so the same non-local response factor dresses both the stiffness kernel and the jump in the stress tensor. Equivalently, in position space, we can express the jump in the stress tensor as the following convolution,
\begin{equation}
    \int_0^{2\pi}\dd x' \mathcal{K}^{\rm shape}_{\rm sep}(x,x') \xi(x')= 4\pi \delta( T^+_{zz}-T^-_{zz})(x)= 4\pi \delta( T^+_{\overline{z}\overline{z}}-T^-_{\overline{z}\overline{z}})(x)\,.
\end{equation}
This relation also makes it clear that the jump in the stress tensor is the linearized response to shape deformations.

Now, we interpret the hyperbolic response computed in (\ref{eq:stressjumphypresp}) or (\ref{eq:shapestiffhypresp}) in terms of the Virasoro coadjoint action on the vacuum orbit.
The kernel of the differential operator $\partial_x^3+\partial_x$ is
\begin{equation}
    (\partial_x^3+\partial_x)\psi(x)=0 \implies \psi(x)=a + b \cos(x)+ c \sin(x)\,.
\end{equation}
The three zero modes generate a PSL$(2,\mathbb{R})$ subgroup, hence we can view $\psi(x)$ as an element in the tangent space of the identity element of $\text{Diff}^+(S^1)/\text{PSL}(2,\mathbb{R})$ which is the vacuum Virasoro coadjoint orbit, denoted $E_1$ in the Hill-equation/monodromy classification \cite{Witten:1988hf, Balog:1997zz}. The jump in the holomorphic stress tensor, being a quadratic differential, is the representative of the welding diffeomorphism in the cotangent space of the vacuum orbit. In the $m_0\to \infty$ limit, the jump in the stress tensor is literally the infinitesimal Virasoro coadjoint action on the vacuum orbit, while at finite $m_0$, the coadjoint action is dressed by a non-local response factor $\frac{m_0}{m_0+2|n|}$ determined by solving the junction conditions. 

Note that in the Bers description of conformal welding, the Schwarzian derivatives of the welding maps define holomorphic quadratic differentials on the two halves of the cylinder \cite{Bers:1960BoundedDomains}. In the present setup, the stress tensors $T^\pm_{zz}$ are Bers-like quadratic differentials since they are dressed by the non-local response factor for finite $m_0$;  in the $m_0\to \infty$ limit, they become the usual Bers differentials. These Bers differentials carry an embedding of the Teichm\"uller space element---namely the wiggle function $\xi(x)$ or equivalently the welding diffeomorphism $\psi(x)$---into the space of holomorphic quadratic differentials. The form of the stress tensors nicely illustrates this embedding in our setup.

Now, we make some comments on the hyperbolic response beyond linear order. We conjecture that in the heavy-shell limit, the full non-linear response to conformal welding is governed by the Schwarzian action or equivalently, the coadjoint action on the vacuum orbit \cite{Witten:1988hf, Alekseev:1989ce}. That is, we conjecture that the jump in the stress tensor in the welding gauge, written as a functional of the welding diffeomorphism $h(x)$, is given by
\begin{equation}
\begin{split}
    \Delta T_{ww}[h](x)= & -\frac{1}{4\pi}\left(\{e^{ih(x)},x\}-\{e^{ix},x\}\right ),\\
    =& -\frac{1}{4\pi}\left(\{h(x),x\}+\frac{1}{2}((h')^2-1)\right)\,.
    \end{split}
\end{equation}
Here $\{f(x),x\}\equiv \frac{f'''}{f'}-\frac{3}{2}\left(\frac{f''}{f'}\right)^2$ is  the Schwarzian derivative. For a shell of finite mass $m_0$, we conjecture that the above universal response is corrected perturbatively in inverse powers of the mass $m_0$ by non-local response terms. Our observation at linear order that the Schwarzian response is dressed by the non-local response factor $\frac{m_0}{m_0+2|n|}$ is consistent with this conjecture.

\subsection{Spectrum of the displacement operator of the shell}

Next, we interpret the stiffness kernel for shape deformations as the 2-point function of a local operator that is conjugate to transverse displacements of the shell. 
This operator coincides with the displacement operator for the Liouville line defect $L_\Sigma$. So, the stiffness kernel that we computed is the 2-point function of the displacement operator denoted by $\mathcal{D}_\perp(x)$,
\begin{equation}
   \mathcal{K}^{\rm shape}_{\rm WH}(x,x')=\langle \mathcal{D}_{\perp}(x)\mathcal{D}_{\perp}(x')\rangle_{\rm Liouv}.
\end{equation}
In the present setup, we have evaluated the 2-point function on a circle, so we can interpret it as the thermal 2-point function of the displacement operator with the temperature set by the circumference of the shell locus. So far, in our analysis we have set the radius to $1$, but now we reinstate the radius $R$ of the shell locus, $x\sim x+2\pi R$. For separated points i.e., with contact terms ignored, the kernel obeys reflection positivity and hence it admits a spectral decomposition with a positive density. We may set $x'=0$ due to translation invariance. Then, the kernel has the following spectral decomposition,
\begin{equation} \label{eq:spectraldecompshape}
    \mathcal{K}^{\rm shape}_{\rm WH}(x,0)=\int_0^{\infty}\dd \omega \rho_D(\omega;R)\frac{\cosh(\omega(\pi R-x))}{\sinh(\pi R \omega)}\,.
\end{equation}
This decomposition makes the Kubo-Martin-Schwinger (KMS) symmetry of thermal correlation functions manifest \cite{Kubo:1957mj, Martin:1959jp}:
\begin{equation}
   \mathcal{K}^{\rm shape}_{\rm WH}(x,0) = \mathcal{K}^{\rm shape}_{\rm WH}(2\pi R - x,0).
\end{equation}
Here $\rho_D(E;R)$ is the spectral density of the displacement operator expressed in terms of the matrix elements of the displacement operator as
\begin{equation}\label{eq:displacement-spectrum}
    \rho_D( \omega;R)=\frac{1}{Z}\int \dd E \rho(E)\rho( \omega+E)e^{-2\pi R E}(1-e^{-2\pi R  \omega})|\!\mel{ \omega+E}{\mathcal{D}_\perp}{E}\!|^2\,.
\end{equation}
Here, $\rho(E)$ is the density of states of the shell defect Hilbert space and $Z$ is the thermal partition function. Eq.~\eqref{eq:displacement-spectrum} is manifestly positive and could also be referred to a Lehmann density for the displacement correlator. In the $R\to \infty$ limit, the spectral decomposition (\ref{eq:spectraldecompshape}) reduces to the usual Kallen-Lehmann spectral decomposition of the vacuum two-point function.
The integration kernel in (\ref{eq:spectraldecompshape}) is the Boltzmann factor, summed over thermal images of the point $x$,
\begin{equation}
    \frac{\cosh( \omega(\pi R-x))}{\sinh(\pi R  \omega)}=\sum_{k\in \mathbb{Z}}e^{- \omega|x+2\pi R k|}.
\end{equation}
The spectral decomposition of the 2-point function can be visualized geometrically using the figure below where the displacement operators are inserted at the marked points on the undeformed shell locus and expanded using a complete set of states inserted along the dotted infinite line that span the defect Hilbert space for the shell,
\begin{equation} \label{eq:noncompactlocus}
        \vcenter{\hbox{
\begin{overpic}[grid=false, scale=0.12]{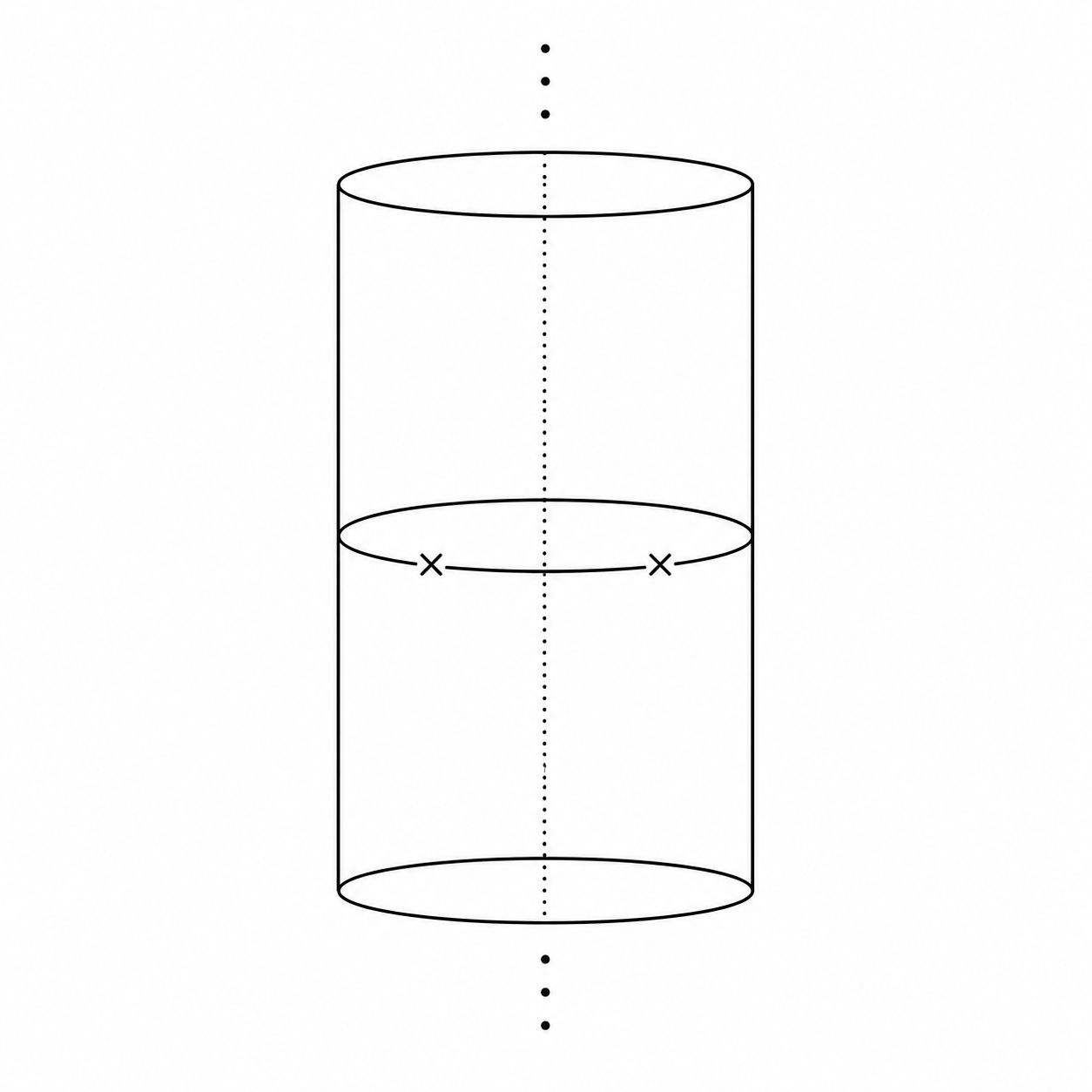}
\end{overpic}}} 
    \end{equation}
Since the Hilbert space is defined on a non-compact slice,  the resulting spectrum is continuous, as we will show below.
Because we have evaluated the kernel in Fourier space, we can take a Fourier transform of the above spectral decomposition to express the Fourier space kernel in Stieltjes form,
\begin{equation}
    \mathcal{K}^{\rm shape}_{\rm WH}(i\omega_n)=\int_0^{\infty}\dd \omega \rho_D( \omega;R) \frac{2 \omega}{ \omega^2+\omega_n^2}+\text{local subtractions}\,.
\end{equation}
The local subtractions correspond to contact terms at $x=0$ in position space, and $\omega_n=\frac{n}{R}$ are the discrete Matsubara frequencies. We can readily extract the spectral density by analytically continuing the stiffness kernel from Euclidean kinematics to Lorentzian kinematics using a retarded prescription written schematically as $i\omega_n \to \omega + i0^+$ and then evaluating its imaginary part. Doing so gives 
\begin{equation}
    \rho_D( \omega;R)=\frac{1}{\pi}\text{Im}( \mathcal{K}^{\rm shape}_{\rm WH, ret}( \omega;R))=\frac{cm_0^2}{6\pi}\frac{ \omega( \omega^2+R^{-2})}{ \omega^2+\frac{m_0^2}{4}}\,.
\end{equation}
At finite \(R\), the low-energy behavior is linear,
\begin{equation}
\rho_D(\omega;R)=\frac{2c}{3\pi R^2}\,\omega+O(\omega^3).
\end{equation}
If the planar limit \(R\to\infty\) is taken first, then
\begin{equation}
\rho_D(\omega;R\to\infty)=\frac{2c}{3\pi}\omega^3+\cdots,
\qquad \omega\ll m_0 .
\end{equation}
This \(\omega^3\) behavior is the long-distance conformal-displacement tail of the correlator which translates to the familiar position space form,
\begin{equation}
    \langle \mathcal{D}_\perp(x) \mathcal{D}_\perp(0) \rangle_{\rm sep} \sim \frac{4c}{\pi x^4}, \qquad |x|\gg m_0^{-1}\,.
\end{equation}
By contrast, the short-distance singularity is controlled by the large-\(\omega\) behavior
\begin{equation}
\rho_D(\omega;R)\sim \frac{c m_0^2}{6\pi}\omega,
\end{equation}
which gives the universal separated-point singularity (\ref{eq:univsepsing}) of the non-conformal shell defect.
It is useful to express the spectral density as
\begin{equation}\label{eq:specdens}
    \rho_D( \omega;R)=\frac{cm_0^2  \omega}{6\pi}+\frac{c m_0^2}{6\pi}\left(R^{-2}-\frac{m_0^2}{4}\right)\frac{ \omega}{ \omega^2+\frac{m_0^2}{4}}\,.
\end{equation}
The first term which is linearly divergent at high energies contributes to the OPE singularity in position space. Now, we use the spectral density in the planar limit $R\to \infty$ to compute the retarded Green's function, 
\begin{equation}
\begin{split}
     \mathcal{K}^{\rm shape}_{\rm WH, ret}(t)=& -i\Theta(t)\langle [\mathcal{D}_{\perp}(t),\mathcal{D}_{\perp}(0)]\rangle \\
    =& -2\Theta(t)\int_0^{\infty}\dd \omega \rho_D( \omega)\sin( \omega t)\,.
    \end{split}
\end{equation}
The UV-singularity piece of the spectral density gives
\begin{equation}
\begin{split}
      \mathcal{K}^{\rm shape}_{\rm WH, ret}(t;R)\supset & -\frac{cm_0^2}{3\pi}\lim_{\epsilon \to 0^+}\int_0^{\infty}\dd \omega \, \omega e^{-\epsilon  \omega}\sin( \omega t),\\
     =& \frac{cm_0^2}{3} \Theta(t) \delta'(t)\,.
     \end{split}
\end{equation}
We introduced a regulator to evaluate the otherwise divergent integral and showed that the regulated integral converges distributionally to $\delta'(t)$. This means for any $t>0$, this term does not contribute to the retarded Green's function. Now, we turn to the second term in the spectral density \eqref{eq:specdens}, which gives a non-zero contribution:
\begin{equation}
    \begin{split}
        &  \mathcal{K}^{\rm shape}_{\rm WH, ret}(t)=\frac{cm_0^4}{12\pi}\Theta(t)\int_0^{\infty}\dd \omega \frac{ \omega}{ \omega^2+\frac{m_0^2}{4}}\sin( \omega t)=\frac{cm_0^4}{24}\Theta(t)e^{-\frac{m_0}{2}t}.
    \end{split}
\end{equation}
We evaluated the integral by closing the contour in the lower half plane thereby picking up the pole at $\omega_*=-\frac{i m_0}{2}$. We will refer to it as a relaxation pole/ quasinormal pole since it governs the decay of the 2-point function at late times. The relaxation time is therefore given by
\begin{equation}
    t_{\rm rel}=\frac{2}{m_0}\,.
\end{equation}
In Liouville CFT, we can interpret this relaxation time as the time scale that governs the dissipation of a transient transverse disturbance to the localized impurity obtained by analytically continuing the direction along the Liouville defect/shell locus to Lorentzian time,
\begin{equation}
    \delta \langle \mathcal{D}_\perp \rangle (t)= \int_{-\infty}^{\infty}\dd t' \mathcal{K}^{\rm shape}_{\rm WH, ret}(t,t')\delta \xi(t')\,,
\end{equation}
as explained in section \ref{sec:linearresponse}.
For a transient transverse disturbance $\delta \xi(t) \sim \delta (t)$, the response decays exponentially at a rate determined by the relaxation time,
\begin{equation}
    \delta \langle \mathcal{D}_\perp \rangle(t) \sim e^{-t/t_{\rm rel}}\,.
\end{equation}
In this interpretation, the mass parameter $m_0$ is treated as a coupling of the impurity worldline to Liouville CFT. In 3d gravity, such an analytic continuation of the sphere 1-point wormhole in the absence of the shape deformation gives a traversable wormhole, sourced by the timelike membrane obtained by analytically continuing the shell. The wormhole is traversable since the stress-energy on the membrane violates the Null Energy Condition as observed in \cite{Wang:2025eow}.\footnote{ This should be contrasted with the standard traversable-wormhole mechanisms of Gao--Jafferis--Wall, where a double-trace coupling between the two boundaries of an eternal black hole produces negative averaged null energy and renders the Einstein--Rosen bridge traversable \cite{Gao:2016bin}, and Maldacena--Qi, where an eternal nearly-$\mathrm{AdS}_2$ traversable wormhole is realized by coupling two SYK systems \cite{Maldacena:2018lmt}. In the present construction, the traversability is instead sourced by the analytically continued thin shell.} In fact, it is easy to see that the boundary time it takes for a light-ray to traverse from the left boundary to the right is $\Delta t=\frac{\pi m_0}{2}$, which is finite. In this traversable wormhole interpretation, the relaxation time is the time scale for the wormhole to relax to a stationary configuration after a transient deformation of the timelike membrane sourcing the wormhole.

\subsection{Stiffness towards mass deformations}
\label{subsec:sphere-one-point-mass-deformation}

We now study a deformation of the mass density of the shell, keeping the
shape of the shell fixed. Thus the shell locus remains at $y=0$ but we allow the mass parameter to be position dependent $ m(x)=m_0+\epsilon \mu(x)$.
In addition, we impose that the total mass is fixed, so
\begin{equation}
    \int_0^{2\pi} \dd x\,\mu(x)=0.
\end{equation}
Equivalently, in Fourier space, the zero mode of the perturbation is set to zero,
\begin{equation}
    \mu(x)=\sum_{n\in \mathbb Z}\mu_n e^{inx},
    \qquad 
    \mu_{-n}=\mu_n^\ast,
    \qquad
    \mu_0=0.
\end{equation}

We now impose the linearized junction conditions. Since the shell remains at
\(y=0\), the continuity condition is simply
\begin{equation}
    \Phi_+(x,0)=\Phi_-(x,0).
\end{equation}
At first order in \(\epsilon\), this gives
\begin{equation}
    \varphi_+(x,0)=\varphi_-(x,0).
\end{equation}
The exact normal-derivative jump condition is
\begin{equation}
    \partial_y\Phi_+(x,0)-\partial_y\Phi_-(x,0)
    =
    -2m(x).
\end{equation}
Using \(m(x)=m_0+\epsilon\mu(x)\), the zeroth-order condition is
\begin{equation}
    \Phi'_{0,+}(0)-\Phi'_{0,-}(0)=-2m_0,
\end{equation}
while the first-order condition is
\begin{equation}
    \partial_y\varphi_+(x,0)-\partial_y\varphi_-(x,0)
    =
    -2\mu(x).
\end{equation}
Solving the linearized junction conditions for the mode coefficients, we obtain the linearized solution:
\begin{equation}
    \varphi_\pm(x,y)
    =
    \sum_{n\in \mathbb{Z}}
    \mu_n e^{inx}
    \frac{
    e^{-|n||y|}
    \left(
    |n|+\coth(A_0+|y|)
    \right)
    }
    {
    |n|\left(|n|+\frac{m_0}{2}\right)+\frac{m_0^2-4}{4}
    }.
\end{equation}

\paragraph{Momentum flux and zero modes.} Now we will compute the stress energy associated with the mass deformations and show that there is a longitudinal momentum flux along the shell. We will analyze the zero modes and understand their origin. There are subtle differences with the corresponding analysis for the shape deformations since those were transverse deformations. 

The Fourier modes of the holomorphic components of the stress tensor just above and below the shell locus are
\begin{equation}
\begin{split}
    & \delta T^+_{zz,n}=\frac{1}{8\pi}\frac{(n+|n|)(n^2-1)}{|n|\left(|n|+\frac{m_0}{2}\right)+\frac{m_0^2-4}{4} } \mu_n\\
    & \delta T^-_{zz,n}=\frac{1}{8\pi}\frac{(|n|-n)(n^2-1)}{|n|\left(|n|+\frac{m_0}{2}\right)+\frac{m_0^2-4}{4} } \mu_n\,. 
    \end{split}
\end{equation}
Similarly, the anti-holomorphic components of the stress tensor give
\begin{equation}
    \begin{split}
    & \delta T^+_{\overline{z}\overline{z},n}=\frac{1}{8\pi}\frac{(|n|-n)(n^2-1)}{|n|\left(|n|+\frac{m_0}{2}\right)+\frac{m_0^2-4}{4} } \mu_n\\
    & \delta T^-_{\overline{z}\overline{z},n}=\frac{1}{8\pi}\frac{(n+|n|)(n^2-1)}{|n|\left(|n|+\frac{m_0}{2}\right)+\frac{m_0^2-4}{4} } \mu_n\,.
    \end{split}
\end{equation}
The jump in the two components across the shell locus is
\begin{equation}
    \begin{split}
        & \delta (T^+_{zz,n}-T^-_{zz,n})=-\delta (T^+_{\overline{z}\overline{z},n}-T^-_{\overline{z}\overline{z},n})=\frac{1}{4\pi}\frac{n(n^2-1)}{|n|\left(|n|+\frac{m_0}{2}\right)+\frac{m_0^2-4}{4}}\mu_n\,.
    \end{split}
\end{equation}
We see that the jump is odd under chirality reversal $n \to -n$ and the jump in the holomorphic and antiholomorphic components is equal and opposite. As a result, there is a net longitudinal momentum flux along the shell,
\begin{equation}
    \delta T_{xy,n}= \frac{i}{2\pi}\frac{n(n^2-1)}{|n|\left(|n|+\frac{m_0}{2}\right)+\frac{m_0^2-4}{4}}\mu_n\,.
\end{equation}
These are some features which contrast the mass deformations from the shape deformations.
Notice that there are three zero modes $n=0,\pm 1$ which are associated to PSL$(2,\mathbb{R})$ reparametrizations of the equatorial circle. So these zero modes are associated with conformal motion `on' the circle and should be contrasted with the PSL$(2,\mathbb{R})$ redundancy for the shape deformations where the zero modes are associated to conformal motion `of' the circle on the sphere.

\paragraph{The stiffness kernel.}

Having derived the linearized solution, we now compute the quadratic correction to the on-shell action. As shown in Appendix \ref{sec:appA}, the quadratic action evaluates to
\begin{align}
    S_2[\mu]=
    -\frac{1}{8\pi}
    \int_0^{2\pi}\dd x\,\mu(x)\varphi(x,0).
\end{align}
Plugging in the solution $\varphi_\pm$ derived above, we find
\begin{equation}
    S_2[\mu]
    =
    -\frac14
    \sum_{n\in \mathbb{Z}}
    \frac{|n|+\frac{m_0}{2}}
    {
    |n|\left(|n|+\frac{m_0}{2}\right)+\frac{m_0^2-4}{4}
    }
    \mu_n\mu_{-n}.
\end{equation}
Thus the quadratic kernel for fixed-total-mass deformations is
\begin{equation}
  S_2=-\frac{1}{2}\sum_{n\in \mathbb{Z}}\mathcal{K}^{\rm mass}_n \mu_n\mu_{-n}\implies \mathcal{K}^{\rm mass}_n
    =
    \frac{1}{2}
    \frac{|n|+\frac{m_0}{2}}
    {
    |n|\left(|n|+\frac{m_0}{2}\right)+\frac{m_0^2-4}{4}
    }.
\end{equation}
It is important to note that, unlike the shape kernel, the above mass kernel is independent of the choice of conformal frame since the locus on which the shell is placed is not being deformed in the present case.
At large frequencies, the kernel has a universal form which is independent of the mass parameter $m_0$,
\begin{equation}
   \mathcal{K}^{\rm mass}_n\to \frac{1}{2|n|}, \qquad |n|\to \infty.
\end{equation}
This translates to a universal logarithmic divergence of the kernel in position space.
For \(m_0>2\), both the numerator and denominator are positive. Hence
\begin{equation}
   \mathcal{K}^{\rm mass}_n>0.
\end{equation}
Therefore, the stiffness kernel for fixed-total-mass density fluctuations is
positive definite for every nontrivial mass-density deformation. So, the action or free energy of the spherically symmetric wormhole is a local maximum in the space of mass-density deformations of the shell that keep the total mass fixed.

\subsection{Spectrum of the mass-density operator of the shell}

Just like how the displacement operator generates transverse deformations of the shell, we can define a mass-density operator that is also local on the shell that generates local mass deformations/ inhomogeneities. Then, we can interpret the stiffness kernel as the two-point function of the mass density operator,
\begin{equation}
    \langle \mathcal{M}(x)\mathcal{M}(x')\rangle_{\rm Liouv} =\mathcal{K}^{\rm mass}_{\rm WH}(x,x')=\frac{c}{6\pi}\sum_{n \in \mathbb{Z}}\mathcal{K}^{\rm mass}_n e^{in(x-x')}\,.
\end{equation}
The mass deformation kernel computed above has no contact terms in position space. It has a logarithmic short distance OPE singularity since we saw that $\mathcal{K}_n^{\rm mass}$ decays to zero linearly at high frequencies. The spectral density for the mass-density operator can be read off by expressing the Fourier-space kernel in the Matsubara form,
\begin{equation}
    \mathcal{K}^{\rm mass}_{\rm WH}(i\omega_n)=\int_0^{\infty} \dd \omega \rho_{M}(\omega)\frac{2\omega}{\omega^2+\omega_n^2}\, .
\end{equation}
The integral is convergent so there is no need for local subtractions. $\omega_n$ are the discrete Matsubara frequencies. The spectral density $\rho_M(\omega)$ can thus be computed to give
\begin{equation}
    \rho_M(\omega;R)=\frac{8c\,\omega(\omega^2+R^{-2})}{3\pi\left((4\omega^2-m_0^2+4R^{-2})^2+4m_0^2\omega^2\right)}
\end{equation}
where we reinstated the radius $R$ of the shell locus. The asymptotic form of the above spectral density at high energies is
\begin{equation}
    \rho_M(\omega;R)\sim \frac{c}{6\pi \omega}, \quad \omega \to \infty\,.
\end{equation}
On the other hand, at low energies after first taking the planar limit $R\to \infty$, we have
\begin{equation}
    \rho_M(\omega;R\to \infty)\sim \frac{8c\,\omega^3}{3\pi m_0^4}, \qquad \omega \ll m_0. 
\end{equation}
which in position space translates to the following form for the 2-point function of the mass-density operator at separated points,
\begin{equation}
    \langle \mathcal{M}(x)\mathcal{M}(0) \rangle_{\rm sep} \sim \frac{16 c}{\pi m_0^4 x^4}, \qquad |x|\gg m_0^{-1}\,.
\end{equation}
Naively, it appears that $\mathcal{M}$ flows to a dimension-2 operator in the IR, but note however that in the strict IR limit $m_0\to \infty$, the above 2-point function goes to zero. This is consistent with the fact that there is no conserved 1d defect stress tensor for a conformal defect \cite{Cuomo:2021rkm}. 

We now use the spectral density in the planar limit $R\to \infty$ to compute the retarded Green's function,
\begin{equation}
    \mathcal{K}^{\rm mass}_{\rm WH, ret}(t)=-2\Theta(t)\int_0^{\infty}\dd\omega \rho_M(\omega)\sin(\omega t)\,.
\end{equation}
The relaxation time is governed by the locations of the poles of the spectral density in the lower half plane,
\begin{equation}
    \omega_*=-\frac{im_0}{4}\pm \frac{\sqrt{3}m_0}{4}\,.
\end{equation}
The retarded Green's function is given by
\begin{equation}
     \mathcal{K}^{\rm mass}_{\rm WH, ret}(t)=-\frac{c}{6}\Theta(t)e^{-\frac{m_0}{4}t}\left[\cos\left(\tfrac{\sqrt{3}}{4}m_0t\right)+\tfrac{1}{\sqrt{3}}\sin\left(\tfrac{\sqrt{3}}{4}m_0t\right)\right].
\end{equation}
Thus, the retarded Green's function is oscillatory with a decaying envelope. The relaxation time is read off to be
\begin{equation}
    t_{\rm rel}=\frac{4}{m_0}\,.
\end{equation}
In Liouville CFT, we can interpret this as the time it takes the system to relax when the impurity coupling $m_0$ is transiently changed,
\begin{equation}
    \delta \langle \mathcal{M}\rangle(t) = \int_{-\infty}^{\infty}\dd t' \mathcal{K}^{\rm mass}_{\rm WH, ret}(t,t')\delta \mu(t') \propto e^{-\frac{m_0}{4}t}\left[\cos\left(\tfrac{\sqrt{3}}{4}m_0t\right)+\tfrac{1}{\sqrt{3}}\sin\left(\tfrac{\sqrt{3}}{4}m_0t\right)\right]
\end{equation}
Similarly, in the traversable wormhole interpretation, it is the time scale for the wormhole to relax to its stationary configuration when the coupling to the timelike membrane that sources the wormhole is transiently changed.

\section{The torus 1-point wormhole}\label{sec:torus1point}

The second example that we study is the torus 1-point wormhole. In hyperbolic slicing, the wormhole is again described by a metric ansatz of the form (\ref{eq:MMmetric}). We will follow the same logic and similar technical steps in computing the stiffness kernels as we did for the sphere 1-point wormhole. Conceptually, the new observation in this setup is that because we are working at finite temperature, the direction transverse to the shell is compact. As a result, the spectrum of the displacement and mass-density operators is discrete with a spacing set by the temperature. Due to the discreteness in the spectrum, perturbations of the traversable wormhole obtained by analytically continuing the direction along the shell locus to Lorentzian time do not relax at late times. 

Now, we set up the linearized Liouville analysis to compute the stiffness kernels.
Let's work with the fundamental domain $y \in [-\frac{\beta}{2},\frac{\beta}{2}]$ on the cylinder with the ends identified. For the background solution, the shell is placed at $y=0$ and the Liouville solution takes the form \cite{Chandra:2024vhm}
\begin{equation}
    \Phi_0(y)=-2\log\left[\frac{1}{r_H}\cos\left(r_H(\tfrac{\beta}{2}-|y|)\right)\right].
\end{equation}
The parameter $r_H$ is determined in terms of the mass of the shell by
\begin{equation}
    m_0=2r_H \tan\left(r_H \tfrac{\beta}{2}\right)\,.
\end{equation}
Geometrically, $2\pi r_H$ is the length of the minimal geodesic at $y=\pm \frac{\beta}{2}$.
The linearized Liouville equation around this background saddle takes the form 
\begin{equation}
   \big(\partial_x^2+\partial_y^2-2r_H^2 \sec^2(r_H(|y|-\tfrac{\beta}{2}))\big)\varphi_\pm(x,y)=0
\end{equation}
where $\varphi_\pm$ are linearized Liouville solutions in the two regions $0<y<\frac{\beta}{2}$ and $-\frac{\beta}{2}<y<0$, respectively. With $x\sim x+2\pi$, we can
do the Fourier decomposition,
\begin{equation}
    \varphi_{\pm}(x,y)=\sum_{n \in \mathbb{Z}}\varphi_{\pm, n}(y)e^{inx}.
\end{equation}
Plugging this into the linearized Liouville equation, we get for $0<y<\frac{\beta}{2}$,
\begin{equation}
\varphi_{+,n}''(y)-(n^2+2r_H^2 \sec^2(r_H(y-\tfrac{\beta}{2}))\varphi_{+,n}(y)=0.
\end{equation}
A convenient basis of solutions to this equation is 
\begin{equation}
    \begin{split}
        & u_n(y)=e^{|n|(y-\frac{\beta}{2})}\left[|n|+r_H \tan(r_H(y-\tfrac{\beta}{2}))\right],\\[1em]
        & v_{n}(y)=e^{-|n|(y-\frac{\beta}{2})}\left[|n|-r_H \tan(r_H(y-\tfrac{\beta}{2}))\right].
    \end{split}
\end{equation}
So, we parametrise the linearized Liouville field as a linear combination of the above basis,
\begin{equation}
    \begin{split}
        & \varphi_{+,n}(y)=a_n u_n(y) + b_n v_n(y), \\
        & \varphi_{-,n}(y)=c_n u_n(-y) + d_n v_n(-y).
    \end{split}
\end{equation}
The 4 coefficients are determined by solving the 4 gluing conditions, 2 across the shell and 2 associated with smooth gluing at $y=\pm \frac{\beta}{2}$. The two conditions associated with smooth gluing of the ends at $y=\pm \frac{\beta}{2}$ are
\begin{equation}
    \varphi_+(x,\tfrac{\beta}{2})=\varphi_-(x,-\tfrac{\beta}{2}), \qquad \partial_y \varphi_+(x,\tfrac{\beta}{2})=\partial_y \varphi_-(x,-\tfrac{\beta}{2})\,.
\end{equation}
This relates the coefficients pairwise to each other, namely $a_n=d_n$ and $b_n=c_n$. The junction conditions across the shell will be analyzed separately in the next subsection.

\begin{figure}
    \centering
    \includegraphics[width=0.5\linewidth]{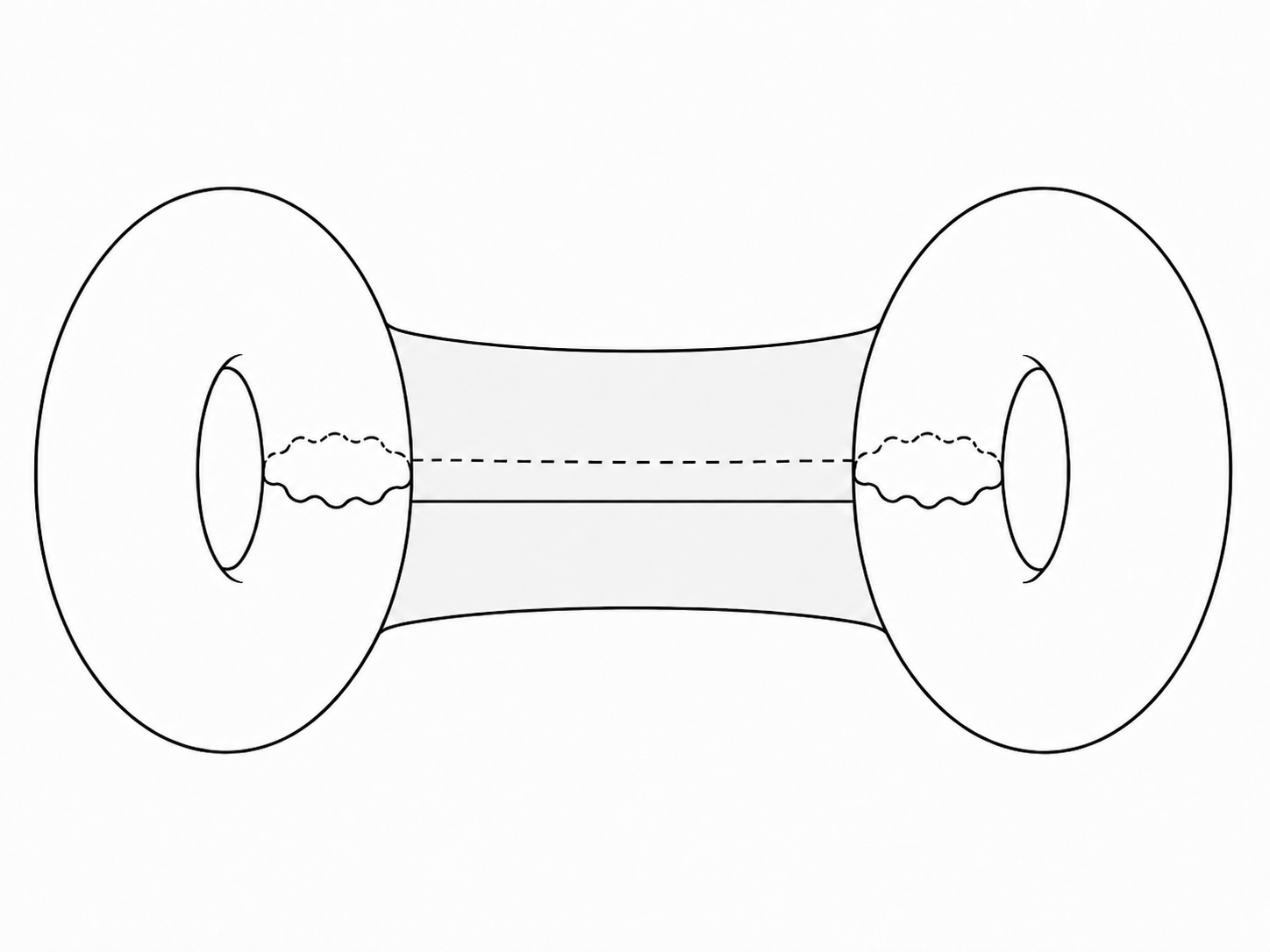}
    \caption{The figure shows a sketch of the torus 1-point wormhole used to compute the stiffness kernel for shape deformations of the shell.}
    \label{fig:torus1pointwh}
\end{figure}

\subsection{Stiffness toward shape deformations}

First we analyze the shape deformations of the shell locus, $y=\epsilon \xi(x)$. See Fig.~\ref{fig:torus1pointwh} for an illustration of the set-up.
The linearized junction conditions across the shell take the form
\begin{equation}
    \varphi_+(x,0)-\varphi_-(x,0)=2m_0 \xi(x), \qquad \partial_y \varphi_+(x,0)=\partial_y \varphi_-(x,0)\,.
\end{equation}
Solving these conditions for the mode coefficients, the solution then takes the form
\begin{equation}
    \varphi_\pm(x,y)=\pm m_0\sum_{n \in \mathbb{Z}}\xi_n e^{inx} \left[\frac{|n|\sinh(|n|\rho_\pm)+r_H \tan(r_H \rho_\pm)\cosh(|n|\rho_\pm)}{|n|\sinh(|n|\frac{\beta}{2})+\frac{m_0}{2}\cosh(|n|\frac{\beta}{2})}\right].
\end{equation}
Here $\rho_\pm$ are shifted coordinates given by $\rho_+=\frac{\beta}{2}-y$ for $0<y<\frac{\beta}{2}$ and $\rho_-=\frac{\beta}{2}+y$ for $-\frac{\beta}{2}<y<0$. Using the linearized solution, we evaluate the quadratic action derived in Appendix~\ref{sec:appA}:
\begin{equation}
    S_2[\xi]=-\frac{m_0}{8\pi}\int_0^{2\pi}\dd x \left( \xi \partial_y \varphi(x,0)+\Phi_0''(0)\xi^2+\Phi_0(0)(\xi'^2)\right)
\end{equation}
to get
\begin{equation} \label{eq:torushapekernel}
\begin{split}
     S_2[\xi]&=-\frac{1}{2}\sum_{n \in \mathbb{Z}} \mathcal{K}^{\rm shape}_n(\beta)\xi_n \xi_{-n}\\
    \mathcal{K}^{\rm shape}_n(\beta)&=-m_0 |n| \left[\frac{m_0|n|-2r_H^2 \tanh(|n|\frac{\beta}{2})}{m_0+2|n|\tanh(|n|\frac{\beta}{2})}\right ]+\frac{1}{2}m_0 n^2 \log\left(r_H^2+\frac{m_0^2}{4}\right)\,.
    \end{split}
\end{equation}
We see that the kernel has the expected zero mode associated with rigid translations of the shell along the torus. The above stiffness kernel has the following high-temperature expansion,
\begin{equation}
    \mathcal{K}^{\rm shape}_n(\beta) = \frac{m_0n^2}{2}\log \left(\frac{m_0}{\beta}\right)+\frac{m_0^3n^2}{720}\beta^2 + O(\beta^3)\,.
\end{equation}
At low temperatures, the kernel has the following expansion
\begin{equation}
    \mathcal{K}_n^{\rm shape}(\beta)=-\frac{m_0^2n^2}{(2|n|+m_0)}+m_0n^2\log\left(\frac{m_0}{2}\right)+\frac{2\pi^2}{\beta^2}\frac{|n|(2n^2+|n|m_0+m_0^2)}{m_0(2|n|+m_0)}+O(\beta^{-3})\,.
\end{equation}
The Fourier-space kernel can be converted to position space as written below with the normalization factors restored,
\begin{equation}
    \mathcal{K}_{\rm WH}^{\rm shape}(x,x';\beta)=\frac{c}{6\pi}\sum_{n \in \mathbb{Z}} \mathcal{K}_n^{\rm shape}(\beta)e^{in(x-x')}\,.
\end{equation}
The non-local part of the kernel gives rise to the short distance singularity
\begin{equation}
    \mathcal{K}_{\rm WH}^{\rm shape}(x,x';\beta) \sim \frac{cm_0^2}{6\pi(x-x')^2}+\text{contact terms}, \qquad |x-x'|\to 0\,.
\end{equation}
This is the same short distance singularity as observed for the sphere 1-point wormhole.

\subsection{Relation to annular sewing and a hyperbolic Virasoro coadjoint orbit}

Now, we relate the shape deformations of the shell on the torus to the problem of
self-sewing the two circular boundary components of an annulus/cylinder with a
diffeomorphism. This is the annular analogue of conformal welding and is closely
related to the sewing operations on bordered or rigged Riemann surfaces used in
the Segal formulation of conformal field theory
\cite{Radnell:2005sewing,MaibachPeltola:2026complex}. 

First, we construct conformal maps that straighten the deformed shell locus $y=\epsilon \xi(x)$. Expressing the maps as $w_\pm(z) = z +\epsilon \alpha_{\pm}(z)+O(\epsilon^2)$ where $z=x+iy$ is the global coordinate on the torus, we have
\begin{equation}
    \begin{split}
        \alpha_+(z)=-i\xi_0-i\sum_{n \neq 0}\xi_n \frac{e^{in(z-i\frac{\beta}{2})}}{\cosh(|n|\frac{\beta}{2})}, \qquad & 0< \text{Im}z<\frac{\beta}{2},\\
        \alpha_-(z)=-i\xi_0-i\sum_{n \neq 0}\xi_n \frac{e^{in(z+i\frac{\beta}{2})}}{\cosh(|n|\frac{\beta}{2})}, \qquad & -\frac{\beta}{2}< \text{Im}z< 0 \,.
    \end{split}
\end{equation}
We can verify explicitly that the ends $z=x\pm i\frac{\beta}{2}$ are glued smoothly under these maps. They also satisfy the straightening condition,
\begin{equation}
    \text{Im}w_\pm |_{z=x+i\epsilon \xi(x)}=\epsilon \xi(x) +\epsilon \text{Im} \alpha_\pm(x)=0\,.
\end{equation}
Restricting these maps to the straightened shell locus, we can relate the shape deformation $\xi(x)$ to the infinitesimal welding diffeomorphism $ x_+=x_-+\epsilon \psi(x_-)+O(\epsilon^2)$,
\begin{equation}
    \psi(x)= 2H_\beta \xi(x), \qquad (H_\beta \xi)_n= -i \text{sgn}(n)\tanh\left(\frac{|n|\beta}{2}\right)\xi_n\,.
\end{equation}
Here, $H_\beta$ is the finite-temperature generalization of the Hilbert transform. 

\begin{figure}
    \centering
    \includegraphics[width=0.8\linewidth]{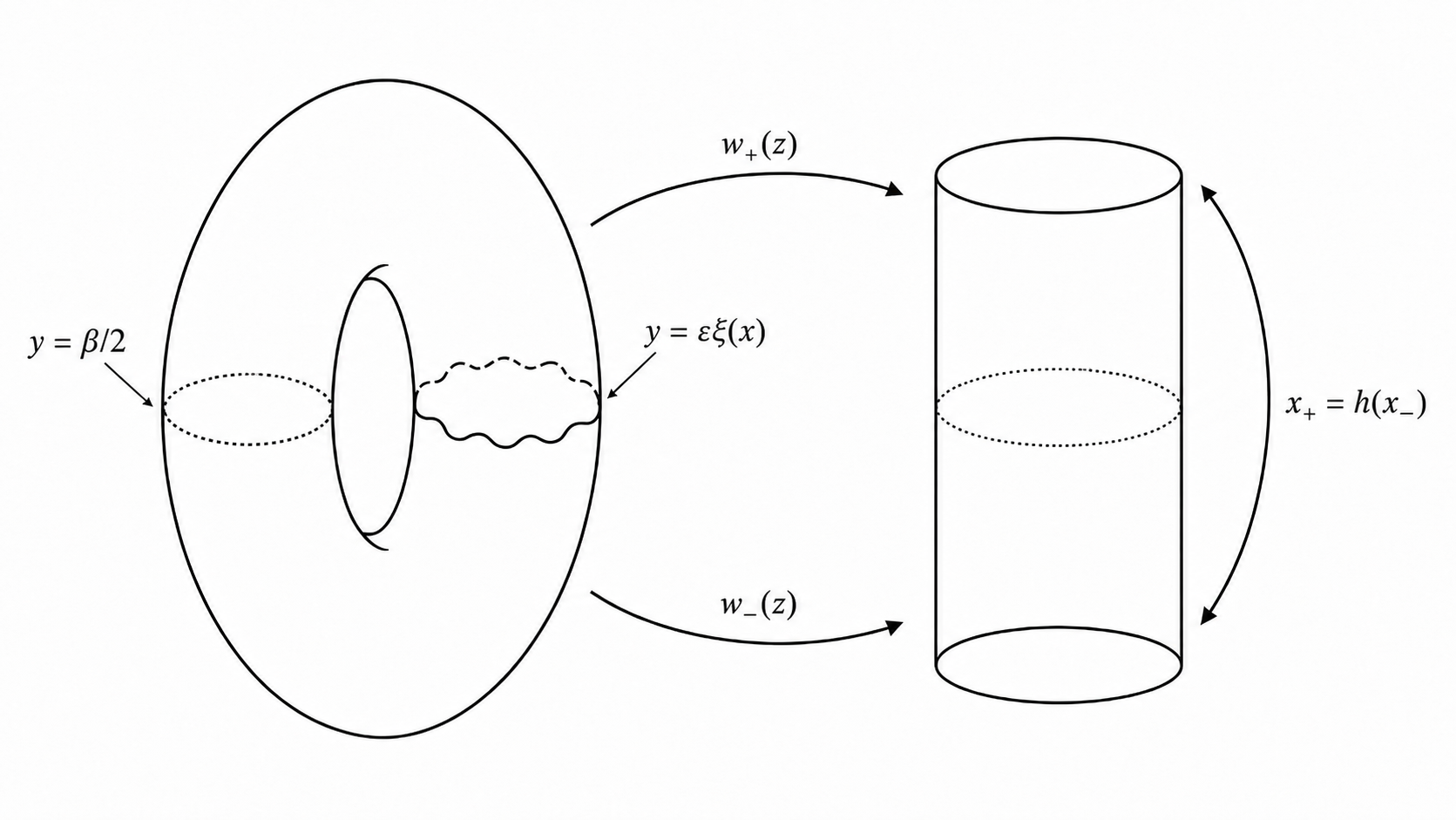}
    \caption{This figure describes the annular sewing across the deformed shell locus $y=\epsilon \xi(x)$. The conformal maps $w_\pm$ solve the sewing condition $h(x_-)=(w_+ \circ w_-^{-1})(x_-)$ given the diffeomorphism $h$, which is expressible in terms of the deformed locus by a finite-temperature version of the Hilbert transform. In addition, they are smooth across the dotted circle.}
    \label{fig:annsewing}
\end{figure}

Now, we can study the hyperbolic response to the above annular-sewing problem by computing the jump in the stress-tensor,
\begin{equation}
    \begin{split}
        \delta (T_{zz}^+-T_{zz}^-)_n=& \frac{m_0}{4\pi}\frac{|n|(n^2+r_H^2)\tanh\left(\frac{|n|\beta}{2}\right)}{|n|\tanh\left(\frac{|n|\beta}{2}\right)+\frac{m_0}{2}}\xi_n ,\\
        =& -\frac{1}{4\pi}\frac{m_0}{m_0+2|n|\tanh\left(\frac{|n|\beta}{2}\right)}[(\partial_x^3-r_H^2\partial_x)\psi(x)]_n\,.
    \end{split}
\end{equation}
Recall that $r_H$ is the proper radius of the geodesic at $y=\pm \frac{\beta}{2}$ and is given by the solution to $m_0=2r_H\tan\left(r_H\frac{\beta}{2}\right)$.
In the second line where we have expressed the stress-tensor jump in terms of the welding diffeomorphism; we see that it is given by a finite-temperature analogue of the Schwarzian action dressed by a non-local response factor. In the heavy shell limit, the non-local factor goes away and we get a universal local response,
\begin{equation}
     \delta (T_{zz}^+-T_{zz}^-)(x) \,\,\longrightarrow\,\, -\frac{1}{4\pi}(\partial_x^3-r_H^2\partial_x)\psi(x), \qquad m_0\to \infty \,.
\end{equation}
The same hyperbolic structure is visible directly in the stiffness kernel once local contact terms are separated. Indeed, the shape kernel in (\ref{eq:torushapekernel}) can be rearranged as
\begin{equation}
\mathcal{K}^{\rm shape}_n(\beta)
=
m_0 n^2
\left[
\frac{1}{2}\log\left(r_H^2+\frac{m_0^2}{4}\right)-1
\right]
+
\frac{
2m_0|n|(n^2+r_H^2)
\tanh\left(\frac{|n|\beta}{2}\right)
}{
m_0+2|n|\tanh\left(\frac{|n|\beta}{2}\right)
}.
\end{equation}
The first term is local in position space, so the separated-point part of the stiffness kernel is
\begin{equation}
\mathcal{K}^{\rm shape}_{n,\mathrm{sep}}(\beta)
=
\frac{
2m_0|n|(n^2+r_H^2)
\tanh\left(\frac{|n|\beta}{2}\right)
}{
m_0+2|n|\tanh\left(\frac{|n|\beta}{2}\right)
}.
\end{equation}
Using the relation between the shape deformation and the sewing diffeomorphism, we obtain
\begin{equation}  \label{eq:torushypresp}
\left(\mathcal{K}^{\rm shape}_{\mathrm{sep}}\xi\right)_n
=
-
\frac{m_0}{
m_0+2|n|\tanh\left(\frac{|n|\beta}{2}\right)
}
\left[
(\partial_x^3-r_H^2\partial_x)\psi
\right]_n.
\end{equation}
Thus, at finite \(m_0\),
\begin{equation} 
\left(\mathcal{K}^{\rm shape}_{\mathrm{sep}}\xi\right)_n
=
4\pi\,\delta(T^+_{zz}-T^-_{zz})_n,
\end{equation}
so the same non-local response factor dresses both the stiffness kernel and the jump in the stress tensor. Equivalently, in position space, the stress tensor jump is given by the convolution
\begin{equation}
    \int_0^{2\pi}\dd x' \mathcal{K}^{\rm shape}_{\rm sep}(x,x') \xi(x')= 4\pi \delta( T^+_{zz}-T^-_{zz})(x)= 4\pi \delta( T^+_{\overline{z}\overline{z}}-T^-_{\overline{z}\overline{z}})(x)\,.
\end{equation}
This generalizes the analysis of section \ref{sec:conformalweldingcylinder} for the sphere 1-point wormhole to finite temperature.

Now, we relate the hyperbolic response (\ref{eq:torushypresp}) to the Virasoro coadjoint action on a hyperbolic orbit.
The differential operator $\partial_x^3-r_H^2\partial_x$ is the infinitesimal Virasoro coadjoint-action operator around a constant hyperbolic representative; in the Virasoro orbit classification this belongs to the $B_0$ family \cite{Witten:1988hf,Balog:1997zz}.
In Fourier space, the differential operator $\partial_x^3-r_H^2 \partial_x$ has only the rigid translation mode $n=0$ (equivalently, the rigid rotation mode in welding gauge) as the zero mode.
The stabilizer of the corresponding constant hyperbolic representative is therefore $U(1)$, so the orbit is locally $\text{Diff}^+(S^1)/U(1)$. In the Hill-equation/monodromy classification of Virasoro coadjoint orbits, this is a $B_0$ hyperbolic orbit, with the continuous hyperbolic parameter set by the constant representative, here equivalently by $r_H$ \cite{Witten:1988hf,Balog:1997zz}. We conjecture that the general non-linear response in the heavy-shell limit is governed by the coadjoint action on this $B_0$ hyperbolic orbit, equivalently by the corresponding Virasoro geometric action \cite{Witten:1988hf,Alekseev:1989ce,Balog:1997zz},
\begin{equation}
    \Delta T_{ww}[h](x)=-\frac{1}{4\pi}\left(\{h(x),x\}-\frac{r_H^2}{2}((h')^2-1)\right)\,.
\end{equation}
At finite $m_0$, there are non-local response functions added to the above universal response which are suppressed by inverse-powers of $m_0$.

\subsection{Stiffness toward mass deformations}

Next, we analyze mass deformations of the shell, with $m(x) = m_0 + \epsilon \mu(x)$. The linearized junction conditions across the shell take the form
\begin{equation}
    \varphi_+(x,0)=\varphi_-(x,0), \qquad \partial_y \varphi_+(x,0)-\partial_y \varphi_-(x,0)=-2\mu(x)\,.
\end{equation}
As in the sphere one-point case, we keep the total shell mass fixed, so the zero mode is removed $\mu_0=0$.
Solving these conditions for the mode coefficients, we obtain the linearized solution,
\begin{equation}
    \varphi_\pm(x,y)=\sum_{n \in \mathbb{Z}}\mu_n e^{inx} \left[\frac{|n|\cosh(|n|\rho_\pm)+r_H\tan(r_H\rho_\pm)\sinh(|n|\rho_\pm)}{(n^2+r_H^2+\frac{m_0^2}{4})\sinh(|n|\frac{\beta}{2})+\frac{m_0|n|}{2}\cosh(|n|\frac{\beta}{2})}\right].
\end{equation}
Here $\rho_\pm$ are shifted coordinates given by $\rho_+=\frac{\beta}{2}-y$ for $0<y<\frac{\beta}{2}$ and $\rho_-=\frac{\beta}{2}+y$ for $-\frac{\beta}{2}<y<0$. The quadratic deformation of the action is given by
\begin{equation}\label{eq:1pointWHmasskernel}
    \begin{split}
       & S_2[\mu]=-\frac{1}{8\pi}\int_0^{2\pi}\dd x \mu(x)\varphi(x,0)=-\frac{1}{2}\sum_n \mathcal{K}^{\rm mass}_n(\beta)\mu_n\mu_{-n},\\
       & \mathcal{K}^{\rm mass}_n(\beta)= \frac{1}{2}\frac{2|n|+m_0\tanh(|n|\frac{\beta}{2})}{2(n^2+r_H^2+\frac{m_0^2}{4})\tanh(|n|\frac{\beta}{2})+m_0|n|}\,.
    \end{split}
\end{equation}
We see that the quadratic kernel for the mass deformations is positive-definite just like in the sphere 1-point wormhole case. At any fixed temperature, we also recover the universal high-frequency behaviour of the kernel,
\begin{equation}
    \mathcal{K}^{\rm mass}_n(\beta) \to \frac{1}{2|n|}, \qquad |n| \to \infty\,,
\end{equation}
which is independent of the mass parameter $m_0$ and temperature and produces a logarithmic short distance singularity in position space. The high temperature expansion takes the form
\begin{equation}
    \mathcal{K}^{\rm mass}_n(\beta)=\frac{1}{2m_0}-\frac{\beta}{4}\left(\frac{n^2}{m_0^2}-\frac{1}{3}\right)-\frac{\beta^2}{8}\left(\frac{m_0}{15}-\frac{n^4}{m_0^3}\right)+O(\beta^3)\,.
\end{equation}
To derive the leading piece, it is useful to note that at high temperatures, the geodesic length parameter $r_H\sim\sqrt{\frac{m_0}{\beta}}$. Notice that the leading piece is independent of the Fourier mode frequency as well as temperature. 
The low temperature expansion takes the form
\begin{equation}
     \mathcal{K}^{\rm mass}_n(\beta) = \frac{1}{2}\frac{|n|+\frac{m_0}{2}}{n^2+\frac{m_0}{2}|n|+\frac{m_0^2}{4}}-\frac{8\pi^2}{\beta^2}\frac{|n|+\frac{m_0}{2}}{\left(n^2+\frac{m_0}{2}|n|+\frac{m_0^2}{4}\right)^2}+O(\beta^{-3})\,.
\end{equation}
We see that the kernel approaches a positive value both at high and low temperatures that is independent of the temperature. 

We can use the Fourier components to write down the kernel in position space,
\begin{equation}
    \mathcal{K}_{\rm WH}^{\rm mass}(x,x';\beta)=\frac{c}{6\pi}\sum_{n\in \mathbb{Z}}\mathcal{K}_n^{\rm mass}(\beta)e^{in(x-x')}\,.
\end{equation}
In the next subsection, we analyze the spectral decomposition of this position-space stiffness kernel.

\subsection{Spectrum of the displacement and mass-density operators}

\begin{figure}
    \centering
    \includegraphics[width=1\linewidth]{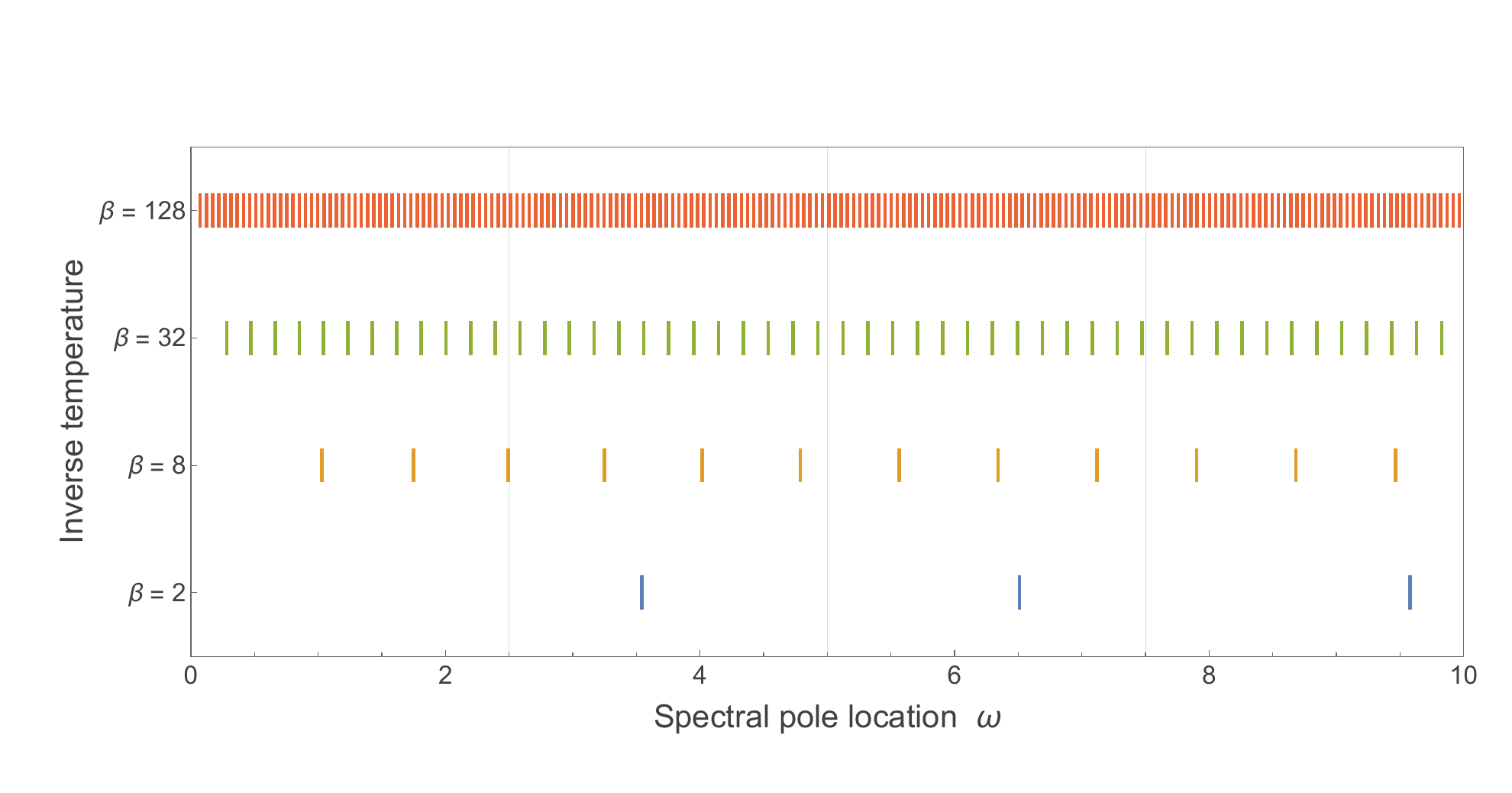}
    \caption{A plot of the displacement operator spectrum for $m_0=3$ for different temperatures. Each vertical line represents a non-spurious spectral pole. The spectrum becomes dense as $\beta$ increases i.e., as we lower the temperature. }
    \label{fig:torus displacement poles}
\end{figure}

Just like in the sphere 1-point wormhole case, we can extract the spectrum of the displacement and mass-density operators from their respective stiffness kernels,
\begin{equation}
    \langle \mathcal{D}_{\perp}(x)\mathcal{D}_{\perp}(x')\rangle_{\rm Liouv} = \mathcal{K}^{\rm shape}_{\rm WH}(x,x';\beta), \qquad \langle \mathcal{M}(x)\mathcal{M}(x')\rangle_{\rm Liouv} = \mathcal{K}^{\rm mass}_{\rm WH}(x,x';\beta)\,.
\end{equation}
The novel feature that we will observe is that the kernels have isolated poles so the spectrum is discrete. This is due to the fact that the Hilbert space is defined on the dual cycle as shown in the figure below, which has finite length $\beta$ and hence the defect spectrum is discrete:\vspace{-5mm}
\begin{equation}
        \vcenter{\hbox{
\begin{overpic}[grid=false, scale=0.107]{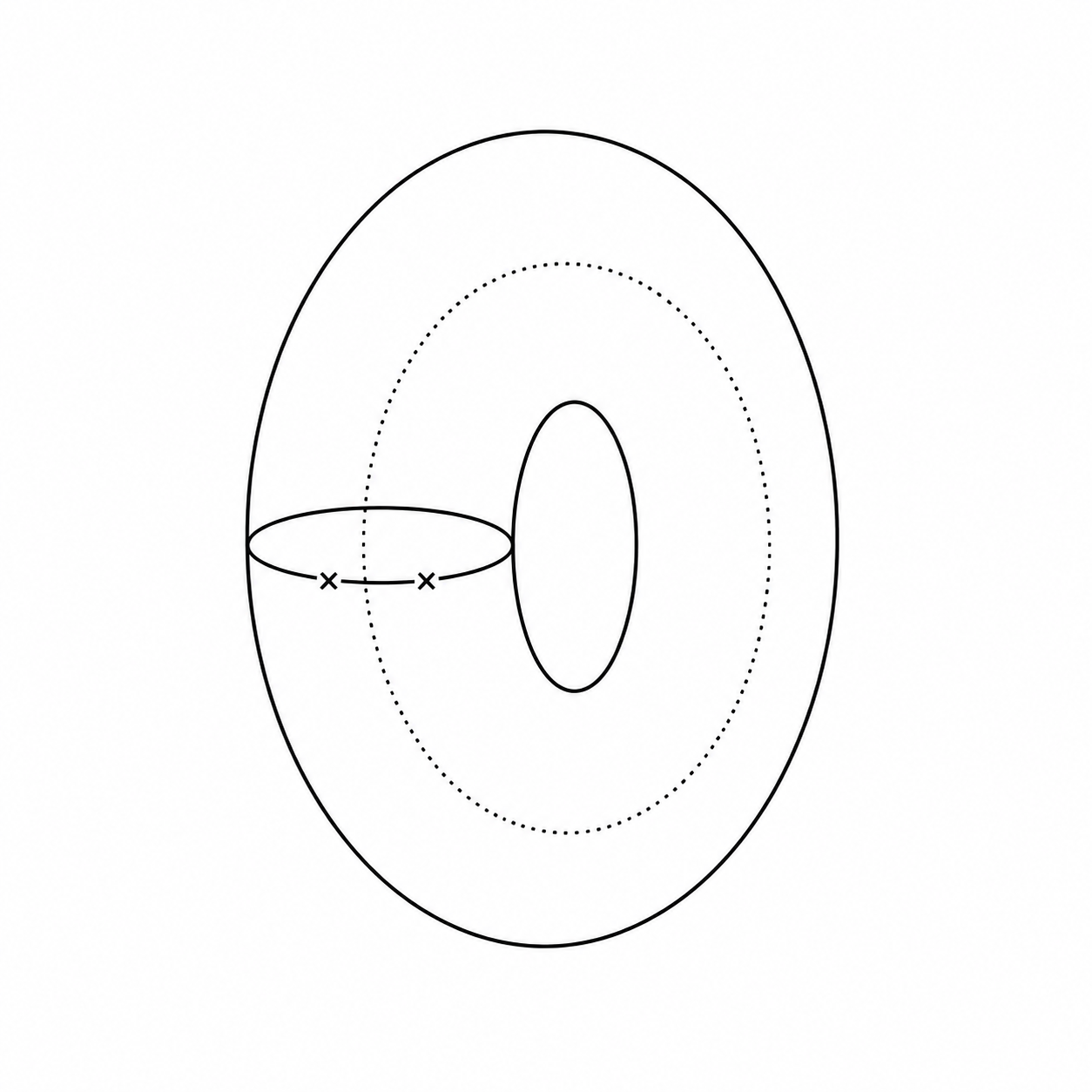}
\end{overpic}}} 
\vspace{-5mm}
    \end{equation}
The displacement or mass-density operators are inserted at the marked points on the shell locus shown as a solid circle and their 2-point functions are expanded using a complete set of states inserted along the dual cycle shown by a dotted circle.  
Since the poles only come from the non-contact piece of the kernel, recall that the separated-point part of the shape kernel is given by
\begin{equation}
    \mathcal{K}^{\rm shape}_{\rm sep}(i\omega_n;\beta)=-m_0 \omega_n\left[\frac{m_0 \omega_n-2r_H^2\tanh(\omega_n\frac{\beta}{2})}{m_0+2\omega_n\tanh(\omega_n\frac{\beta}{2})}\right]\,.
\end{equation}
Here, $\omega_n$ are the discrete Matsubara frequencies. Treated as a complex valued function of $\omega$ after analytically continuing $i\omega_n \to \omega$, it has discrete poles at 
\begin{equation} \label{eq:torusshapepoleeqn}
    m_0=2\omega \tan(\omega \frac{\beta}{2})\,.
\end{equation}
There are infinitely many solutions denoted as $\omega_{j,D}$ to the above transcendental equation, one in each of the intervals
\begin{equation}
    \omega_{j,D} \in \left(\frac{2\pi j}{\beta}, \frac{(2j+1)\pi}{\beta}\right), \qquad j=0,1,2\dots
\end{equation}
Note however that the $j=0$ pole for which $\omega=r_H$ is spurious since the residue vanishes, hence the non-trivial poles contributing to the spectral density start from $j=1$. Hence the spectral density of the displacement operator can be expressed as
\begin{equation}
    \rho_D(\omega;\beta)=\sum_{j=1}^{\infty}\rho_{j,D} \delta (\omega-\omega_{j,D}), \qquad \rho_{j,D}=\frac{c\, m_0^2(\omega_j^2-r_H^2)}{6\left(\tan(\frac{\beta \omega_j}{2})+\frac{\beta \omega_j}{2}\sec^2(\frac{\beta \omega_j}{2})\right)}\, .
\end{equation}
Here we reinstated the normalization factor of $\frac{c}{3}$. Note that
the spectral coefficients $\rho_j$ are all positive since $\omega_j > r_H$. At any temperature, the high-energy spectrum (i.e. the spectral poles at large $j$) are uniformly distributed with the corresponding spectral coefficients growing linearly,
\begin{equation} \label{eq:highTspectorusshape}
    \omega_j \sim \frac{2\pi j}{\beta}, \qquad \rho_j \sim \frac{2\pi c m_0^2}{3\beta^2}j\,.
\end{equation}
Below, we provide a Lorentzian interpretation for this uniformly spaced high energy spectrum in terms of lightcone singularities of the Lorentzian OPE of the displacement operator.
As we lower the temperature i.e., increase $\beta$, the poles get denser as illustrated by the plot in Fig.~\ref{fig:torus displacement poles} and eventually condense into a branch cut at zero temperature, reproducing the continuous spectral density for the sphere 1-point wormhole in the planar limit. In the high temperature limit, the spectrum is sparse and is uniformly distributed with the same spacings $\frac{2\pi}{\beta}$ as (\ref{eq:highTspectorusshape}),
\begin{equation}
    \omega_j = \frac{2\pi j}{\beta}+\frac{m_0}{2\pi j} + O(\beta)\,.
\end{equation}

\begin{figure}
    \centering
    \includegraphics[width=1\linewidth]{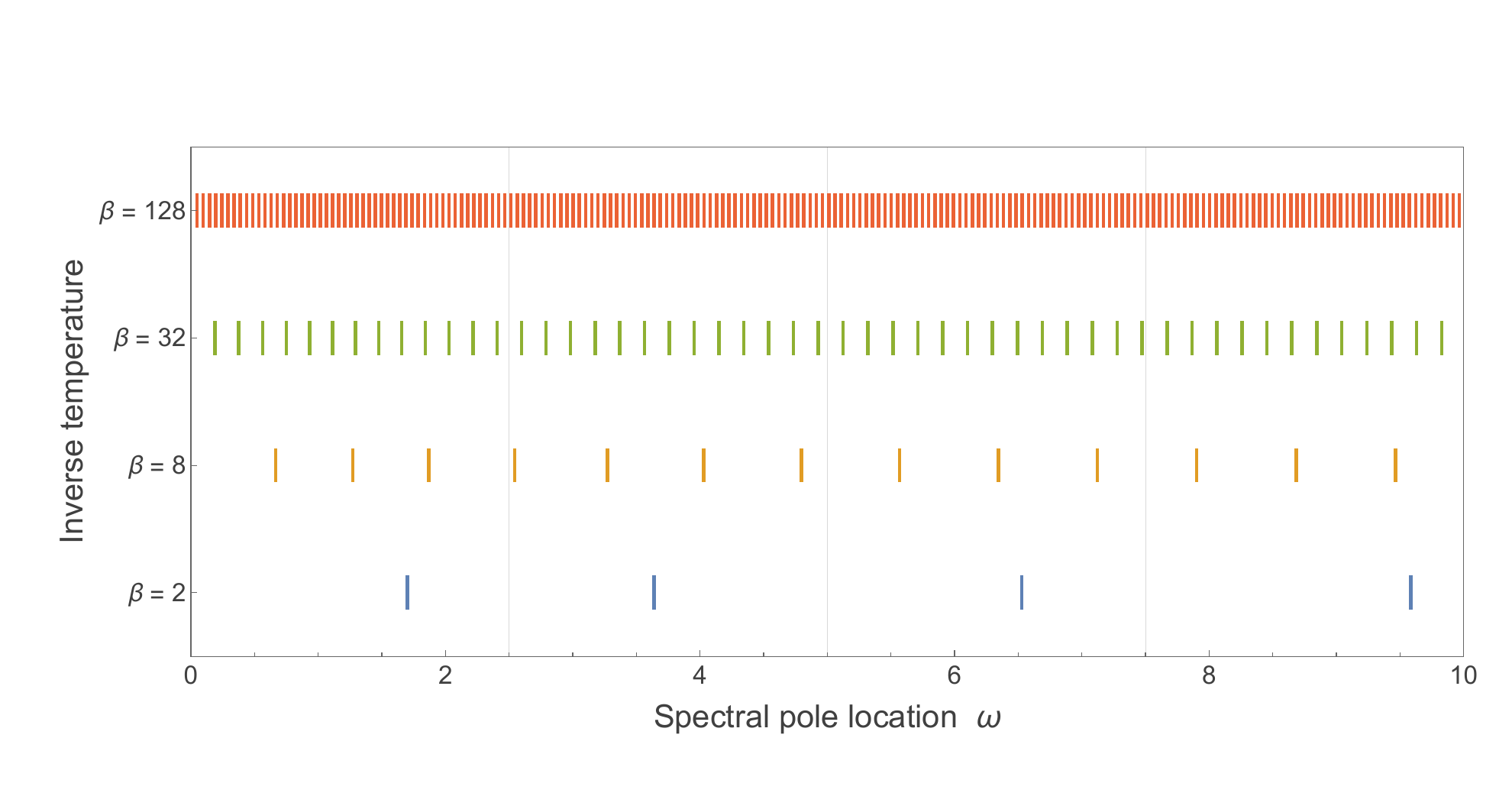}
    \caption{A plot of the mass-density operator spectrum for $m_0=3$ for different temperatures. Each vertical line represents a non-spurious spectral pole. The spectrum becomes dense as $\beta$ increases i.e., as we lower the temperature.}
    \label{fig:torus mass poles}
\end{figure}

Turning to the mass kernel, when expressed in terms of the Matsurbara frequencies, it takes the form,
\begin{equation}
    \mathcal{K}^{\rm mass}(i\omega_n;\beta)=\frac{1}{2}\frac{2\omega_n + m_0 \tanh\left(\omega_n \frac{\beta}{2}\right)}{2(\omega_n^2+r_H^2+\frac{m_0^2}{4})\tanh\left(\omega_n \frac{\beta}{2}\right)+m_0\omega_n}\,.
\end{equation}
After the analytic continuation $i\omega_n \to \omega$, we see that the spectral poles are determined by the following transcendental equation,
\begin{equation} \label{eq:torusmasspoleeqn}
    m_0 \omega + 2(r_H^2+\frac{m_0^2}{4}-\omega^2)\tan\left(\omega \frac{\beta}{2}\right)=0\,.
\end{equation}
Note that this differs from the spectral pole condition for shape deformations (\ref{eq:torusshapepoleeqn}), which means that the mass-density and displacement operators are in general supported on distinct discrete spectra. The spectral density of the mass-density operator (after reinstating the normalization factor of $\frac{c}{3}$) is given by
\begin{equation} \label{eq:torusmassspecdensity}
    \rho_M(\omega;\beta)=\sum_{j=1}^{\infty}\rho_{j,M}\delta(\omega-\omega_{j,M}), \qquad \rho_{j,M}= \frac{c\left(2\omega_j+m_0 \tan\left(\frac{\beta \omega_j}{2}\right)\right)}{6\left(4\omega_j\tan\left(\frac{\beta \omega_j}{2}\right)+\beta(\omega_j^2-r_H^2-\frac{m_0^2}{4})\sec^2\left(\frac{\beta \omega_j}{2}\right)-m_0\right)}
\end{equation}
where $\omega_{j,M}$ are positive roots of (\ref{eq:torusmasspoleeqn}) ordered increasingly in energy.
Again, all the spectral coefficients are positive. At high energies, the spectral poles are uniformly spaced just like in the displacement spectrum but the spectral coefficients now decay linearly,
\begin{equation}  \label{eq:highTspectorusmass}
    \omega_j \sim \frac{2\pi j}{\beta}, \qquad \rho_j\sim \frac{c}{6\pi j}\,.
\end{equation}
As illustrated by the plot in Fig.~\ref{fig:torus mass poles}, the poles become denser as the temperature is lowered and eventually condense into a branch cut at zero temperature.

Just as in the sphere 1-point wormhole case, we can use the spectral densities for the shape and mass deformations to compute the retarded Green's function,
\begin{equation}
    \begin{split}
        & \mathcal{K}^{\rm shape}_{\rm WH, ret}(t)=-2\Theta(t)\sum_{j=1}^{\infty}\rho_{j,D}\sin(\omega_{j,D}t),\\
        & \mathcal{K}^{\rm mass}_{\rm WH, ret}(t)=-2\Theta(t)\sum_{j=1}^{\infty}\rho_{j,M}\sin(\omega_{j,M}t)\,.
    \end{split}
\end{equation}
A notable feature of the retarded Green's function is the tower of lightcone singularities\footnote{We thank Ahmed Abdalla for raising the question about lightcone singularities in our setup.} since we are now computing the 2-point function of the displacement or mass density operators at points separated in time but at a finite spatial volume determined by $\beta$,
\begin{equation} \label{eq:toruslightcone}
    \begin{split}
        & \mathcal{K}^{\rm shape}_{\rm WH, ret}(t) \supset \frac{cm_0^2}{3}\Theta(t)\sum_{k \in \mathbb{Z}}\delta'(t-k \beta),\\
        & \mathcal{K}^{\rm mass}_{\rm WH, ret}(t) \supset \Theta(t)\left(-\frac{c}{6}+\frac{c}{3\beta}(t-k \beta)\right), \qquad k \beta <t <(k+1)\beta\,.
    \end{split}
\end{equation}
There is an infinite tower of lightcone singularities in the retarded Green's function at the times $t=k \beta$ where $k$ is a non-negative integer. They arise from the high energy spectra (\ref{eq:highTspectorusshape}) and (\ref{eq:highTspectorusmass}). The singularities are sharper for the shape deformations because of the linear growth in the spectral coefficients as opposed to linear decay in the case of mass deformations. As a result, the lightcone singularities are realized merely as discontinuities in the retarded Green's function.

Due to the discreteness of the spectrum, the retarded Green's function has an oscillatory behaviour that does not decay to zero at late times. Therefore, the impurity in Liouville CFT or the traversable wormhole in 3d gravity do not relax to a stationary configuration at late times. This is not surprising since the length of the thermal cycle determines the spatial volume in the Lorentzian continuation. At finite spatial volume, we do not expect to observe relaxation since there are not enough degrees of freedom for the perturbation to get dissipated into.

\section{The one-sided black hole} \label{sec:one-sided}

\begin{figure}
    \centering
\tikzset{every picture/.style={line width=1.2pt}} 
\begin{tikzpicture}[x=0.75pt,y=0.75pt,yscale=-1.2,xscale=1.2]
\draw  [dash pattern={on 1.5pt off 1.5pt on 1.5pt off 1.5pt}]  (450,182.5) -- (450,20) ;
\draw  [draw opacity=0][dash pattern={on 1.5pt off 1.5pt on 1.5pt off 1.5pt}] (410,160) .. controls (410,154.48) and (427.91,150) .. (450,150) .. controls (470.73,150) and (487.78,153.94) .. (489.8,158.99) -- (450,160) -- cycle ; \draw  [dash pattern={on 1.5pt off 1.5pt on 1.5pt off 1.5pt}] (410,160) .. controls (410,154.48) and (427.91,150) .. (450,150) .. controls (470.73,150) and (487.78,153.94) .. (489.8,158.99) ;  
\draw  [draw opacity=0][fill={rgb, 255:red, 245; green, 166; blue, 35 }  ,fill opacity=0.46 ] (491,69.74) .. controls (491.63,69.43) and (468.75,84.75) .. (470.25,105.08) .. controls (471.75,125.42) and (495.11,129.11) .. (486.11,137.11) .. controls (477.11,145.11) and (474.42,138.17) .. (471.67,138.67) .. controls (468.92,139.17) and (468.44,143.42) .. (461.44,142.67) .. controls (454.44,141.92) and (448.89,140.94) .. (439.89,144.44) .. controls (430.89,147.94) and (433,137.93) .. (425.5,137.38) .. controls (418,136.83) and (409.89,135.4) .. (410,134.46) .. controls (410.12,133.52) and (430,116.58) .. (429.5,103.08) .. controls (429,89.58) and (407.75,73.33) .. (411.5,69.83) .. controls (415.25,66.33) and (417.67,70) .. (425.5,67.08) .. controls (433.33,64.17) and (433.67,58) .. (439.89,60.22) .. controls (446.11,62.44) and (455.22,62.44) .. (462.33,61.78) .. controls (469.44,61.11) and (466.33,66.22) .. (471.22,66.22) .. controls (476.11,66.22) and (475.27,63.93) .. (479.44,64.89) .. controls (483.62,65.85) and (490.38,70.06) .. (491,69.74) -- cycle ;
\draw  [color={rgb, 255:red, 202; green, 202; blue, 202 }  ,draw opacity=1 ][fill={rgb, 255:red, 202; green, 202; blue, 202 }  ,fill opacity=1 ] (198,30) -- (250,30) -- (250,169.8) -- (198,169.8) -- cycle ;
\draw  [draw opacity=0][fill={rgb, 255:red, 202; green, 202; blue, 202 }  ,fill opacity=1 ] (250,60) .. controls (250,60) and (250,60) .. (250,60) .. controls (272.09,60) and (290,77.91) .. (290,100) .. controls (290,122.09) and (272.09,140) .. (250,140) -- (250,100) -- cycle ; \draw   (250,60) .. controls (250,60) and (250,60) .. (250,60) .. controls (272.09,60) and (290,77.91) .. (290,100) .. controls (290,122.09) and (272.09,140) .. (250,140) ;  
\draw   (252.6,100.2) -- (256.4,100.2)(254.5,98.2) -- (254.5,102.2) ;
\draw    (250,60) -- (250,30) ;
\draw    (250,140) -- (250,169.8) ;
\draw    (188.73,115) -- (188.73,87.4) ;
\draw [shift={(188.73,84.4)}, rotate = 90] [fill={rgb, 255:red, 0; green, 0; blue, 0 }  ][line width=0.08]  [draw opacity=0] (5.36,-2.57) -- (0,0) -- (5.36,2.57) -- cycle    ;
\draw    (209.47,180) -- (236.47,180) ;
\draw [shift={(239.47,180)}, rotate = 180] [fill={rgb, 255:red, 0; green, 0; blue, 0 }  ][line width=0.08]  [draw opacity=0] (5.36,-2.57) -- (0,0) -- (5.36,2.57) -- cycle    ;
\draw  [dash pattern={on 1.5pt off 1.5pt on 1.5pt off 1.5pt}]  (198,169.8) -- (198,30) ;
\draw [color={rgb, 255:red, 245; green, 101; blue, 35 }  ,draw opacity=1 ][line width=1.5]    (250,60) .. controls (229.2,89.8) and (229.6,110.6) .. (250,140) ;
\draw [shift={(250,140)}, rotate = 55.24] [color={rgb, 255:red, 245; green, 101; blue, 35 }  ,draw opacity=1 ][fill={rgb, 255:red, 245; green, 101; blue, 35 }  ,fill opacity=1 ][line width=1.5]      (0, 0) circle [x radius= 1.74, y radius= 1.74]   ;
\draw [shift={(250,60)}, rotate = 124.91] [color={rgb, 255:red, 245; green, 101; blue, 35 }  ,draw opacity=1 ][fill={rgb, 255:red, 245; green, 101; blue, 35 }  ,fill opacity=1 ][line width=1.5]      (0, 0) circle [x radius= 1.74, y radius= 1.74]   ;
\draw  [draw opacity=0] (490,160) .. controls (490,160) and (490,160) .. (490,160) .. controls (490,160) and (490,160) .. (490,160) .. controls (490,165.52) and (472.09,170) .. (450,170) .. controls (427.91,170) and (410,165.52) .. (410,160) .. controls (410,159.65) and (410.07,159.3) .. (410.21,158.96) -- (450,160) -- cycle ; \draw   (490,160) .. controls (490,160) and (490,160) .. (490,160) .. controls (490,160) and (490,160) .. (490,160) .. controls (490,165.52) and (472.09,170) .. (450,170) .. controls (427.91,170) and (410,165.52) .. (410,160) .. controls (410,159.65) and (410.07,159.3) .. (410.21,158.96) ;  
\draw   (410,40) .. controls (410,34.48) and (427.91,30) .. (450,30) .. controls (472.09,30) and (490,34.48) .. (490,40) .. controls (490,45.52) and (472.09,50) .. (450,50) .. controls (427.91,50) and (410,45.52) .. (410,40) -- cycle ;
\draw    (410,40) -- (410,160) ;
\draw    (490,40) -- (490,160) ;
\draw [color={rgb, 255:red, 245; green, 101; blue, 35 }  ,draw opacity=1 ][line width=1.5]    (410,71.31) .. controls (418.25,65.9) and (419.67,68.22) .. (425.5,67.08) .. controls (431.33,65.94) and (434.21,56.34) .. (443.75,61.58) .. controls (453.29,66.82) and (458.43,57.83) .. (467.57,64.34) .. controls (476.71,70.86) and (471,56.86) .. (490,70) ;
\draw [color={rgb, 255:red, 245; green, 108; blue, 35 }  ,draw opacity=1 ][line width=1.5]  [dash pattern={on 1.5pt off 1.5pt on 1.5pt off 1.5pt}]  (410,71.31) .. controls (414.5,71.01) and (425.5,80.45) .. (433.75,77.31) .. controls (442,74.16) and (444.75,81.63) .. (454.25,79.67) .. controls (463.75,77.7) and (463.88,74.55) .. (474.25,75.93) .. controls (484.63,77.31) and (480.25,70.62) .. (490,70) ;
\draw [color={rgb, 255:red, 245; green, 101; blue, 35 }  ,draw opacity=1 ][line width=1.5]    (410,133.15) .. controls (418.25,138.56) and (419.5,136) .. (425.5,137.38) .. controls (431.5,138.75) and (434.22,148.12) .. (443.75,142.88) .. controls (453.29,137.64) and (458.43,146.63) .. (467.57,140.11) .. controls (476.72,133.6) and (471,147.6) .. (490,134.46) ;
\draw [color={rgb, 255:red, 245; green, 108; blue, 35 }  ,draw opacity=1 ][line width=1.5]  [dash pattern={on 1.5pt off 1.5pt on 1.5pt off 1.5pt}]  (410,133.15) .. controls (414.5,133.44) and (425.5,124.01) .. (433.75,127.15) .. controls (442,130.3) and (444.75,122.83) .. (454.25,124.79) .. controls (463.75,126.76) and (463.88,129.9) .. (474.25,128.53) .. controls (484.63,127.15) and (480.25,133.84) .. (490,134.46) ;
\draw [color={rgb, 255:red, 245; green, 101; blue, 35 }  ,draw opacity=1 ][line width=1.5]    (410,134.46) .. controls (435,108.21) and (435,100.94) .. (410,71.31) ;
\draw [color={rgb, 255:red, 245; green, 101; blue, 35 }  ,draw opacity=1 ][line width=1.5]    (490,70) .. controls (463.67,95.71) and (464,112.13) .. (490,134.46) ;
\draw    (501.05,156.1) .. controls (507.15,173.39) and (482.25,176.65) .. (472.26,178.16) ;
\draw [shift={(469.33,178.67)}, rotate = 347.2] [fill={rgb, 255:red, 0; green, 0; blue, 0 }  ][line width=0.08]  [draw opacity=0] (5.36,-2.57) -- (0,0) -- (5.36,2.57) -- cycle    ;

\draw (175.13,91.2) node [anchor=north west][inner sep=0.75pt]    {$y$};
\draw (218.47,182.5) node [anchor=north west][inner sep=0.75pt]    {$r$};
\draw (301,87.4) node [anchor=north west][inner sep=0.75pt]  [font=\large]  {$\times \ S^{1}$};
\draw (255.93,145.07) node [anchor=north west][inner sep=0.75pt]  [font=\small]  {$y=-\tau _{0}$};
\draw (231.25,146.9) node [anchor=north west][inner sep=0.75pt]    {$+$};
\draw (268.75,109.4) node [anchor=north west][inner sep=0.75pt]    {$-$};
\draw (255.93,45.6) node [anchor=north west][inner sep=0.75pt]  [font=\small]  {$y=\tau _{0}$};
\draw (501,126.73) node [anchor=north west][inner sep=0.75pt]  [font=\small]  {$y=-\tau _{0}$};
\draw (501,62.07) node [anchor=north west][inner sep=0.75pt]  [font=\small]  {$y=\tau _{0}$};
\draw (158.67,153.4) node [anchor=north west][inner sep=0.75pt]    {$a)$};
\draw (370,154.07) node [anchor=north west][inner sep=0.75pt]    {$b)$};
\draw (502,168.07) node [anchor=north west][inner sep=0.75pt]    {$x$};
\end{tikzpicture}
    \caption{The one-sided black hole set-up. $a)$ The standard spherically symmetric geometry, formed by gluing a portion of the Euclidean BTZ ($-$) to global EAdS$_3$ ($+$) across a thin shell (orange). $b)$ The same geometry, viewed in the cylinder frame, where now we allow the shell to wiggle along the transverse $S^1$ direction.}
    \label{fig:one_sided_bh}
\end{figure}
Having analyzed the wormhole examples, we now turn to black hole geometries.
Recall from \cite{Chandra:2024vhm} that the one-sided Euclidean thin-shell black hole geometry is constructed by gluing a portion of a BTZ black hole to a portion of global AdS$_3$, across a thin shell. For simplicity, we work with the non-rotating thin-shell black hole geometry. The trajectory of the shell starts and ends on the same asymptotic boundary. In the bulk, we take the shell to lie entirely behind the horizon, as illustrated in Figure \ref{fig:one_sided_bh}.

In the dual CFT, the one-sided thin-shell black hole is dual to a pure state $\ket{\Psi}$. The semiclassical gravitational path integral $e^{-I_{\text{BH}}}$ computes the norm $\overline{\braket{\Psi}{\Psi}}$ after suitable coarse-graining \cite{Chandra:2022fwi}. Equivalently, the norm is interpreted as a cylinder two-point function 
\begin{equation}
   \braket{\Psi}{\Psi} = \expval{D_\Sigma^\dagger (-\tau_0)D_\Sigma(\tau_0)}_{\text{cyl}}
\end{equation}
 where $D_\Sigma(\tau_0)$ is the line operator creating the shell at Euclidean time $\tau_0$. For spherically symmetric shells, the trajectory of the shell is known in closed form, and the on-shell action was computed in \cite{Chandra:2022fwi,Chandra:2024vhm}. From this, various properties of the state $\ket{\Psi}$ were derived, such as its coarse-grained Von Neumann entropy $S(\bar \rho)$. This was shown to reproduce the area law
\begin{equation}
    S(\bar \rho) = \frac{A(\gamma)}{4G_N}
\end{equation}
where $\bar \rho = \overline{\ketbra{\Psi}{\Psi}}$ and $A(\gamma) = 2\pi r_H$ is the length of the apparent horizon.

In this section, we consider deformations of the shell in the one-sided black hole background that explicitly break the angular $U(1)$ isometry of the background geometry. This produces a family of states $\ket{\Psi_\xi}$ labeled by the deformation $\xi(x)$, where we consider both shape and mass deformations. We will then study the response of the state to these deformations, measured by stiffness kernels as in the previous sections, as well as the change in apparent-horizon entropy $\Delta S$ induced by the deformation.

To set up the problem, let us work in cylindrical coordinates $x\sim x+2\pi$ and $y\in (-\infty,\infty)$, and take a bulk radial coordinate $r\in [0,\infty)$, with $r=0$ the center of global AdS$_3$ (the dashed vertical line in Fig.~\ref{fig:one_sided_bh}). In the spherically symmetric set-up, the shell operator is inserted at $y = \pm\tau_0$ and the mass of the shell is $m_0$. These two parameters uniquely specify the spherically symmetric solution, including the inverse temperature $\beta$ associated to the black hole horizon. 

This solution was constructed in \cite{Chandra:2024vhm} by solving the junction conditions along the shell.  These in turn translate to a jump in the normal derivative of the auxiliary Liouville field $\Phi$ that defines the gravity solution. The Liouville field lives on the half $y>0$ with ZZ boundary conditions at $y=0$. For the spherically symmetric background, the Liouville solution takes the form
\begin{equation}\label{eq:backgroundLV}
\Phi_0(y)=
\begin{cases}
\displaystyle
\Phi_{0,-}(y) = -2\log\left({\tfrac{1}{r_H}}\sin(r_H y)\right), & 0<y<\tau_0, \\[6pt]
\displaystyle \Phi_{0,+}(y) = 
-2\log\sinh(y-\tau_0+A), & y>\tau_0 .
\end{cases}
\end{equation}
We denoted the black hole region by $-$ and the global AdS$_3$ region by $+$. Here $r_H$ is the horizon radius and the parameter $A$ is related to the mass $m_0$ and proper radius $r_0$ of the shell locus by
\begin{equation}\label{eq:params}
m_0=\coth A-r_H\cot(r_H\tau_0), \quad r_0 =\frac{r_H}{\sin(r_H\tau_0)}=\frac{1}{\sinh A}.
\end{equation}
Combining the above equations we can also write $m_0 = \sqrt{1+r_0^2}+\sqrt{r_0^2-r_H^2}$, where we chose the branch of the square root that corresponds to a heavy shell behind the horizon \cite{Chandra:2024vhm}, which has $\frac{\pi}{2}<r_H\tau_0 <\pi$ (so that $\cot(r_H\tau_0)<0$).
Now we want to study the linearized perturbations around this background. We parametrize the linearized Liouville field as
\begin{equation}
\Phi_\pm(x,y)=\Phi_{0,\pm}(y)+\epsilon \varphi_\pm(x,y).
\end{equation}
We proceed in two steps: first we 
write down the linearized Liouville equations in the $\pm$ regions,
\begin{equation}
\left(\partial_x^2+\partial_y^2-2r_H^2\csc^2(r_Hy)\right)\varphi_-=0,
\qquad
0<y<\tau_0,
\end{equation}
\begin{equation}
\left(\partial_x^2+\partial_y^2-2\csch^2(y-\tau_0+A)\right)\varphi_+=0,
\qquad
y>\tau_0 .
\end{equation}
A basis of solutions in each region is found to be
\begin{equation}
    u_\pm(x,y) = \sum_{n\in \mathbb{Z}} u_{\pm, n}(y) e^{inx},
\end{equation}
with 
\begin{equation}
    u_{-,n}(y) = \cosh(|n|y)-{r_H\over |n|}\cot(r_Hy)\sinh(|n|y)
\end{equation}
and
\begin{equation}
u_{+,n}(y)=
e^{-|n|(y-\tau_0)}
\left(|n|+\coth(y-\tau_0+A)\right).
\end{equation}
Note that $u_{-,n}(y)$ is regular at the ZZ boundary $y=0$. Also note that $u_{+,n}(y)$ is the decaying solution in the $y>\tau_0$ region. It is convenient to introduce normalized basis functions
\begin{equation}
\widehat u_{\pm,n}(y):={u_{\pm,n}(y)\over u_{\pm,n}(\tau_0)} ,
\qquad
\widehat u_{\pm,n}(\tau_0)=1.
\end{equation}
Next, we expand $\varphi_\pm(x,y)$ in this basis. To determine the expansion coefficients, we have to solve the linearized junction conditions. We will distinguish as before between shape deformations and mass deformations, which we analyze separately below.

\paragraph{Shape deformations.}
For shape deformations, we deform the shell locus to $y = \tau_0 + \epsilon \xi(x)$, for some arbitrary wiggle function $\xi$ in the transverse $S^1$. The linearized junction conditions are
\begin{equation}\label{eq:linearizedjunction}
    \varphi_+(x,\tau_0)-\varphi_-(x,\tau_0)=2m_0\xi(x), \qquad \partial_y \varphi_+(x,\tau_0)=\partial_y \varphi_-(x,\tau_0)\,.
\end{equation}
These are solved by
\begin{equation}\label{eq:varph1}
    \varphi_\pm(x,y)= \pm 2m_0\sum_{n\in\mathbb Z}
{\frac{\lambda_{\mp,n}}{\lambda_{+,n}+\lambda_{-,n}}}
\xi_n e^{inx}\,\widehat u_{\pm,n}(y),
\end{equation}
where we defined the Dirichlet-to-Neumann eigenvalues $\lambda_{\pm,n}$ by the normal derivatives of $\widehat{u}_\pm$ at $\tau_0$:
\begin{equation}
    \lambda_{\pm,n} = \mp \,\partial_y  \widehat{u}_{\pm,n}(\tau_0).
\end{equation}

\paragraph{Mass deformations.}
For mass deformations $m = m_0 + \epsilon \mu(x)$, the junction conditions are
\begin{equation}\label{eq:linearizedjunctionmass}
    \varphi_+(x,\tau_0)=\varphi_-(x,\tau_0), \qquad \partial_y \varphi_+(x,\tau_0)-\partial_y \varphi_-(x,\tau_0)=-2\mu(x)\,.
\end{equation}
These are solved by
\begin{equation}\label{eq:massdefsol}
\varphi_\pm(x,y)
=
\sum_{n\in\mathbb Z}
{2\mu_n\over \lambda_{+,n}+\lambda_{-,n}}
e^{inx}\,\widehat u_{\pm,n}(y),
\end{equation}
with the same $\lambda_{\pm,n}$ as above.

\subsection{Stiffness kernels and their spectrum}\label{eq:1sidedstiffnesskernels}

We will now use the above solutions to the linearized Liouville equations to compute the stiffness kernels for the shape and mass deformations of the one-sided black hole. This will allow us to analyze the displacement operator two-point function and its spectral density.
\vspace{-2mm}
\subsubsection{Shape deformations}
To compute the stiffness kernels for shape deformations, we use the general formula for the quadratic effective action for shape deformations derived in appendix \ref{sec:appA}:
\begin{equation}\label{eq:effectiveactionshape}
S_2[\xi]
=
-{m_0\over 8\pi}
\int_0^{2\pi}dx
\left(
\xi(x)\,\partial_y\varphi(x,\tau_0)
+
\Phi_0''(\tau_0)\xi^2
+
\Phi_0(\tau_0)(\xi')^2
\right) 
\end{equation}
Evaluating the normal derivative $\partial_y\varphi(x,\tau_0)$ using \eqref{eq:varph1} (where we can take either one of $\varphi_\pm$, since their normal derivatives match along the shell), we have
\begin{equation}
    \partial_y\varphi(x,\tau_0) = -2m_0\sum_{n\in \mathbb{Z}}{\lambda_{+,n}\lambda_{-,n}\over \lambda_{+,n}+\lambda_{-,n}} \xi_ne^{inx},
\end{equation}
where explicitly
\begin{equation}\label{eq:lambda1}
\lambda_{-,n}
=
{(n^2+r_0^2)\tanh(|n|\tau_0)-|n|r_H\cot(r_H\tau_0)
\over
|n|-r_H\cot(r_H\tau_0)\tanh(|n|\tau_0)},
\end{equation}
and
\begin{equation}\label{eq:lambda2}
\lambda_{+,n}
=
{|n|(|n|+\coth A)+\csch^2 A\over |n|+\coth A}
=
{|n|(|n|+\sqrt{1+r_0^2})+r_0^2
\over
|n|+\sqrt{1+r_0^2}}.
\end{equation}
The $n=0$ term should be understood as the smooth limit of the above expressions as $n\to 0$. The undeformed shell solution with $m_0>1$ only exists in the region $0<r_H \tau_0 <\pi$, and in this regime we can explicitly check that the eigenvalues $\lambda_{\pm,n}$ are positive for all $n\in \mathbb{Z}$:\footnote{To see this, use that $r_H\tau_0 \cot (r_H\tau_0) <1$, $r_0>\frac{1}{\tau_0}$ and $(1+x^2)\tanh(x) - x>0$ for $x>0$.}
\begin{equation}
    \lambda_{\pm, n} > 0.
\end{equation}
Using furthermore that along the shell $\Phi_0''(\tau_0)=2r_0^2$ and $\Phi_0(\tau_0)=2\log r_0$, and expanding in Fourier modes, we find for the quadratic action in the cylinder frame
\begin{equation}
  S_2[\xi]  = -\frac{1}{2}\sum_{n\in \mathbb{Z}} \mathcal{K}^{\rm shape}_n \xi_n\xi_{-n},
\end{equation}
with
\begin{equation}\label{eq:shapekerBH}
\mathcal{K}^{\rm shape}_n
=
m_0
\left(r_0^2
+n^2\log r_0-
m_0{\lambda_{+,n}\lambda_{-,n}\over \lambda_{+,n}+\lambda_{-,n}}
\right).
\end{equation}
Recall that $r_H$ and $r_0$ are functions of $m_0$ and $\tau_0$ through \eqref{eq:params}, so the shape deformation kernel is entirely determined by the parameters $(m_0,\tau_0)$ of the undeformed black hole solution. 

\paragraph{Displacement two-point function.} As in the previous sections, we can  interpret the shape deformation kernel as the two-point function of the displacement operator of the Liouville line defect:  
\begin{equation}
    \expval{\mathcal{D}_{\perp}(x)\mathcal{D}_{\perp}(x')}_{\rm Liouv} = \mathcal{K}^{\rm shape}_{\rm BH}(x,x') = \frac{c}{6\pi}\sum_{n\in\mathbb{Z}}\mathcal{K}^{\rm shape}_n e^{in(x-x')}.
\end{equation}
Translation symmetry allows us to set $x' = 0$ w.l.o.g. Since the Liouville line defect is placed on a cylinder with ZZ boundary conditions, the defect Hilbert space is defined on a non-compact slice, so we expect the displacement spectrum $\rho_D(\omega)$ to be continuous. Indeed, following the same Matsubara trick, evaluating the Fourier transform $\widetilde{\mathcal{K}}^{\rm shape}(i\omega_n)$ we can read off the spectral density from the discontinuity of the branch cut on the positive real $\omega$ axis.\footnote{Concretely, this entails replacing $|n|$ by $-i(\omega+i0^+)$ in \eqref{eq:shapekerBH} and taking the imaginary part.} We find
\begin{equation}\label{eq:one-sided-BH-displacement-spectral-density}
    \rho_D(\omega)= \frac{cm_0^2}{3\pi}\frac{Y(\omega)B(\omega)^2}{(X(\omega)+B(\omega))^2+Y(\omega)^2}\,,
\end{equation}
with the three functions given by
\begin{equation}\label{eq:threefunctions}
    \begin{split}
        & B(\omega)=\frac{(r_0^2-\omega^2)\sin(\omega \tau_0)-r_H\omega\cot(r_H\tau_0)\cos(\omega \tau_0)}{\omega \cos(\omega \tau_0)-r_H \cot(r_H\tau_0)\sin(\omega \tau_0)},\\[1em]
        & Y(\omega) =\frac{\omega (\omega^2+1)}{\omega^2+1+r_0^2},\quad X(\omega)=\frac{r_0^2\sqrt{1+r_0^2}}{\omega^2+1+r_0^2}\,.
    \end{split}
\end{equation}
In Figure \ref{fig:shapespectrumBH}, we have plotted $\rho_D(\omega)$ for $\tau_0 = 1,r_H=2$ (lying in the heavy range $\frac{\pi}{2}<r_H\tau_0<\pi$). Note that for fixed $\tau_0$, the horizon radius $r_H$ is a function of $m_0$ by \eqref{eq:params}. From the plot, we see that the spectral density is positive and continuous, as expected. For large $\omega$, the spectral density has the following UV behaviour
\begin{equation}\label{eq:uvbehavior}
    \rho_D(\omega) \sim \tfrac{cm_0^2}{3\pi} \,\omega \sin^2(\omega \tau_0).
\end{equation}
 The linear growth in $\omega$ signals the OPE singularity in position space $\langle \mathcal{D}_\perp(x)\mathcal{D}_\perp(0)\rangle \sim \frac{cm_0^2}{6\pi x^2}$, $x\to 0$ which is universal for all the examples we have studied. The oscillating piece signals a `lightcone' singularity in the retarded correlator when we go to real time:
 \begin{equation}
    \mathcal{K}_{\rm ret}(t)=-2\Theta(t) \int_0^{\infty} d\omega \rho_D(\omega)\sin(\omega t)\, \supset \tfrac{cm_0^2}{3} \delta'(t)-\tfrac{cm_0^2}{6}\,\delta'(t-2\tau_0).
\end{equation}
This lightcone singularity in the displacement OPE precisely occurs at the time it takes light to travel from the insertion of $\mathcal{D}_\perp$ at $t=0$ to reflect off the ZZ boundary placed at a distance of $\tau_0$ from the impurity and reach the impurity again at time $t=2\tau_0$.

Superimposed on the oscillatory linear growth (\ref{eq:uvbehavior}), the
spectral density $\rho_D(\omega)$ exhibits alternating local maxima and
minima as illustrated in figure \ref{fig:shapespectrumBH}. The local extrema occur at the frequencies solving $\rho_D'(\omega)=0$. Since the spectral density is non-negative and is a smooth function of $\omega$, the local minima occur precisely at $\rho_D(\omega)=0$ i.e.,
the zeros of $B(\omega)$\footnote{Note that naively, $\omega=r_H$ solves the transcendental equation for the local minimum. However, this is spurious since the pole of the $B(\omega)$ function coincides with the zero at this value. In fact, we can check that the function has a finite non-zero limit at this point,
 \begin{equation}
     B(\omega \to r_H)=r_H\frac{r_H \tau_0 \cos(r_H\tau_0)-\sin(r_H\tau_0)(1+\sin^2(r_H\tau_0))}{\sin(r_H\tau_0)(\sin(r_H\tau_0)\cos(r_H\tau_0)-r_H\tau_0)}.
 \end{equation}
Hence, $\omega=r_H$ is not a local minimum.},
\begin{equation}
\text{local minima:} \quad (r_0^2-\omega^2)\tan(\omega\tau_0)
=
r_H\omega\cot(r_H\tau_0).
\end{equation}
At large frequencies, the local maxima are asymptotically given by the locations of the poles of the function $B(\omega)$,
\begin{equation}
    \text{local maxima:} \quad \tan(\omega \tau_0)=\frac{\omega}{r_H\cot(r_H\tau_0)}, \qquad \omega \to \infty\,.
\end{equation}
Physically, these extrema in the spectral density occur because the space in between the shell and the ZZ boundary behaves like a cavity that supports standing waves with constructive or destructive interference.

We can also interpret the lightcone singularity in the retarded Green's function and the local extrema in the spectral density from the dual holographic CFT. The lightcone singularity occurs at the time it takes light to travel from one defect to the other. So it shows up in the 2-point function of the displacement operator with one operator inserted on one defect and the other inserted on the second defect. Similarly, the local extrema arise due to interference of deformation modes in the region between the two defects.

\paragraph{Relaxation analysis in the planar limit.}
Lastly, there is further information encoded in the analytic structure of $\rho_D(\omega)$ in the complex plane. For instance, the relaxation time is encoded in the location of the poles of $\rho_D(\omega)$ with $\mathrm{Im}(\omega)<0$. The pole with the smallest negative imaginary part determines the relaxation time. 

For simplicity, we compute the relaxation time in the planar limit, in which we reinstate the radius $x\sim x+2\pi R$ of the shell locus, and take $R\to \infty$ (keeping $m_0$ and $\tau_0$ fixed). In this limit, the junction conditions for the background saddle give the simplified relation
\begin{equation}
    m_0 = r_H \tan(\frac{r_H\tau_0}{2})\,.
\end{equation}
Here, $\frac{\pi}{2}<r_H \tau_0<\pi$ so a solution to the above junction condition exists only if $\tau_0 > \frac{\pi}{2m_0}$.
In the planar limit, the Matsubara frequencies $\omega$ become continuous and the Dirichlet-to-Neumann factors that determine the stiffness kernel in \eqref{eq:shapekerBH} take a simpler form:
\begin{equation}\label{eq:reltime}
    \lambda_+=\frac{\omega^2+r_0 \omega + r_0^2}{\omega +r_0}, \qquad \lambda_-=\frac{(\omega^2+r_0^2)\tanh(\omega\tau_0)-\omega r_H \cot(r_H\tau_0)}{\omega - r_H \cot(r_H \tau_0)\tanh(\omega \tau_0)}\,.
\end{equation}
The relaxation time is determined by the pole of the stiffness kernel that is closest to the imaginary axis. In this planar limit, the condition determining the pole location is given by
\begin{equation}\label{eq:polelocation}
    2\omega(\omega^2+m_0\omega+m_0r_0)e^{2\omega \tau_0}+m_0r_0^2(e^{2\omega \tau_0}-1)=0\,.
\end{equation}
It is important to note that this equation also contains spurious poles for which the residue vanishes so we should isolate the physical relaxation pole after removing all the spurious poles. Numerically, we have extracted the relaxation time for various values of the parameters $m_0$ and $\tau_0$ and also observed that it increases monotonically with $\tau_0$ at fixed $m_0$. This suggests that when the impurities are closer, the system relaxes faster as shown in figure \ref{fig:relaxation_time}.

\subsubsection{Mass deformations} Next, we compute the stiffness kernel for mass deformations of the one-sided black hole. The quadratic action for mass deformations derived in appendix \ref{sec:appA} is
\begin{equation}\label{eq:effectiveactionmass}
S_2[\mu]
=
-{1\over 8\pi}
\int_0^{2\pi}dx\,\mu(x)\varphi(x,\tau_0).
\end{equation}
Substituting the linearized solution $\varphi$ obtained in \eqref{eq:massdefsol}, we obtain:\footnote{Note that $n=0$ does not appear because we removed the zero mode from $\mu(x)$, i.e.~$\mu_0 = 0$.}
\begin{equation}
S_2[\mu]
=
-\frac{1}{2}\sum_{n\in \mathbb{Z}}\mathcal{K}^{\rm mass}_n\mu_n\mu_{-n}, \quad \mathcal{K}^{\rm mass}_n
=
{1\over \lambda_{+,n}+\lambda_{-,n}}.
\end{equation}
From this we read off the stiffness kernel for mass deformations:
\begin{equation}
    \mathcal{K}_{\rm BH}^{\rm mass}(x,x') = \frac{c}{6\pi} \sum_{n\in\mathbb{Z}}\mathcal{K}^{\rm mass}_n e^{in(x-x')}.
\end{equation}
Since $\lambda_{+,n}>0$ and $\lambda_{-,n}>0$, the mass deformation kernel
is positive-definite just like in the wormhole examples.
From the expressions \eqref{eq:lambda1}, \eqref{eq:lambda2} for $\lambda_{\pm,n}$, we infer that
the stiffness kernel has the universal high-frequency behavior
\begin{equation}
\mathcal{K}^{\rm mass}_n\to {1\over 2|n|},
\qquad
|n|\to\infty .
\end{equation}
As in the previous sections, this signals a universal logarithmic divergence in the 2-point function of the mass density operator $\langle \mathcal{M}(x)\mathcal{M}(x')\rangle$ in Liouville CFT.

The derivation of the spectral density for mass deformations is entirely analogous to the previous part, so we simply quote the result:
\begin{equation}
    \rho_M(\omega)=\frac{c}{3\pi}\frac{Y(\omega)}{(X(\omega)+B(\omega))^2+Y(\omega)^2}\,.
\end{equation}
Here $Y,X$ and $B$ are the same functions as in \eqref{eq:threefunctions}. At large frequencies, the UV tail is now damped as $1/\omega$:
\begin{equation}
    \rho_M(\omega) \sim \tfrac{c}{3\pi} \,\omega^{-1} \cos^2(\omega \tau_0).
\end{equation}
Similarly to shape deformations, $\rho_M$ has alternating maxima and minima as illustrated in the plot \ref{fig:massspectrumBH} occuring at $\rho_M'(\omega)=0$. The minima occur exactly at $\rho_M(\omega)=0$ i.e., at the locations of the poles of the function $B(\omega)$\footnote{$\omega=r_H$ is a spurious pole since the residue vanishes.},
\begin{equation}
    \text{Local minima:} \quad  \tan(\omega \tau_0)=\frac{\omega}{r_H\cot(r_H\tau_0)}\,.
\end{equation}
At large frequencies, the locations of the maxima asymptote to the locations of zeroes of $B(\omega)$,
\begin{equation}
    \text{Local maxima:} \quad (r_0^2-\omega^2)\tan(\omega\tau_0)=r_H\omega\cot(r_H\tau_0), \qquad \omega \to \infty\,.
\end{equation}
Notice interestingly by comparing with the local extrema conditions derived for shape deformations that at large frequencies, the extrema occur approximately at complementary frequencies for the two deformations i.e., the local maxima for shape deformations approximately occur at the local minima for mass deformations and vice versa. This feature is also illustrated by comparing the plots in figures \ref{fig:shapespectrumBH} and \ref{fig:massspectrumBH}.

Moreover, it turns out that the relaxation pole is identical for both the shape and mass deformations (both are determined by equation \eqref{eq:polelocation}) thereby giving the same relaxation time for both the types of deformations, plotted in figure \ref{fig:relaxation_time}.

\subsection{Corrections to apparent-horizon entropy from shape deformations}\label{sec:ahshape}

As described in the beginning of this section, the one-sided black hole is dual to a pure state $\ket{\Psi}$ in a single copy of the CFT at the boundary \cite{Chandra:2022fwi}. The semiclassical gravitational partition function computes the norm of $\ket{\Psi}$ after coarse-graining over an ensemble of microstates. Hence, the coarse-grained density matrix $\bar\rho = \overline{\ketbra{\Psi}{\Psi}}$ can be mixed, and may thus carry a Von Neumann entropy $S(\bar \rho)$. 

In the dual geometry, this Von Neumann entropy is computed by the length of the apparent horizon $L(\gamma)$.
In the spherically symmetric, non-rotating case, this is the minimal geodesic in the spatial slice at $y_H = \frac{\pi}{2r_H}$, whose length is determined just by the horizon radius $L(\gamma) = 2\pi r_H$. In our case, which is still non-rotating but allows for a shape deformation profile $\xi(x)$ of the shell, the length of the apparent horizon varies with $\xi$. At linear order in the deformation, there is no correction to the length $L(\gamma)$, because the horizon geodesic is locally extremal in length (and we dropped the rigid translation zero mode of $\xi$).    

At quadratic order, there are two sources of corrections to the length of the apparent horizon: $1)$ the displacement of the geodesic from its original locus, and $2)$ the change in the hyperbolic metric on the original locus. Unlike the computation of the stiffness kernel where it was sufficient to compute the linearized solution to the Liouville equation, we will also need the solution to the Liouville solution at quadratic order for the present computation.
We work on the real $r_H$ branch with heavy shells $m_0>1$, with the undeformed horizon located at constant $y_H = \frac{\pi}{2r_H}$. 

To find the deformed geodesic, we will extremise the length functional
\begin{equation}\label{eq:lengthfunctional}
    L[y(x)] = \int_0^{2\pi}\dd x \,e^{\frac{1}{2}\Phi(x,y(x))}\sqrt{1+y'(x)^2}
\end{equation}
to quadratic order in the deformation parameter $\epsilon$. To this end, we define the following expansions for the curve $y(x)$ and the Liouville field:
\begin{equation}
    \begin{split}
        & y(x)= y_H+\epsilon \hspace{0.2mm}\eta(x), \\[1em] 
        & \Phi(x,y)=\Phi_0(y)+\epsilon \hspace{0.2mm}\varphi(x,y)+\epsilon^2 \chi(x,y).
    \end{split}
\end{equation}
Here $\Phi_0(y)$ is the background Liouville solution \eqref{eq:backgroundLV} in the BTZ region (called $\Phi_{0,-}(y)$ there). Using that $\Phi_0(y_H) = 2\log r_H$, $\Phi_0'(y_H) = 0$ and $\Phi_0''(y_H) = 2r_H^2$, we have
\begin{equation}
    \Phi(x,y(x)) = 2\log r_H + \epsilon\hspace{0.2mm} \varphi(x,y_H)  + \epsilon^2 \Big(r_H^2\eta(x)^2 + \eta(x) \partial_y \varphi(x,y_H) + \chi(x,y_H) \Big)+ \mathcal{O}(\epsilon^3).
\end{equation}
Plugging this into $L[y(x)]$ and expanding to quadratic order in $\epsilon$ (noting that $\int_0^{2\pi}\dd x \,\varphi(x,y_H) = 0$ because we removed the zero mode), we find
an effective action for the displacement field $\eta(x)$:
\begin{equation}\label{eq:effectiveaction}
    \Delta L[\eta(x)] = \epsilon^2 \frac{r_H}{2} \int_0^{2\pi}dx\Big[\chi(x,y_H)+\tfrac{1}{4}\varphi(x,y_H)^2+\eta'(x)^2+r_H^2\eta(x)^2 +\eta(x)\partial_y\varphi(x,y_H)\Big]
\end{equation}
where $\Delta L$ is the change in the length from the undeformed value, i.e.~$L = 2\pi r_H + \Delta L$. Extremising this functional with respect to $\eta(x)$, we get the Euler-Lagrange equation
\begin{equation}\label{eq:E-L}
    (-\partial_x^2+r_H^2)\eta(x)=-\frac{1}{2}\partial_y \varphi(x,y_H)\,.
\end{equation}
We can solve this by doing the Fourier expansions $\eta(x) = \sum_{n} \eta_n e^{inx}$ and $\varphi(x,y) = \sum_{n} \varphi_n(y) e^{inx}$, where $\varphi_n(y)$ are the Fourier coefficients of the first order solution $\varphi_-(x,y)$ found in \eqref{eq:varph1},
\begin{equation}
    \varphi_n(y) = - 2m_0
{\frac{\lambda_{+,n}}{\lambda_{+,n}+\lambda_{-,n}}}
\xi_n \,\widehat u_{-,n}(y).
\end{equation}
Solving \eqref{eq:E-L} for $\eta_n$,
and plugging this back into the effective action \eqref{eq:effectiveaction}, we find
\begin{equation}
    \Delta L[\xi] = \frac{\epsilon^2}{2} (2\pi r_H) \sum_{n\neq 0} \mathcal{Q}^{\rm shape}_n \xi_n \xi_{-n},\vspace{-2mm}
\end{equation}
where
\begin{equation}
\mathcal{Q}^{\rm shape}_n
=
c_n(y_H)
+{1\over 4}f_n(y_H)^2
-{1\over 4}{f'_n(y_H)^2\over n^2+r_H^2}.
\end{equation}
Here we defined $\varphi_n(y) \equiv f_n(y)\, \xi_n$ to bring out the dependence on $\xi_n$. The $c_n(y)$ are the expansion coefficients of the zero mode $\chi_0(y) = \frac{1}{2\pi} \int_0^{2\pi} \dd x\, \chi(x,y)$ in the basis of the first-order solution,
\begin{equation}
    \chi_0(y) = \sum_{n\neq 0} c_n(y)\, \xi_n\xi_{-n}.
\end{equation}
We will have to determine these coefficients $c_n(y)$ by solving the Liouville equation expanded around $\Phi_{0,-}(y)$ to second order. This gives the following inhomogeneous ODE:
\begin{equation}\label{eq:2ndorderLiouville}
   \big(\partial_y^2-2r_H^2 \csc^2(r_Hy)\big)\,c_n(y)=r_H^2 \csc^2(r_Hy) f_n(y)^2 \,.
\end{equation}
We can solve it formally by integrating the RHS against the Green's function for the differential operator on the LHS. The boundary conditions are determined by the $2^{\rm nd}$ order junction conditions and amount to a Robin boundary condition at $y=\tau_0$, as further explained in appendix \ref{app:entropy}. Relegating the details to the appendix, the result is
\begin{equation}
    c_n(y_H)=
\int_0^{ r_H\tau_0} \dd s\,G_\gamma(t_H,s)\csc^2(s)\,f_n(\tfrac{s}{r_H})^2+
\zeta_n^{\rm shape} K_\gamma(t_H), \quad t_H = r_H y_H.
\end{equation}
Here $G_\gamma(t,s)$ and $K_\gamma(t)$ are the Robin Green's function and bulk-to-boundary propagator, respectively, derived in appendix \ref{app:entropy}. $\zeta^{\rm shape}_n$ is given in \eqref{eq:zetashape} and $\gamma$ in \eqref{eq:gammashape}. We stress that we have thus determined a fully explicit expression for the Fourier coefficients $\mathcal{Q}^{\rm shape}_n$ of the leading variation $\Delta L$ of the apparent-horizon length. 

This allows us to obtain the coarse-grained entropy associated to the state $\rho_\xi$ in the dual CFT:
\begin{equation}
    S(\overline{\rho_\xi}) = \frac{2\pi r_H}{4G_N} + \Delta S^{\rm shape}[\xi]\,,
\end{equation}
where
\begin{equation}
    \Delta S^{\rm shape}[\xi] = \frac{\Delta L[\xi]}{4 G_N} = \frac{\epsilon^2}{2} \frac{2\pi r_H}{4 G_N} \sum_{n\neq 0} \mathcal{Q}^{\rm shape}_n \xi_n \xi_{-n}.
\end{equation}
In figure \ref{fig:AHentropy}, we have plotted the change in apparent-horizon entropy due to the $n^{\rm th}$ Fourier mode $\xi_n$ for several values of $n$, as a function of $\tau_0$ and $r_H$. In the domain $\frac{\pi}{2}<r_H\tau_0 <\pi$, we observe that the entropy correction is always positive.

\subsection{Corrections to apparent-horizon entropy from mass deformations}\label{sec:ahmass}
 
Now we repeat the previous section for mass deformations. We have a family of CFT states $\ket{\Psi_\mu}$ dual to the one-sided black hole with inhomogeneous mass distribution along the shell, labeled by the mass deformation function $\mu(x)$. Their coarse-grained entanglement entropy $S(\overline{\rho_\mu})$ is computed holographically by $\frac{1}{4G_N}$ times the length of the apparent horizon $L = 2\pi r_H + \Delta L[\mu]$. 

Following the same steps as above, the correction to the apparent-horizon length takes the form
\begin{equation}
\Delta L[\mu]
=
2\pi r_H\frac{\epsilon^2}{2}
\sum_{n\neq 0}\mathcal{Q}^{\rm mass}_n\mu_n\mu_{-n},
\end{equation}
where
\begin{equation}
\mathcal{Q}^{\rm mass}_n
=
d_n(y_H)
+{1\over 4}g_n(y_H)^2
-{1\over 4}{g'_n(y_H)^2\over n^2+r_H^2}.
\end{equation}
Here $g_n(y)$ are the Fourier coefficients of the first-order solution $\varphi_-(x,y)$ on the black hole side, found in
\eqref{eq:massdefsol}:
\begin{equation}
    \varphi_-(x,y) = \sum_{n\neq 0} g_n(y)\mu_n e^{inx},
\qquad
g_n(y)=\frac{2}{\lambda_{+,n}+\lambda_{-,n}}\widehat u_{-,n}(y).
\end{equation}
We also defined $d_n(y)$ to be the expansion coefficients of the second order zero mode solution, in the basis of the first-order solution:
\begin{equation}
\chi_{-,0}(y)=\sum_{n\neq 0}d_n(y)\mu_n\mu_{-n}.
\end{equation}
It remains to derive $d_n(y_H)$ from the Liouville equation expanded to second order. In appendix \ref{app:entropy}, we analyze the boundary conditions and show that at $y=\tau_0$ we again have Robin boundary conditions. The coefficients $d_n$ can then be expressed in terms of the Robin Green's function $G_\gamma$ and bulk-to-boundary propagator $K_\gamma$ as
\begin{equation}
d_n(y_H)
=
\int_0^{r_H\tau_0} \dd s\,G_\gamma(t_H,s)\csc^2(s)\,g_n(\tfrac{s}{r_H})^2
+
\zeta^{\rm mass}_n K_\gamma(t_H),\quad t_H = r_Hy_H.
\end{equation}
Here $\zeta^{\rm mass}_n$ and  $\gamma$ are given in Eq.~\eqref{eq:zetamass}.

 With these expressions in hand, we obtain the correction to the apparent-horizon entropy due to mass deformations of the shell:
\begin{equation}
    \Delta S^{\rm mass}[\mu] = \frac{\Delta L[\mu]}{4 G_N} = \frac{\epsilon^2}{2} \frac{2\pi r_H}{4 G_N} \sum_{n\neq 0} \mathcal{Q}^{\rm mass}_n \mu_n \mu_{-n}.
\end{equation}
We have plotted the Fourier coefficients $\Delta S^{\rm mass}_n$ as a function of $r_H$ and $\tau_0$ in figure \ref{fig:AHentropy_mass}. In the heavy regime $\frac{\pi}{2}<r_H\tau_0<\pi$, we now observe that the correction to the apparent-horizon entropy due to mass deformations is always negative.

\section{The two-sided PETS black hole} \label{sec:PETS}

\begin{figure}
    \centering
\tikzset{every picture/.style={line width=1.2pt}} 
\begin{tikzpicture}[x=0.75pt,y=0.75pt,yscale=-1.1,xscale=1.1]
\draw  [dash pattern={on 1.5pt off 1.5pt on 1.5pt off 1.5pt}]  (477.21,60) -- (477.21,180) ;
\draw  [dash pattern={on 1.5pt off 1.5pt on 1.5pt off 1.5pt}]  (450.79,60) -- (450.79,180) ;
\draw  [draw opacity=0][fill={rgb, 255:red, 202; green, 202; blue, 202 }  ,fill opacity=1 ] (172.69,119.67) .. controls (172.69,93.04) and (194.25,71.44) .. (220.84,71.44) .. controls (247.44,71.44) and (269,93.04) .. (269,119.67) .. controls (269,146.31) and (247.44,167.9) .. (220.84,167.9) .. controls (194.25,167.9) and (172.69,146.31) .. (172.69,119.67) -- cycle ;
\draw  [draw opacity=0] (244.32,161.79) .. controls (237.36,165.6) and (229.36,167.76) .. (220.84,167.76) .. controls (193.99,167.76) and (172.23,146.23) .. (172.23,119.67) .. controls (172.23,93.11) and (193.99,71.58) .. (220.84,71.58) .. controls (229.3,71.58) and (237.26,73.72) .. (244.18,77.48) -- (220.84,119.67) -- cycle ; \draw   (244.32,161.79) .. controls (237.36,165.6) and (229.36,167.76) .. (220.84,167.76) .. controls (193.99,167.76) and (172.23,146.23) .. (172.23,119.67) .. controls (172.23,93.11) and (193.99,71.58) .. (220.84,71.58) .. controls (229.3,71.58) and (237.26,73.72) .. (244.18,77.48) ;  
\draw  [draw opacity=0][dash pattern={on 1.5pt off 1.5pt on 1.5pt off 1.5pt}] (424,180) .. controls (424,174.48) and (441.91,170) .. (464,170) .. controls (484.73,170) and (501.78,173.94) .. (503.8,178.99) -- (464,180) -- cycle ; \draw  [dash pattern={on 1.5pt off 1.5pt on 1.5pt off 1.5pt}] (424,180) .. controls (424,174.48) and (441.91,170) .. (464,170) .. controls (484.73,170) and (501.78,173.94) .. (503.8,178.99) ;  
\draw  [draw opacity=0][fill={rgb, 255:red, 245; green, 166; blue, 35 }  ,fill opacity=0.46 ] (505,89.74) .. controls (505.63,89.43) and (482.75,104.75) .. (484.25,125.08) .. controls (485.75,145.42) and (509.11,149.11) .. (500.11,157.11) .. controls (491.11,165.11) and (488.42,158.17) .. (485.67,158.67) .. controls (482.92,159.17) and (482.44,163.42) .. (475.44,162.67) .. controls (468.44,161.92) and (462.89,160.94) .. (453.89,164.44) .. controls (444.89,167.94) and (447,157.93) .. (439.5,157.38) .. controls (432,156.83) and (423.89,155.4) .. (424,154.46) .. controls (424.12,153.52) and (444,136.58) .. (443.5,123.08) .. controls (443,109.58) and (421.75,93.33) .. (425.5,89.83) .. controls (429.25,86.33) and (431.67,90) .. (439.5,87.08) .. controls (447.33,84.17) and (447.67,78) .. (453.89,80.22) .. controls (460.11,82.44) and (469.22,82.44) .. (476.33,81.78) .. controls (483.44,81.11) and (480.33,86.22) .. (485.22,86.22) .. controls (490.11,86.22) and (489.27,83.93) .. (493.44,84.89) .. controls (497.62,85.85) and (504.38,90.06) .. (505,89.74) -- cycle ;
\draw  [draw opacity=0][fill={rgb, 255:red, 202; green, 202; blue, 202 }  ,fill opacity=1 ] (244.78,78.01) .. controls (248.75,76.82) and (252.96,76.18) .. (257.32,76.18) .. controls (281.48,76.18) and (301.06,95.8) .. (301.06,120) .. controls (301.06,144.19) and (281.48,163.81) .. (257.32,163.81) .. controls (252.77,163.81) and (248.38,163.11) .. (244.26,161.82) -- (257.32,120) -- cycle ; \draw   (244.78,78.01) .. controls (248.75,76.82) and (252.96,76.18) .. (257.32,76.18) .. controls (281.48,76.18) and (301.06,95.8) .. (301.06,120) .. controls (301.06,144.19) and (281.48,163.81) .. (257.32,163.81) .. controls (252.77,163.81) and (248.38,163.11) .. (244.26,161.82) ;  
\draw   (256.22,120.21) -- (260.38,120.21)(258.3,118.02) -- (258.3,122.4) ;
\draw [color={rgb, 255:red, 245; green, 101; blue, 35 }  ,draw opacity=1 ][line width=1.5]    (244.8,78) .. controls (232.16,111.01) and (231.29,128.1) .. (244.27,161.83) ;
\draw [shift={(244.27,161.83)}, rotate = 68.94] [color={rgb, 255:red, 245; green, 101; blue, 35 }  ,draw opacity=1 ][fill={rgb, 255:red, 245; green, 101; blue, 35 }  ,fill opacity=1 ][line width=1.5]      (0, 0) circle [x radius= 1.74, y radius= 1.74]   ;
\draw [shift={(244.8,78)}, rotate = 110.95] [color={rgb, 255:red, 245; green, 101; blue, 35 }  ,draw opacity=1 ][fill={rgb, 255:red, 245; green, 101; blue, 35 }  ,fill opacity=1 ][line width=1.5]      (0, 0) circle [x radius= 1.74, y radius= 1.74]   ;
\draw  [draw opacity=0] (504,180) .. controls (504,180) and (504,180) .. (504,180) .. controls (504,180) and (504,180) .. (504,180) .. controls (504,185.52) and (486.09,190) .. (464,190) .. controls (441.91,190) and (424,185.52) .. (424,180) .. controls (424,179.65) and (424.07,179.3) .. (424.21,178.96) -- (464,180) -- cycle ; \draw   (504,180) .. controls (504,180) and (504,180) .. (504,180) .. controls (504,180) and (504,180) .. (504,180) .. controls (504,185.52) and (486.09,190) .. (464,190) .. controls (441.91,190) and (424,185.52) .. (424,180) .. controls (424,179.65) and (424.07,179.3) .. (424.21,178.96) ;  
\draw   (424,60) .. controls (424,54.48) and (441.91,50) .. (464,50) .. controls (486.09,50) and (504,54.48) .. (504,60) .. controls (504,65.52) and (486.09,70) .. (464,70) .. controls (441.91,70) and (424,65.52) .. (424,60) -- cycle ;
\draw    (424,60) -- (424,180) ;
\draw    (504,60) -- (504,180) ;
\draw [color={rgb, 255:red, 245; green, 101; blue, 35 }  ,draw opacity=1 ][line width=1.5]    (424,91.31) .. controls (432.25,85.9) and (433.67,88.22) .. (439.5,87.08) .. controls (445.33,85.94) and (448.21,76.34) .. (457.75,81.58) .. controls (467.29,86.82) and (472.43,77.83) .. (481.57,84.34) .. controls (490.71,90.86) and (485,76.86) .. (504,90) ;
\draw [color={rgb, 255:red, 245; green, 108; blue, 35 }  ,draw opacity=1 ][line width=1.5]  [dash pattern={on 1.5pt off 1.5pt on 1.5pt off 1.5pt}]  (424,91.31) .. controls (428.5,91.01) and (439.5,100.45) .. (447.75,97.31) .. controls (456,94.16) and (458.75,101.63) .. (468.25,99.67) .. controls (477.75,97.7) and (477.88,94.55) .. (488.25,95.93) .. controls (498.63,97.31) and (494.25,90.62) .. (504,90) ;
\draw [color={rgb, 255:red, 245; green, 101; blue, 35 }  ,draw opacity=1 ][line width=1.5]    (424,153.15) .. controls (432.25,158.56) and (433.5,156) .. (439.5,157.38) .. controls (445.5,158.75) and (448.22,168.12) .. (457.75,162.88) .. controls (467.29,157.64) and (472.43,166.63) .. (481.57,160.11) .. controls (490.72,153.6) and (485,167.6) .. (504,154.46) ;
\draw [color={rgb, 255:red, 245; green, 108; blue, 35 }  ,draw opacity=1 ][line width=1.5]  [dash pattern={on 1.5pt off 1.5pt on 1.5pt off 1.5pt}]  (424,153.15) .. controls (428.5,153.44) and (439.5,144.01) .. (447.75,147.15) .. controls (456,150.3) and (458.75,142.83) .. (468.25,144.79) .. controls (477.75,146.76) and (477.88,149.9) .. (488.25,148.53) .. controls (498.63,147.15) and (494.25,153.84) .. (504,154.46) ;
\draw [color={rgb, 255:red, 245; green, 101; blue, 35 }  ,draw opacity=1 ][line width=1.5]    (424,154.46) .. controls (449,128.21) and (449,120.94) .. (424,91.31) ;
\draw [color={rgb, 255:red, 245; green, 101; blue, 35 }  ,draw opacity=1 ][line width=1.5]    (504,90) .. controls (477.67,115.71) and (478,132.13) .. (504,154.46) ;
\draw    (509.55,176.1) .. controls (515.68,193.48) and (488.4,195.35) .. (478.09,196.57) ;
\draw [shift={(475.25,197)}, rotate = 347.2] [fill={rgb, 255:red, 0; green, 0; blue, 0 }  ][line width=0.08]  [draw opacity=0] (5.36,-2.57) -- (0,0) -- (5.36,2.57) -- cycle    ;
\draw   (217.73,120.21) -- (221.88,120.21)(219.8,118.02) -- (219.8,122.4) ;
\draw   (450.79,60) .. controls (450.79,58.02) and (456.71,56.42) .. (464,56.42) .. controls (471.29,56.42) and (477.21,58.02) .. (477.21,60) .. controls (477.21,61.98) and (471.29,63.58) .. (464,63.58) .. controls (456.71,63.58) and (450.79,61.98) .. (450.79,60) -- cycle ;
\draw   (450.79,180) .. controls (450.79,178.02) and (456.71,176.42) .. (464,176.42) .. controls (471.29,176.42) and (477.21,178.02) .. (477.21,180) .. controls (477.21,181.98) and (471.29,183.58) .. (464,183.58) .. controls (456.71,183.58) and (450.79,181.98) .. (450.79,180) -- cycle ;
\draw (310.74,106.72) node [anchor=north west][inner sep=0.75pt]  [font=\large]  {$\times \ S^{1}$};
\draw (219.09,170.13) node [anchor=north west][inner sep=0.75pt]  [font=\small]  {$D^{\dagger }( -\tau _{0})$};
\draw (228.4,53.64) node [anchor=north west][inner sep=0.75pt]  [font=\small]  {$D( \tau _{0})$};
\draw (514,148.57) node [anchor=north west][inner sep=0.75pt]  [font=\small]  {$D_\xi^\dagger(-\tau _{0})$};
\draw (514,82.07) node [anchor=north west][inner sep=0.75pt]  [font=\small]  {$D_\xi(\tau _{0})$};
\draw (143.42,175.16) node [anchor=north west][inner sep=0.75pt]    {$a)$};
\draw (382,174.73) node [anchor=north west][inner sep=0.75pt]    {$b)$};
\draw (504,193.07) node [anchor=north west][inner sep=0.75pt]    {$x$};
\draw (177.23,109.16) node [anchor=north west][inner sep=0.75pt]    {$\ell _{+}$};
\draw (280.91,109.59) node [anchor=north west][inner sep=0.75pt]    {$\ell _{-}$};
\end{tikzpicture}
    \caption{The two-sided black hole set-up. $a)$ The standard spherically symmetric geometry, gluing two portions of the non-rotating Euclidean BTZ black hole across a thin shell (orange). Here $\ell_- = 2\tau_0$, $\ell_+ = \beta-2\tau_0$ with the shell operator inserted at $\pm\tau_0$. $b)$ The same topology, viewed as $A\times I$, where now we allow the shell to wiggle along the transverse $S^1$ direction.}
    \label{fig:two_sided_bh}
\end{figure}
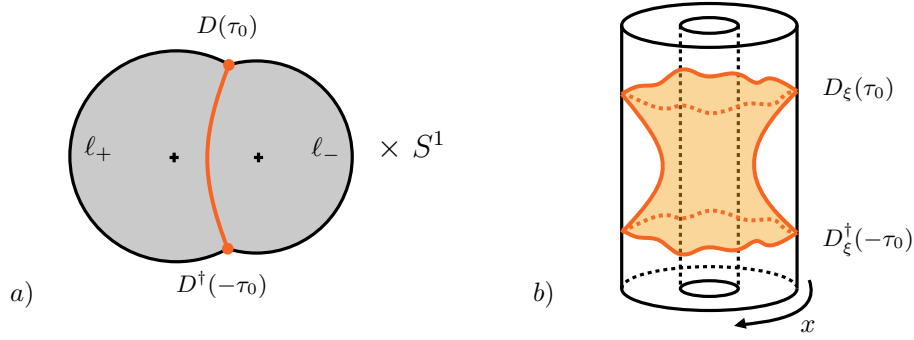

In this section, we study deformations of the two-sided black hole with a thin shell behind the horizon. Its geometry was constructed in \cite{Sasieta:2022ksu, Chandra:2024vhm} by gluing two portions of a non-rotating BTZ black hole across a thin shell of heavy matter. Topologically, it is an annulus times an interval $A\times I$, with a circular defect along the non-contractible cycle of the annulus (see Fig.~\ref{fig:two_sided_bh}). 

The dual CFT state is a partially entangled thermal state (PETS) \cite{Goel:2018ubv} living in $\mathcal{H}_L\otimes \mathcal{H}_R$, which is obtained by acting on the purifying side of the thermofield double (at inverse temperature $\beta$) with an operator $\hat{D}_\Sigma(\tau_0) = D_\Sigma(\tau_0)\otimes I$ inserted at Euclidean time $\tau_0$:
\begin{equation}\label{eq:PETS}
    \ket{\rm PETS_{\beta,\tau_0}} = \hat{D}_\Sigma(\tau_0) \ket{\rm TFD_\beta}, \quad \ket{\rm TFD_\beta} = \sum_{n} e^{-\beta E_n/2}\ket{E_n}_L\ket{E_n}_R.
\end{equation}
The semiclassical gravitational path integral evaluated on the Euclidean two-sided thin-shell black hole $Z_{\rm BH} =e^{-I_{\rm BH}}$ computes the norm of this state, again after appropriate coarse-graining \cite{deBoer:2023vsm}. The norm of the PETS state can also be rewritten as the torus two-point function of $D_\Sigma$:
\begin{equation}
   \braket{\rm PETS_{\beta,\tau_0}}{\rm PETS_{\beta,\tau_0}} = \expval{D_\Sigma^\dagger (-\tau_0)D_\Sigma(\tau_0)}_{\beta}.
\end{equation}
We work in the conformal frame where the boundary torus is flat, with coordinate $z = x+iy$, and $z\sim z+2\pi \sim z+ i\beta$.  
We take $y \in (-\frac{\beta}{2},\frac{\beta}{2})$, with the shell inserted at $y = \pm \tau_0$. The full 3d geometry can be foliated by radial slices that are hyperbolic cylinders $e^\Phi |\dd z|^2$ with finite height $y \in (0,\tfrac{\beta}{2})$. We impose ZZ boundary conditions at $y = 0$ and $y=\frac{\beta}{2}$. 

In this hyperbolic slicing, the Einstein equations reduce to the Liouville equation $\partial\bar\partial \Phi = \frac{1}{2}e^{\Phi}$ away from the shell locus. 
The shell itself obeys a set of junction conditions. For the spherically symmetric shell, the relevant background Liouville solution is \cite{Chandra:2024vhm}
\begin{equation}\label{eq:PETSbackground}
    \Phi_0(y)=\begin{cases}
         \Phi_{0,-}(y) = -2\log \left(\frac{1}{r_H}\sin(r_H y)\right), \qquad & 0<y<\tau_0,\\
        \Phi_{0,+}(y) = -2\log \left(\frac{1}{r_H'}\sin\left(r_H'(\frac{\beta}{2}-y)\right)\right), \qquad &\tau_0<y<\frac{\beta}{2}\,.
    \end{cases}
\end{equation}
Here $r_H, r_H'$ are the radii of the two black hole horizons. They are related to the two moduli $\beta,\tau_0$ by the Israel junction conditions, which in this case read
\begin{equation}
   r_H \csc(r_H\tau_0)=r_H'\csc (r_H'(\tfrac{\beta}{2}-\tau_0)), \quad m_0 =-r_H \cot(r_H\tau_0)-r_H'\cot(r_H'(\tfrac{\beta}{2}-\tau_0)).
\end{equation}
The proper radius of the shell locus, calculated in the hyperbolic metric, is 
$  r_0 =  r_H \csc(r_H\tau_0)$. So, if we fix $(\beta,\tau_0,m_0)$, then $(r_H,r_H',r_0)$ are uniquely determined. From now on we also take the shell to be in the heavy regime $m_0^2 > r_H^2 - (r_H')^2$.

Having set up the background solution, we want to study deformations that explicitly break spherical symmetry. As in the previous section, we focus on time-reflection-symmetric shape deformations $\xi(x)$ of the shell locus (as illustrated in Fig.~\ref{fig:two_sided_bh}), as well as deformations of the mass distribution $\mu(x)$ of the shell. The response to these deformations is measured by stiffness kernels, which we compute in section \ref{sec:PETS_stiffness}. In the holographic dual, we obtain a family of CFT states $\ket{\Psi_{\xi}}, \ket{\Psi_\mu}$ that are continuously connected to the PETS state. In section \ref{sec:PETS_ETH}, we study how the ETH microstate statistics of these deformed states behaves as we vary $\xi$ and $\mu$. Finally, in section \ref{sec:PETS_entanglement}, we study how the PETS entanglement entropy between the two CFT Hilbert spaces is corrected by the deformation moduli.

\subsection{Stiffness kernels and their spectrum}\label{sec:PETS_stiffness}

To set up the deformation problem, we proceed as in the previous section and linearize the Liouville equation to first order around the background saddle \eqref{eq:PETSbackground}, giving
\begin{equation}
    \begin{split}
         \left(\partial_x^2+\partial_y^2-2r_H^2 \csc^2(r_Hy)\right)\varphi_-=0, \qquad & 0<y<\tau_0 \\[0.8em]
        \left(\partial_x^2+\partial_y^2-2(r'_H)^2 \csc^2(r'_H(\tfrac{\beta}{2}-y))\right)\varphi_+=0, \qquad & \tau_0<y<\frac{\beta}{2}\,.
    \end{split}
\end{equation}
As before we expanded $\Phi_\pm(x,y) = \Phi_{0,\pm}(y) + \epsilon \varphi_{\pm}(x,y)$. Solving these equations in Fourier space, a basis of solutions that are regular at the two ZZ boundaries is
\begin{equation}
    \begin{split}
        u_{-,n}(y) &= \cosh(|n|y)-\frac{r_H}{|n|}\cot(r_Hy)\sinh(|n|y), \\
        u_{+,n}(y) &= \cosh(|n|(\tfrac{\beta}{2}-y))-\frac{r'_H}{|n|}\cot(r'_H(\tfrac{\beta}{2}-y))\sinh(|n|(\tfrac{\beta}{2}-y))
    \end{split}
\end{equation}
with $u_\pm(x,y) = \sum_{n\in\mathbb{Z}}u_{\pm,n}(y)e^{inx}$. The $n=0$ mode should be understood as the smooth $n\to 0$ limit of the above expressions. As before, we introduce the normalized basis functions and the Dirichlet-to-Neumann eigenvalues by
\begin{equation}
\widehat u_{\pm,n}(y)= {u_{\pm,n}(y)\over u_{\pm,n}(\tau_0)} ,
 \qquad  \lambda_{\pm,n} = \mp \,\partial_y  \widehat{u}_{\pm,n}(\tau_0).
\end{equation}
They can be written in closed form as
\begin{equation}\label{eq:PETSlambda}
    \lambda_{\pm,n}= \frac{(n^2+r_0^2)\tanh(|n|\frac{\ell_\pm}{2})-|n|a_\pm }{|n|-a_\pm \tanh(|n|\frac{\ell_\pm}{2})}\,,
\end{equation}
where we introduced some compact notation:
\begin{equation}
    \ell_-= 2\tau_0,\quad \ell_+=\beta-2\tau_0, \quad a_-=r_H\cot(\tfrac{r_H\ell_-}{2}),\quad a_+=r_H'\cot(\tfrac{r_H'\ell_+}{2})\,.
\end{equation}
Next, we want to expand $\varphi_\pm$ in the basis $\widehat{u}_{\pm,n}$. The expansion coefficients are determined by the linearized junction conditions, which are structurally identical to those of the one-sided black hole, written in \eqref{eq:linearizedjunction}. We can thus readily write down the solutions for the two types of deformations.

\subsubsection{Stiffness towards shape deformations}
 The shell locus is parametrized as $y = \tau_0 + \epsilon \xi(x)$, where $\xi(x)$ is the shape deformation modulus along the $S^1$ direction. From the linearized junction conditions in \eqref{eq:linearizedjunction}, we derive
\begin{equation}\label{eq:linshape}
    \varphi_\pm(x,y)= \pm 2m_0\sum_{n\in\mathbb Z}
{\frac{\lambda_{\mp,n}}{\lambda_{+,n}+\lambda_{-,n}}}
\xi_n e^{inx}\,\widehat u_{\pm,n}(y).
\end{equation}
This is structurally precisely the same as for the one-sided black hole; the only difference is that the Dirichlet-to-Neumann eigenvalues $\lambda_{\pm,n}$ are now given by \eqref{eq:PETSlambda}.

Following the same steps as in section \ref{eq:1sidedstiffnesskernels}, we evaluate the effective action  \eqref{eq:effectiveactionshape} for shape deformations $S_2[\xi]$ on the first-order solution \eqref{eq:linshape}. Using that for the PETS black hole, the undeformed solution \eqref{eq:PETSbackground} satisfies $\Phi_0''(\tau_0)=2r_0^2$ and $\Phi_0(\tau_0)=2\log r_0$, the quadratic on-shell action is
\begin{equation}
  S_2[\xi]  = -\frac{1}{2}\sum_{n\in \mathbb{Z}} \mathcal{K}^{\rm shape}_n \xi_n\xi_{-n},
\end{equation}
with
\begin{equation}
\mathcal{K}^{\rm shape}_n
=
m_0
\left(r_0^2
+n^2\log r_0-
m_0{\lambda_{+,n}\lambda_{-,n}\over \lambda_{+,n}+\lambda_{-,n}}
\right).
\end{equation}

From now on, we tune the background solution to the symmetric point $\tau_0 = \frac{\beta}{4}$, in which case $r_H = r_H'$ and $m_0 = -2r_H \cot(\frac{r_H \beta}{4})$ with $\frac{\pi}{2}<\frac{r_H\beta}{4}<\pi$. This simplifies many of the expressions below (although in principle we could keep the full $\tau_0$ dependence), because $\lambda_{+,n}=\lambda_{-,n}$. For this value of $\tau_0$, the shape deformation kernel for the PETS black hole takes the explicit form:
\begin{equation}\label{eq:symmetricK}
    \mathcal{K}^{\rm shape}_n(\beta)=m_0\left[\frac{n^2}{2}\log\left(r_H^2+\frac{m_0^2}{4}\right) -\frac{m_0n^2\tanh\left(\frac{|n|\beta}{4}\right)-2|n|r_H^2}{2|n|+m_0\tanh\left(\frac{|n|\beta}{4}\right)}\right]\,.
\end{equation}
The above stiffness kernel has the following behaviour at high temperatures:
\begin{equation}
    \mathcal{K}_n^{\rm shape}(\beta)=\frac{4\pi^2 m_0}{\beta^2}-\frac{m_0^2(\pi^2-8)}{2\beta}-m_0 n^2\log\left(\frac{\beta}{2\pi}\right)+O(\beta^0)\,.
\end{equation}
We have only included the singular terms in the above high temperature expansion. The leading divergence comes from the $r_H^2$ term in the kernel since $r_H\sim \frac{2\pi}{\beta}$ at high temperatures. Notice that both the power law divergent terms are independent of the Fourier mode $n$.

At low temperatures, the kernel can be expanded as
\begin{equation}
    \mathcal{K}_n^{\rm shape}(\beta)=-\frac{m_0^2n^2}{2|n|+m_0}+m_0n^2\log\left(\frac{m_0}{2}\right)+\frac{16\pi^2}{\beta^2}\frac{2|n|(2n^2+|n|m_0+m_0^2)}{m_0(2|n|+m_0)}+O(\beta^{-3})\,.
\end{equation}
For $n\neq 0$, we see that the kernel approaches a constant value which is independent of temperature. For the stationary mode $n=0$, the low temperature behaviour is much softer, 
\begin{equation}
    \mathcal{K}_0^{\rm shape}(\beta)=\frac{128\pi^2}{\beta^3}+O(\beta^{-4}).
\end{equation}

\paragraph{Saddle-point energy of PETS state.} The energy expectation value in the PETS state (at the symmetric point $\tau_0 = \frac{\beta}{4}$) can be computed by taking the $\beta$-derivative of the norm,
\begin{equation}
   \overline{\langle H \rangle}_{\rm PETS,\beta} = -\frac{\partial}{\partial \beta}\log Z_{\rm BH}.
\end{equation}
Here $H \equiv \frac{1}{2}(H_L+H_R)$ is the total Hamiltonian, and the overline denotes an appropriate microcanonical averaging over the matrix elements of the shell operator $|\mel{m}{D_\Sigma}{n}|^2$ in energy eigenstates \cite{Sasieta:2022ksu}.
Using the quadratic correction to the on-shell action due to shape deformations that we found above, $-\log Z_{\rm BH}[\xi] = I_0(\beta) + \frac{c\epsilon^2}{3}S_2[\xi]$, we can calculate the change in saddle-point energy caused by the shape deformation of the shell:
\begin{equation}
     \Delta E_{\rm shape}(\beta)=-\frac{c}{6}\epsilon^2 \sum_{n \neq 0} \partial_\beta \mathcal{K}_n^{\rm shape}(\beta)\xi_n\xi_{-n}\,.
\end{equation}
In the above expression, we used that the rigid translation of the shape deformation $\xi$ is set to zero, so that the two saddlepoint energies remain the same even after the shape deformation.
Using the high and low temperature expansions for the shape deformation kernel, we can write down corresponding expansions for the correction to the saddlepoint energy,
\begin{equation}
    \begin{split}
       \text{high }T:\quad  & \Delta E_{\rm shape}(\beta)= \frac{c}{6}\epsilon^2\sum_{n \neq 0}\left(\frac{8\pi^2 m_0}{\beta^3}-\frac{m_0^2(\pi^2-8)}{2\beta^2}+ \frac{ m_0n^2}{\beta}+O(\beta^0)\right)\xi_n\xi_{-n},\\
       \text{low }T: \quad & \Delta E_{\rm shape}(\beta)= \frac{c}{6}\epsilon^2\sum_{n \neq 0}\left(\frac{32\pi^2}{\beta^3}\frac{2|n|(2n^2+|n|m_0+m_0^2)}{m_0(2|n|+m_0)}+O(\beta^{-4})\right)\xi_n\xi_{-n}\,.
    \end{split}
\end{equation}
In both limits, we see that the leading correction is positive for each Fourier mode so the saddlepoint energy increases due to the shape deformation.

\paragraph{Displacement two-point function.}
As in the previous sections, we can also interpret the stiffness kernel for shape deformations as a 2-point function of the displacement operator in Liouville CFT:
\begin{equation}\label{eq:displacementPETS}
    \expval{\mathcal{D}_{\perp}(x)\mathcal{D}_{\perp}(x')}_{\rm Liouv} = \mathcal{K}^{\rm shape}_{\rm BH}(x,x';\beta) = \frac{c}{6\pi}\sum_{n\in\mathbb{Z}}\mathcal{K}^{\rm shape}_n(\beta) e^{in(x-x')}.
\end{equation}
Now, in the case of the PETS black hole, the topology is a solid torus, and hence the $y$-coordinate is periodic with period $\beta$. So the Liouville Hilbert space lives on a compact space, and hence we expect the spectral decomposition of \eqref{eq:displacementPETS} to be discrete.

Indeed, we find a discrete spectrum. Following the general method outlined in section \ref{sec:linearresponse}, we obtain the spectral density $\rho(\omega)$ by taking $\mathcal{K}_n^{\rm shape}$ in  \eqref{eq:symmetricK} with Matsubara frequencies $\omega_n$, and analytically continuing $i\omega_n\to\omega$. We extract the spectrum from the poles in the $\omega$-plane, which satisfy
\begin{equation}
    2\omega + m_0 \tan(\frac{\omega \beta}{4})=0.
\end{equation}
There are infinitely many solutions to the above transcendental equation. The solution corresponding to $\omega=r_H$ is spurious since the residue at the pole vanishes. So there is one non-trivial solution each in the range $\omega_j\in (\frac{(4j+2)\pi}{\beta}, \frac{(4j+4)\pi}{\beta})$ with $j\geq 1$. Hence the spectral density can be expressed as the following infinite discrete sum:
\begin{equation}\label{eq:PETS-displacement-spectral-density}
    \rho_D(\omega)=\sum_{j =1}^\infty \rho_{j,D} \delta(\omega-\omega_{j}), \qquad \rho_{j,D}= \frac{c}{3}\frac{2m_0 \omega_j (\omega_j^2-r_H^2)}{\frac{m_0\beta}{4}\sec^2(\frac{\beta \omega_j}{4})+2}\,.
\end{equation}
The weights $\rho_{j,D}$ are the residues of \eqref{eq:symmetricK} at the poles $\omega_j$. They are all positive because $\omega_j > r_H$.
As in the case of the torus wormhole from section \ref{sec:torus1point}, the discrete spectrum approaches a continuum in the low temperature limit $\beta\to \infty$, which acts as a decompactification limit. At high energies, the spectrum takes the form:
\begin{equation}
    \omega_{j}\sim \left(j+\frac{1}{2}\right)\frac{4\pi}{\beta}, \qquad \rho_{j,D}\sim \frac{8\pi cm_0^2}{3\beta^2}\left(j+\frac{1}{2}\right)\,.
\end{equation}
The above asymptotics give rise to lightcone singularities in the retarded Green's function,
\begin{equation}
    \mathcal{K}^{\rm shape}_{\rm ret}(t;\beta)\supset \frac{cm_0^2}{3}\Theta(t)\sum_{k \in \mathbb{Z}}(-1)^{k}\delta'(t-k\tfrac{\beta}{2})\,.
\end{equation}
This is exactly analogous to the torus 1-point case (\ref{eq:toruslightcone}) with the only difference being that the singularities are now spaced by $\frac{\beta}{2}$, since this is the spatial volume set by the size of the annulus. In addition, there is a relative phase of $(-1)^{k}$ which arises because the high-energy spectrum is now half-integer spaced.

\subsubsection{Stiffness towards mass deformations}

For deformations of the mass density $m = m_0 + \epsilon \mu(x)$, the junction conditions were written in \eqref{eq:linearizedjunctionmass}, with solution
\begin{equation}\label{eq:massdefPETSsol}
\varphi_\pm(x,y)
=
\sum_{n\in\mathbb Z}
{2\mu_n\over \lambda_{+,n}+\lambda_{-,n}}
e^{inx}\,\widehat u_{\pm,n}(y).
\end{equation}
The difference with the one-sided black hole is that the eigenvalues $\lambda_{\pm,n}$ are now given by \eqref{eq:PETSlambda}. Evaluating the quadratic effective action for mass deformations \eqref{eq:effectiveactionmass} on the above solution, we find as before
\begin{equation}
    \begin{split}
          S_2[\mu]=& -\frac{1}{2}\sum_{n \in \mathbb{Z}} \mathcal{K}^{\rm mass}_n\mu_n\mu_{-n}, \\
        \mathcal{K}^{\rm mass}_n= & \frac{1}{\lambda_{+,n}+\lambda_{-,n}}.
    \end{split}
\end{equation}
The mass deformation kernel is manifestly positive-definite, because $\lambda_{\pm,n} >0$.\footnote{A simple way to show this is that near the ZZ boundary, 
\begin{equation}
    u_{-,n}(y\to 0)\to \frac{n^2+r_H^2}{3}y^2 > 0, \qquad u'_{-,n}(y\to 0) \to \frac{2(r_H^2+n^2)}{3}y > 0.
\end{equation}
In addition, the second derivative is positive whenever $u_{-,n}$ is positive because
\begin{equation}
    u''_{-,n}(y)=(n^2+2r_H^2\csc^2(r_Hy))u_{-,n}(y)\,.
\end{equation}
This means $u_{-,n}(y)$ is positive and monotonically increasing everywhere in the domain. So, the logarithmic derivative $\partial_y \log u_{-,n}(y)$ is positive everywhere, including on the shell locus $y=\tau_0$, so that $\lambda_{-,n}>0$. By a similar argument, we can show $\lambda_{+,n}>0$.
} From \eqref{eq:PETSlambda}, we see that the high-frequency behaviour is $\lambda_{\pm, n} \sim |n|$, so we have the same universal decay $ \mathcal{K}^{\rm mass}_n \to \frac{1}{2|n|}$ as $n\to\infty$. At the symmetric point $\tau_0 = \frac{\beta}{4}$, we can write down the explicit expression
\begin{equation}\label{eq:masskernelPETS}
    \mathcal{K}^{\rm mass}_n(\beta) = \frac{1}{2}\frac{|n|+\frac{m_0}{2}\tanh\left(\frac{|n|\beta}{4}\right)}{(n^2+r_H^2+\frac{m_0^2}{4})\tanh\left(\frac{|n|\beta}{4}\right)+\frac{|n|m_0}{2}}.
\end{equation}
Notice that the above expression is deceptively identical to the expression for the mass deformation kernel for the torus 1-point wormhole, \eqref{eq:1pointWHmasskernel}. The important difference is the functional form of $r_H$ in terms of the temperature. For example, for the present PETS black hole setup, the leading behaviour at high temperatures is $r_H\sim \frac{2\pi}{\beta}$ whereas for the torus 1-point wormhole, $r_H \sim \sqrt{\frac{m_0}{\beta}}$.

As a function of inverse temperature $\beta$, the above kernel is monotonic
\begin{equation}
    \frac{\partial \mathcal{K}^{\rm mass}_n(\beta)}{\partial \beta}>0\,.
\end{equation}
It is also bounded, with a lower bound at the high-temperature point
\(\beta\to 0\) and an upper bound at the low-temperature point
\(\beta\to\infty\). The high-temperature expansion takes the form
\begin{equation}
     \mathcal{K}^{\rm mass}_n(\beta) = \frac{\beta}{2\pi^2}-\frac{m_0(12-\pi^2)}{16\pi^4}\beta^2- \frac{1}{8\pi^4}\left(\frac{(12-\pi^2)n^2}{12}+\frac{(\pi^2-10)m_0^2}{\pi^2}\right)\beta^3+O(\beta^4)\,.
\end{equation}
We see that the kernel decays linearly to zero at high temperatures and the leading term is universal since it does not depend on the mass $m_0$ and is also independent of the Fourier mode. The leading frequency dependence comes at $O(\beta^3)$. 

Similarly, the low temperature expansion takes the form
\begin{equation}
     \mathcal{K}^{\rm mass}_n(\beta) = \frac{1}{2}\frac{|n|+\frac{m_0}{2}}{n^2+\frac{m_0}{2}|n|+\frac{m_0^2}{4}}-\frac{8\pi^2}{\beta^2}\frac{|n|+\frac{m_0}{2}}{\left(n^2+\frac{m_0}{2}|n|+\frac{m_0^2}{4}\right)^2}+O(\beta^{-3})\,.
\end{equation}
So, at low temperatures, the kernel approaches a constant value. Also, notice that the above terms in the low temperature expansion are identical to those appearing in the low temperature expansion of the mass deformation kernel for the torus 1-point wormhole. However, the high temperature behaviour is qualitatively different due to the different scaling of $r_H$ with $\beta$ mentioned above.

\paragraph{Saddle-point energy.}
We now use the stiffness kernel to compute the corrections to the saddlepoint energies in the norm of the deformed PETS state. For the symmetric PETS state, the two saddlepoint energies are equal and their correction due to mass deformations is given by
\begin{equation}
     \Delta E_{\rm mass}(\beta)=-\frac{c}{6}\epsilon^2 \sum_{n} \partial_\beta \mathcal{K}_n^{\rm mass}(\beta)\mu_n\mu_{-n}\,.
\end{equation}
Since we showed above that the mass deformation kernel is monotonic in $\beta$ for any Fourier mode, we see that the correction to the energy from mass deformations is always negative,
\begin{equation}
    \Delta E_{\rm mass}(\beta)<0\,.
\end{equation}

\paragraph{Mass-density 2-point function.} In Liouville CFT, the mass-density operator 2-point function is computed by the stiffness kernel,
$\langle \mathcal{M}(x)\mathcal{M}(x')\rangle_{\rm Liouv} =\mathcal{K}_{\rm BH}^{\rm mass}(x,x') = \frac{c}{6\pi} \sum_{n\in\mathbb{Z}}\mathcal{K}^{\rm mass}_n(\beta)e^{in(x-x')}$. As in all previous examples, the 2-point function admits a decomposition into a positive-definite spectral density $\rho(\omega)$, which is obtained from the residues of the discrete infinite set of poles of \eqref{eq:masskernelPETS}. The spectral poles $\omega_{j}$ are solutions of the following transcendental equation\footnote{The equation also has the removable endpoint solution \(\omega=0\),
and the solution \(\omega=r_H\) is spurious because its residue vanishes.
The spectral sum is therefore over the remaining positive roots.},
\begin{equation} \label{eq:PETSmasspoleeqn}
    m_0\omega + 2\left(r_H^2+\frac{m_0^2}{4}-\omega^2\right)\tan\left(\frac{\beta \omega}{4}\right)=0\,.
\end{equation}
Note that $\omega=r_H$ is a spurious pole since the residue vanishes.
These are distinct from the spectral poles obtained for the shape kernel showing that the displacement and mass-density operators are supported on distinct spectra just as we observed in the case of the torus 1-point wormhole.
The spectral density is found to be:
\begin{equation} \label{eq:PETSmassspec}
    \rho_M(\omega;\beta)=\sum_{j=1}^{\infty}\rho_{j,M}\delta(\omega-\omega_{j}), \quad \rho_{j,M}= \frac{c}{6}\,\frac{2\omega_j+m_0 \tan\left(\frac{\beta \omega_j}{4}\right)}{4\omega_j\tan\left(\frac{\beta \omega_j}{4}\right)+\frac{\beta}{2}(\omega_j^2-r_H^2-\frac{m_0^2}{4})\sec^2\left(\frac{\beta \omega_j}{4}\right)-m_0}.
\end{equation}
Note, interestingly, that the spectral pole condition (\ref{eq:PETSmasspoleeqn}) and the spectral density (\ref{eq:PETSmassspec}) are identical to those of the torus 1-point wormhole after rescaling the temperature parameter $\beta \to \frac{\beta}{2}$. This is unlike the case with shape deformations. 

\subsection{Corrections to black hole microstate statistics from deformations}\label{sec:PETS_ETH}

In the previous section, we computed the quadratic corrections from shape and mass deformations to the gravitational on-shell action, which determines the norm of the PETS state. Microscopically, the norm of this state takes the form
\begin{equation}
   \braket{\rm PETS_{\beta,\tau_0}}{\rm PETS_{\beta,\tau_0}} =  \sum_{n,m} e^{-2\tau_0 E_n - (\beta-2\tau_0)E_m} |\langle E_m | D_\Sigma|E_n\rangle |^2.
\end{equation}
The semiclassical gravitational path integral computes the average of the norm, which by linearity becomes an average of the matrix elements 
 $\overline{|\langle E_m | D_\Sigma|E_n\rangle |^2}$. So the path integral contains information about the microstate statistics (in this case the variance), after performing the inverse Laplace transform. These statistics are closely related to the eigenstate thermalization hypothesis (ETH) \cite{Srednicki_1994,PhysRevA.43.2046,DAlessio:2015qtq}, as worked out in the thin-shell case by \cite{Sasieta:2022ksu}. 
 
 In this section, we will study how the variance of the defect microstates gets corrected to leading order by shape and mass deformations of the shell.
  Of course, in 2d CFT, the relevant matrix elements are labeled by primary conformal weights $h,\bar h \in (\frac{c-1}{24},\infty)$, 
\begin{equation}
    \overline{\big |\langle h,\bar h |D_\Sigma |h',\bar{h'}\rangle\big|^2},
\end{equation}
because the descendant states should not be treated as random variables \cite{Chandra:2022bqq, Chandra:2022fwi, deBoer:2023vsm, Chandra:2024vhm}. This can be taken into account by expanding in torus (defect) conformal blocks. However, the contribution of descendants is a 1-loop effect, while here we focus on the leading saddle only.

At the level of the on-shell action, the inverse Laplace transform becomes a Legendre transform. This is equivalent to computing the on-shell action with \emph{geodesic} boundary conditions instead of asymptotic boundaries. More specifically, the auxiliary Liouville field $\Phi$ that describes the metric in hyperbolic slicing should now obey geodesic boundary conditions at the two ends of the cylinder, instead of the ZZ boundary conditions that we used in the previous section. The proper lengths
of the two geodesic boundaries are fixed by the weights of the external heavy states,
\begin{equation}
    h=\frac{c}{24}(1+r_-^2),\qquad
    h'=\frac{c}{24}(1+r_+^2),
\end{equation}
so the two geodesic boundaries have lengths $2\pi r_-$ and $2\pi r_+$. We are considering non-rotating black holes, so $h=\bar h$ and we label the defect matrix elements as $\langle h | D_\Sigma| h'\rangle$.

Let us first discuss the undeformed, spherically symmetric saddle, which was obtained in \cite{Chandra:2024vhm}. Let \(y_\pm>0\) denote the coordinate distances from the shell to the two geodesic boundaries. We take the cylinder to be
\(-y_-<y<y_+\), with geodesic boundaries at \(y=-y_-\) and \(y=y_+\), and the shell at \(y=0\). Then the background Liouville solution is
\begin{equation}\label{eq:backgroundPETS2}
    \Phi_0(y)=\begin{cases}
         \Phi_{0,-}(y) = -2\log \left(\frac{1}{r_-}\cos(r_- (y+y_-))\right), \qquad & -y_-<y<0,\\
        \Phi_{0,+}(y) = -2\log \left(\frac{1}{r_+}\cos\left(r_+(y_+-y)\right)\right), \qquad &0<y<y_+\,.
    \end{cases}
\end{equation}
The parameters obey
\begin{equation}
    r_0=r_- \sec(r_-y_-)=r_+\sec(r_+y_+),\qquad
    m_0=r_-\tan(r_-y_-)+r_+\tan(r_+y_+).
\end{equation}
The gravitational on-shell action  (equivalently, up to a prefactor, the on-shell Liouville action) of this fixed-geodesic-length solution can be found in Eq.~(5.30) of \cite{Chandra:2024vhm}. Our job is to expand the action to quadratic order around the undeformed saddle \eqref{eq:backgroundPETS2}. 

As before, we write $\Phi_\pm(x,y) = \Phi_{0,\pm}(y)+\epsilon \varphi_\pm(x,y)$ and linearize the Liouville equations,
\begin{equation}
    \begin{split}
         \left(\partial_x^2+\partial_y^2-2r_-^2 \sec^2(r_-(y+y_-)\right)\varphi_-=0, \qquad & -y_-<y<0 \\[0.8em]
        \left(\partial_x^2+\partial_y^2-2r_+^2 \sec^2(r_+(y_+-y)\right)\varphi_+=0, \qquad & 0<y<y_+\,.
    \end{split}
\end{equation}
Let us denote $s_+ = y_+-y$ and $s_- = y + y_-$.
Going to Fourier space, a basis of solutions with the correct geodesic boundary conditions at $y_\pm$ is given by
\begin{equation}
    U_{\pm,n}(s)=\cosh(|n|s)+\frac{r_\pm}{|n|}\tan(r_\pm s)\sinh(|n|s),
    \qquad
    \widehat U_{\pm,n}(s):=\frac{U_{\pm,n}(s)}{U_{\pm,n}(y_\pm)} ,
\end{equation}
with the $n=0$ expression defined by the smooth $|n|\to 0$ limit. The Dirichlet-to-Neumann eigenvalues are defined as $\lambda_{\pm,n} = \partial_{s_\pm}\widehat U_{\pm,n}(s_\pm) \mid_{s_\pm = y_\pm}$, and take the explicit form
\begin{equation}
    \lambda_{\pm,n}
    =
    \frac{
    (n^2+r_0^2)\sinh(|n|y_\pm)
    +|n| p_\pm \cosh(|n|y_\pm)
    }{
    |n|\cosh(|n|y_\pm)+p_\pm\sinh(|n|y_\pm)
    },
    \qquad
    p_\pm:=r_\pm\tan(r_\pm y_\pm).
\end{equation}

For shape deformations of the shell locus, parametrized as $y = \xi(x)$ with $\xi(x)= \sum_n \xi_n e^{inx}$, the first-order solution $\varphi_\pm(x,y) = \sum_{n}\varphi_{\pm,n}(y)e^{inx}$ can be expanded in the $\widehat{U}_{\pm, n}$ basis precisely as in \eqref{eq:linshape}, only with $\hat{u}_{\pm,n}$ replaced by $\widehat{U}_{\pm,n}$. Evaluating the quadratic effective action $S_2[\xi]$ on $\varphi_\pm$, we obtain the quadratic correction to the on-shell action due to the shape deformation $\xi$. Using our holographic dictionary, we find that the variance is corrected due to shape deformations as
\begin{align}
    \frac{
    \overline{\big|\langle h'|D_{\Sigma}[\xi]|h\rangle\big|^2}
    }{
    \overline{\big|\langle h'|D_\Sigma[0]|h\rangle\big|^2}
    }
    &=
    \exp\left[
    -\frac{c}{3}\epsilon^2
    S_2[\xi]
    +O(\epsilon^3)
    \right]
    \nonumber\\
    &=
    \exp\left[
    \frac{c}{6}\epsilon^2
    \sum_{n\neq0}
\left(
   m_0( r_0^2+n^2\log r_0)-m_0^2\frac{\lambda_{+,n}\lambda_{-,n}}{\lambda_{+,n}+\lambda_{-,n}}
    \right)\xi_n\xi_{-n}
    +O(\epsilon^3)
    \right].
\end{align}
Here we denoted the defect operator creating the shell with non-zero shape deformation by $D_\Sigma[\xi]$, and the undeformed defect operator by $D_\Sigma[0]$.

Similarly, for mass density deformations $m(x)=m_0+\epsilon\mu(x)$, with $\mu(x)=\sum_{n}\mu_n e^{inx}$ and $\mu_0=0$, the first-order solution takes the same form as \eqref{eq:massdefPETSsol}, again with $\hat{u}_{\pm,n}$ replaced by $\widehat{U}_{\pm,n}$. Substituting the solutions $\varphi_\pm$ into the quadratic effective action $S_2[\mu]$ yields the second-order contribution of the mass deformation $\mu$ to the on-shell action. Through the holographic dictionary, this translates into the following correction to the variance:
\begin{align}
    \frac{
    \overline{\big|\langle h'|D_\Sigma[\mu]|h\rangle\big|^2}
    }{
    \overline{\big|\langle h'|D_\Sigma[0]|h\rangle\big|^2}
    }
    &=
    \exp\left[
    -\frac{c}{3}\epsilon^2
    S_2[\mu]
    +O(\epsilon^3)
    \right]
    \nonumber\\
    &=
    \exp\left[
    \frac{c}{6}\epsilon^2
    \sum_{n\neq0}
    \frac{\mu_n\mu_{-n}}{\lambda_{+,n}+\lambda_{-,n}}
    +O(\epsilon^3)
    \right].
\end{align}
Since $\lambda_{\pm,n}>0$, the variance is enhanced by the mass deformations.

\subsection{Corrections to PETS entanglement entropy from deformations}\label{sec:PETS_entanglement}
Recall that $\ket{\rm PETS_{\beta,\tau_0}}$, defined in  
\eqref{eq:PETS}, is an entangled state in two copies of the CFT. The entanglement between the two copies, which is quantified by the Von Neumann entropy $S_{\rm EE}(\rho_L)$ of the reduced density matrix $\rho_L = \Tr_R \rho$, is encoded geometrically by the length of the minimal geodesic on the spatial slice of the black hole.\footnote{Since we are using semiclassical gravity, this should be understood in a coarse-grained sense, where $\frac{A}{4G_N}$ accounts for the coarse-grained entanglement entropy $S(\bar \rho_L)$.} For the spherically symmetric PETS black hole, at the symmetric point $\tau_0=\frac{\beta}{4}$, this minimal geodesic is just determined by the horizon radius, as $L(\gamma) = 2\pi r_H = 2\pi r_H'$ \cite{Chandra:2024vhm}, hence the entanglement entropy is 
\begin{equation}
    S_{\rm EE} = \frac{2\pi r_H}{4G_N}.
\end{equation}

In this subsection, we compute the correction to the length of the minimal geodesic due to shape and mass density deformations of the shell that break spherical symmetry. For simplicity, we remain at the $\mathbb{Z}_2$-symmetric point $\tau_0 = \frac{\beta}{4}$. The method will be very similar to that of sections \ref{sec:ahshape} and \ref{sec:ahmass}. Namely, since $L(\gamma)$ is locally extremal in length, we have to go second order in the deformation parameter $\epsilon$ to obtain a non-trivial correction. This will involve solving the linearized Liouville equation to quadratic order. We discuss shape and mass deformations separately below.

\subsubsection{Entanglement entropy correction from shape deformations}

To determine the length of the minimal geodesic on the spatial slice of the PETS black hole, with deformed shell locus $y = \tau_0 + \epsilon \xi(x)$, we extremise the length functional $L[y(x)]$ given in
\eqref{eq:lengthfunctional}. The curve with minimal length will be parametrized as $y(x) = y_H + \epsilon \eta(x)$, where $y_H = \frac{\pi}{2r_H}$ is the location of the geodesic in the undeformed geometry. The Liouville field is expanded to second order as  
\begin{equation}
    \Phi(x,y) = \Phi_0(y)+\epsilon \hspace{0.2mm}\varphi(x,y)+\epsilon^2 \chi(x,y),
\end{equation}
where now $\Phi_0(y)$ is the PETS background \eqref{eq:PETSbackground} at the symmetric point $\tau_0 =\frac{\beta}{4}$, and $\varphi(x,y)$ is the first-order solution \eqref{eq:linshape} found in section \ref{sec:PETS_stiffness}. Expanding the length functional $L[y(x)]$ to quadratic order in $\epsilon$, and solving the Euler-Lagrange equation as in section \ref{sec:ahshape}, we evaluate the quadratic correction to the length of the minimal geodesic, finding
\begin{equation}
    \Delta L[\xi] = \frac{\epsilon^2}{2} (2\pi r_H) \sum_{n\neq 0} \mathcal{Q}^{\rm shape}_n \xi_n \xi_{-n},\vspace{-2mm}
\end{equation}
where
\begin{equation}
\mathcal{Q}^{\rm shape}_n
=
c_n(y_H)
+{1\over 4}f_n(y_H)^2
-{1\over 4}{f'_n(y_H)^2\over n^2+r_H^2}.
\end{equation}
This is structurally identical to the one-sided black hole case, with the difference being that now
the $f_n$ are the Fourier coefficients of the first-order PETS solution $\varphi_-(x,y)$ in \eqref{eq:linshape}, to wit:
\begin{equation}\label{eq:fnPETS}
    f_n(y)= - 2m_0
{\frac{\lambda_{+,n}}{\lambda_{+,n}+\lambda_{-,n}}}
\,\widehat u_{\pm,n}(y),
\end{equation}
and with the Dirichlet-to-Neumann eigenvalues $\lambda_{\pm,n}$ of the PETS black hole given in \eqref{eq:PETSlambda}. At $\tau_0 = \frac{\beta}{4}$, we have $\lambda_{+,n} = \lambda_{-,n}$, which simplifies $f_n(y) = -m_0 \widehat u_{\pm,n}(y)$.

The coefficient $c_n(y)$ is related to the zero mode of the second-order Liouville field as
\begin{equation}
    \chi_0(y) = \frac{1}{2\pi}\int_0^{2\pi}\dd x \chi(x,y)\,, \qquad \chi_{0}(y) = \sum_n c_n(y) \xi_n \xi_{-n}\,.
\end{equation}
To determine $c_n(y)$ we have to solve the second order Liouville equation, which is the same as \eqref{eq:2ndorderLiouville}, now with $f_n(y)$ given by \eqref{eq:fnPETS}. The novelty compared to section \ref{sec:ahshape} are the boundary conditions imposed on $c_n(y)$, which are regularity at the ZZ boundary $y=0$ and 
\begin{equation}\label{eq:neumannbc}
    c_n'\left(\frac{\beta}{4}\right)=\frac{m_0}{2}(n^2+2r_0^2)\,.
\end{equation}
The above condition arises from expanding the junction conditions to second order after imposing reflection symmetry, $c_n(y)=c_n(\tfrac{\beta}{2}-y)$. Note that \eqref{eq:neumannbc} is a Neumann boundary condition. In appendix \ref{app:entropy}, we compute the relevant Green's function on an interval with Neumann boundary conditions at one endpoint, using which we can readily compute $c_n(y)$. We find that its value at the undeformed horizon locus $y=y_H$ is
\begin{multline}
     c_n(y_H)=\frac{m_0}{2r_H}\frac{(n^2+2r_H^2 \csc^2(\Theta))}{R'(\Theta)}\\-\frac{1}{\Theta-\sin(\Theta)\cos(\Theta)}\left[\int_0^{\frac{\pi}{2}}\dd t \csc^2(t)R(t)f_n(t/r_H)^2+\int_{\frac{\pi}{2}}^\Theta \dd t \csc^2(t)N(t)f_n(t/r_H)^2\right]\,,
\end{multline}
where we defined the functions
\begin{equation}\label{eq:functions}
    \Theta=\frac{r_H\beta}{4}, \qquad R(t)=1-t\cot(t), \qquad N(t)=1+(\Theta-\sin(\Theta)\cos(\Theta)-t)\cot(t)\,.
\end{equation}
With these results, we can express the correction to the PETS entanglement entropy due to shape deformations as
\begin{equation}
    \Delta S_{\rm EE}[\xi]=\frac{\epsilon^2}{2} \frac{2\pi r_H}{4G_N}\sum_{n\neq 0}\mathcal{Q}_n^{\rm shape}\xi_n\xi_{-n}.
\end{equation}
In figure \ref{fig:PETSentanglement}, we have plotted $\Delta S_{\rm EE}$ as a function of temperature $T =1/\beta$, for a PETS black hole of mass $m_0 =5$ (the qualitative features of the plot are the same for different values of $m_0>1$).

At high temperatures, the above quadratic correction has the following expansion:
\begin{equation}
    \Delta S_{\rm EE}[\xi]=\frac{c}{3}\epsilon^2 \sum_{n\neq 0}\left(\frac{4\pi^2 m_0}{\beta^2}+\frac{m_0^2}{\beta}\left(\frac{\pi^2(4-\pi^2)}{16}-2\pi I_0\right)+O(\beta^0)\right)\xi_n\xi_{-n}\,,
\end{equation}
where 
\begin{equation}
    I_0=\int_0^{\frac{\pi}{2}}\dd t \csc^2(t)(1-t\cot(t))^3 \approx 0.21.
\end{equation}
At low temperatures, the correction has the asymptotics
\begin{equation}
    \Delta S_{\rm EE}[\xi]=\frac{8\pi^2 c}{3\beta^2}\epsilon^2 \sum_{n\neq 0}\left(m_0+\frac{2n^2}{m_0}-\frac{m_0^2}{2|n|+m_0}\right)\xi_n\xi_{-n}+O(\beta^{-3})\,.
\end{equation}
Note from these asymptotic expressions that the correction to the PETS entanglement entropy due to shape deformations of the shell is positive at both low and high temperatures.

\subsubsection{Entanglement entropy correction from mass deformations}

We can readily adapt the previous analysis to compute the quadratic correction to the PETS entanglement entropy due to mass deformations of the shell. We will keep the analysis brief and only highlight the differences from the analysis of shape deformations. The length of the deformed geodesic is found to be
\begin{equation}
\begin{split}
     \Delta L[\mu] & = \frac{\epsilon^2}{2} (2\pi r_H)  \sum_{n \neq 0} \mathcal{Q}^{\rm mass}_n \mu_n \mu_{-n},\\
     \mathcal{Q}^{\rm mass}_n & = d_n(y_H)+\frac{1}{4}g_n(y_H)^2-\frac{1}{4}\frac{g_n'(y_H)^2}{n^2+r_H^2}\,.
\end{split}
\end{equation}
Here $y_H = \frac{\pi}{2r_H}$ and we defined
\begin{equation}
    g_n(y) = {2\over \lambda_{+,n}+\lambda_{-,n}}
\,\widehat u_{\pm,n}(y), \qquad \chi_0(y) = \sum_{n\in \mathbb{Z}} d_n(y) \mu_n\mu_{-n}.
\end{equation}
The zero-mode coefficients $d_n(y)$ are found from the second-order Liouville equation
\begin{equation}
     (\partial_y^2-2r_H^2 \csc^2(r_Hy))d_n(y)=r_H^2 \csc^2(r_Hy) g_n(y)^2 \,
\end{equation}
by imposing regularity at the ZZ boundary and a Neumann boundary condition at $y= \frac{\beta}{4}$. Solving the above ODE gives the following expression for $d_n$ at the undeformed geodesic locus $y=y_H$,
\begin{multline}
     d_n(y_H)=-\frac{1}{\Theta-\sin(\Theta)\cos(\Theta)}\left[\int_0^{\frac{\pi}{2}}\dd t \csc^2(t)R(t)g_n(t/r_H)^2+\int_{\frac{\pi}{2}}^\Theta \dd t \csc^2(t)N(t)g_n(t/r_H)^2\right]\,,
\end{multline}
where $\Theta$, $N(t)$ and $R(t)$ are the same functions as in \eqref{eq:functions}. 
Therefore, the quadratic correction to the PETS entanglement entropy due to mass deformations of the shell is given by
\begin{equation}
    \Delta S_{\rm EE}[\mu]=\epsilon^2 \frac{2\pi r_H}{4G_N}\sum_{n\neq 0}\mathcal{Q}_n^{\rm mass}\mu_n\mu_{-n},
\end{equation}
which we plotted in figure \ref{fig:PETSentanglement2} as a function of temperature.

At high temperatures, the above correction has the following asymptotics:
\begin{equation}
     \Delta S_{\rm EE}[\mu]=-\frac{c}{6} \epsilon^2 \beta\sum_{n\neq 0}\left(\frac{1}{8}-\frac{1}{2\pi^2}+\frac{4I_0}{\pi^3}\right)\mu_n\mu_{-n}+O(\beta^2)\,,
\end{equation}
where 
\begin{equation}
    I_0=\int_0^{\frac{\pi}{2}}\dd t \csc^2(t)(1-t\cot(t))^3 \approx 0.21 \implies \left(\frac{1}{8}-2\pi^2+\frac{4I_0}{\pi^3}\right) \approx 0.1\,.
\end{equation}
At low temperatures, it has the following asymptotics,
\begin{equation}
    \Delta S_{EE}=-\frac{8c \pi^2}{3 \beta^2} \epsilon^2 \sum_{n\neq 0}I_n \mu_n\mu_{-n}+O(\beta^{-3})\,,
\end{equation}
where
\begin{equation}
    I_n=\frac{1}{3\left(n^2+\frac{m_0|n|}{2}+\frac{m_0^2}{4}\right)^2}\int_0^{\infty}\dd z\, e^{-2|n|z}\left(|n|+\frac{m_0}{m_0z+2}\right)^2\left(1+\frac{16}{(m_0z+2)^3}\right)\,.
\end{equation}
Importantly, we observe from the above asymptotics that the correction to the PETS entanglement entropy due to mass deformations of the shell is negative at both high and low temperatures. This is in contrast to the shape deformations where the correction was positive in these limits.

\section{Discussion}

The central theme of this paper has been that thin-shell black holes and wormholes provide a useful meeting point between three seemingly different structures: gravitational saddles, non-conformal impurities, and the geometry of Riemann surfaces. From the gravitational point of view, the shell is an extended matter source whose shape and mass density define elastic moduli of the black hole or wormhole saddle. From the boundary CFT point of view, the same object is a line defect, and the response to these elastic moduli is governed by defect-local operators such as the displacement operator and the mass-density operator. From the Liouville point of view, the computation becomes a problem in hyperbolic geometry: deforming the shell corresponds to changing the gluing data of hyperbolic surfaces, closely related to the conformal welding problem in mathematics. 

The stiffness kernels computed in this paper therefore simultaneously measure the rigidity of a gravitational saddle, the two-point functions of impurity operators, and the response of a uniformized Riemann surface to changes in its sewing data. Their spectral decomposition and Lorentzian continuation then give a dynamical interpretation of these geometric data, distinguishing situations with genuine relaxation from those with persistent finite-volume oscillations. This perspective suggests that elastic response is a natural observable for organizing how black hole and wormhole geometries encode defect dynamics, impurity relaxation, and the geometry of moduli spaces.

\subsection{Relation to the literature}

\paragraph{Relation to holographic Kondo models.}
The impurity interpretation of our results is reminiscent of the holographic Kondo model. In the ordinary Kondo problem, a localized magnetic impurity coupled to a bath of conduction electrons triggers an impurity RG flow, leading at low temperatures to the screening of the impurity spin \cite{Kondo:1964}. The holographic Kondo model realizes this physics in a strongly coupled large-\(N\) setting: a \((1+1)\)-dimensional CFT is coupled to a \((0+1)\)-dimensional impurity, and the Kondo coupling drives a defect RG flow that is described in the bulk by fields localized on an AdS$_2$ brane inside AdS$_3$ \cite{Erdmenger:2013dpa}. Time-dependent versions of this model study the relaxation of the impurity sector after a quench of the Kondo coupling. In that context, the late-time response is controlled by the lowest quasinormal frequency of the impurity sector: in the bulk this frequency is obtained by solving a Lorentzian fluctuation problem with infalling boundary conditions at the horizon and source-free boundary conditions at the asymptotic boundary of the impurity brane \cite{Erdmenger:2016msd}.

Our setup describes a different relaxation mechanism. The closest analogy is instead at the level of analytic structure: in both systems, the late-time response is governed by the complex-frequency singularity of a retarded impurity correlator closest to the real axis. In the holographic Kondo model, this singularity is a genuine quasinormal mode, obtained from an infalling Lorentzian bulk boundary-value problem. In the present work, by contrast, the relevant singularities arise by analytically continuing the Euclidean stiffness kernels.
We therefore refer to these singularities as relaxation poles rather than quasinormal modes. This distinction is useful because the dissipation mechanism studied here is geometric: relaxation occurs when the defect-local operators \(\mathcal{D}_\perp\) and \(\mathcal{M}\) have a continuous spectrum, allowing dephasing into the non-compact Liouville direction, while finite-volume examples with discrete spectra exhibit persistent oscillatory response. Thus, while both systems describe non-conformal impurities in a large-\(N\) holographic theory, the mechanism studied here is a geometric stiffness/dephasing mechanism for elastic deformations of the impurity, rather than Kondo screening.

\paragraph{Relation to displacement correlators on Wilson lines.}

There is extensive literature on computation of displacement correlators on conformal line defects, especially Wilson lines. For conformal defects, the displacement operator is the defect-local operator that appears in the Ward identity for broken transverse translations, and its two-point coefficient is an intrinsic piece of defect CFT data \cite{Billo:2016cpy}. For supersymmetric Wilson lines and more general half-BPS line defects, this data has been studied using deformations of Wilson loops and perturbative defect-CFT methods \cite{Drukker:2006xg,Cooke:2017qgm}, supersymmetric localization and integrability \cite{Giombi:2018qox}, conformal bootstrap methods \cite{Liendo:2018ukf}, and the dual \(AdS_2\) string worldsheet description at strong coupling \cite{Giombi:2017cqn}.
In particular, for Wilson lines describing heavy external probes, the coefficient of the displacement two-point function is related to Bremsstrahlung radiation and small-angle cusp data \cite{Correa:2012at,Bianchi:2018zpb}. More recently, thermal displacement correlators on Wilson lines were studied holographically by solving for transverse fluctuations of the dual string worldsheet in an AdS black-hole background, revealing bouncing singularities and a high-frequency structure controlled by the defect OPE \cite{Barrat:2024vwu,Giombi:2026line}. 

Our setup is complementary. The line defects considered here are non-conformal and backreact to produce thin-shell black holes and wormholes. Consequently, the displacement two-point function is not fixed by conformal symmetry, but is a geometric stiffness kernel determined by the Liouville/Dirichlet-to-Neumann response of the glued hyperbolic geometry. This leads to a different set of physical questions: the spectrum of \(\mathcal{D}_\perp\) may be continuous or discrete depending on the geometry transverse to the shell, and its Lorentzian continuation diagnoses relaxation or persistent finite-volume oscillations of the corresponding thin-shell saddle. To our knowledge, this is the first explicit computation of displacement spectral densities and Lorentzian relaxation poles for non-conformal, gravitationally backreacting line defects.

\subsection{Future directions}

We now discuss some future directions related to the ideas put forth in this paper.

\paragraph{Stiffness kernels that measure finer correlations on the shell.} In this paper, we computed the stiffness kernels that measure two-point correlations of the shape or mass distribution of the shell. It would be interesting to compute stiffness kernels that measure higher-point correlations on the shell. For this, one would have to solve the Liouville equation around the background saddle to higher orders in the deformation parameter and use it to compute the effective action for the respective deformations. For example, in the sphere 1-point wormhole, based on $y\to -y$ reflection symmetry of the background saddle, we can argue that displacement correlators with odd number of insertions of the displacement operator should vanish since the partition function is an even functional of the deformation function $\xi(x)$. As a consequence, the next non-zero contribution is expected to come from the 4-point function,
    \begin{equation}
        \langle \mathcal{D}_\perp(x_1)\mathcal{D}_\perp(x_2)\mathcal{D}_\perp(x_3)\mathcal{D}_\perp(x_4)\rangle_{\rm WH} \,.
    \end{equation}
    On the other hand, for geometries which do not possess such a reflection symmetry, for example, the one-sided black hole, the next non-zero contribution is expected to come from the displacement 3-point function,
    \begin{equation}
        \langle \mathcal{D}_\perp(x_1)\mathcal{D}_\perp(x_2)\mathcal{D}_\perp(x_3)\rangle_{\rm BH} \,.
    \end{equation}
    The 4-point function in the wormhole background or the 3-point function in the black hole background require solving the Liouville equation to second order around the respective backgrounds. We have computed the zero mode of the second-order Liouville field in Appendix \ref{app:entropy} since it was relevant to compute the quadratic correction to the black hole entropy. One can readily generalize this to compute the other Fourier modes of the second order Liouville field, which would allow us to compute the 3-point function of the displacement operator.
    
\paragraph{Finite-$G_N$ effects on stiffness kernels and relaxation times.} The computation of the stiffness kernels in this paper was done in the $G_N\to 0$ limit (equivalently, in the large central charge limit in the dual holographic CFT, or the $b\to 0$ limit in Liouville CFT). It would be interesting to compute the effects of loop corrections around the deformed saddles, which would in turn help us determine the finite-$G_N$ effects on the retarded response. Relatedly, it would be interesting to understand how the relaxation times that we computed in this paper are affected by finite-$G_N$ or $1/c$-effects.

\paragraph{Stiffness kernels for asymmetric deformations of black holes and wormholes.} In this paper, we have only computed stiffness of black holes and wormholes toward symmetric deformations of the line defects that source the shell. For instance, in the sphere 1-point wormhole and the torus 1-point wormhole examples, we deformed the pair of line defects in the holographic CFT by the same deformation function. More generally, one could compute stiffness toward deformations that deform the pair of line defects asymmetrically. To construct wormhole saddles with these asymmetric boundary conditions, one could use the Uhlenbeck slicing of almost-Fuchsian wormholes \cite{Chandra:2022bqq, Krasnov:2005dm},
   \begin{equation}
       ds^2=d\rho^2+\cosh^2(\rho)e^{\Phi}|dz+\tanh(\rho)e^{-\Phi}\overline{t}(\overline{z})d\overline{z}|^2\,.
   \end{equation}
   Here, $t(z)$ is a holomorphic quadratic differential on the Riemann surface. In the present setup, we can interpret the deformation function to be specifying the complex structure and a metric of the above form describes how the complex structure varies smoothly from one boundary to the other. The quadratic effective action for such a wormhole takes the schematic form,
   \begin{equation}
       \delta \log Z[\xi_1,\xi_2]\sim \mathcal{K}^S_n|\xi_{1,n}+\xi_{2,n}|^2+\mathcal{K}^A_n|\xi_{1,n}-\xi_{2,n}|^2\,,
   \end{equation}
   for each Fourier mode $n$ of the deformation. Here, $\mathcal{K}^S$ and $\mathcal{K}^A$ are stiffness kernels for symmetric and antisymmetric sectors of the deformations. In this paper, we only analyzed the symmetric sector, but one could use the above idea to access the antisymmetric sector. In the CFT$_2$-ensemble, this would enable us to compute the self- and cross correlations between the displacement operators on the same or different boundaries of the wormhole, written schematically as $\overline{\langle \mathcal{D}_\perp(x_1)\mathcal{D}_{\perp}(x_2) \rangle}$ and $\overline{\langle \mathcal{D}_\perp(x_1)\rangle \langle\mathcal{D}_\perp(x_2)\rangle}$ respectively. One can similarly study asymmetric deformations of the pair of line defects that create the black hole and thereby compute the displacement correlator matrix for the pair of holographic line defects. The details of these computations will be reported in a future paper.

   These asymmetric black holes and wormholes are also interesting from the context of closed universes \cite{Antonini:2023hdh,Antonini:2024mci,Antonini:2025ioh,Kudler-Flam:2025cki} and positivity rules for the gravitational path integrals \cite{Liu:2025ikq, Iliesiu:2025ias,DiUbaldo:2026rly}. This is because such wormholes can be interpreted as computing transition amplitudes between different closed universe states. The wormholes discussed above compute inner products between closed universe states labeled by deformation functions $\langle \xi_1\ket{\xi_2}$. A consistency check to have such a Hilbert space interpretation would be that the inner-product matrix must be positive semi-definite.\footnote{We thank Luca Iliesiu for discussions on this point.}
    
\paragraph{Stiffness kernels for black holes sourced by conformal line defects.} The thin-shell geometries analyzed in this paper are sourced by non-conformal line defects in the dual holographic CFT. It would be interesting to repeat the stiffness-kernel computation for black holes sourced instead by conformal line defects. In this case the displacement operator is genuine defect-CFT data: for a conformal line defect in a two-dimensional CFT,
    \begin{equation}
         \langle \mathcal{D}_\perp(x_1)\mathcal{D}_\perp(x_2)\rangle = \frac{C_D}{(x_1-x_2)^4}.
    \end{equation}
  For conformal interfaces, the coefficient $C_D$ controls the free-energy cost of small interface deformations and is tied to energy reflection and transmission across the defect \cite{Billo:2016cpy,Meineri:2019ycm}. In holographic defect CFTs, this point of view is sharpened by the interpretation of the norm of the displacement operator as an inertial tension, which reduces to the Nambu--Goto tension for classical thin probe branes \cite{Bachas:2024hcb}. A natural bulk model is therefore a constant-tension $\mathrm{AdS}_2$ brane/domain wall in $\mathrm{AdS}_3$, for which energy transmission, black-hole phases, and the extension to smooth Janus interfaces have already been studied \cite{Bachas:2020yxv,Bachas:2021fqo,Bachas:2022cba,Bak:2011ga}. In these setups, it would be interesting to compute the full shape-stiffness kernel for a backreacting conformal defect or domain wall and extract the corresponding displacement coefficient $C_D$.
    
\paragraph{Displacement correlators and impurity relaxation in CFTs.} As we have emphasized several times in this paper, the computation of the stiffness kernels using Liouville line defects is also interesting from the perspective of studying relaxation of a many-body system with localized impurities toward deformations of the shape or coupling of the impurities. One could use Liouville CFT as a model to more generally study the properties of non-conformal impurities. Other examples of non-conformal impurities in Liouville CFT are provided by line defects constructed by integrating Liouville vertex operators around a curve \cite{Abdalla:2026wdx}. It would be very interesting to compute relaxation times for these impurities and study their behaviour under Renormalization Group flows.  

    More broadly, we expect this relaxation mechanism to apply to other impurity RG flows in higher-dimensional CFTs. Once a defect-local operator is conjugate to a shape or coupling deformation, its retarded two-point function should diagnose relaxation toward corresponding transient deformations. A concrete setup to analyze this is the localized magnetic-field defect in the critical $O(N)$ model, where the defect RG flow and its defect data can be studied using large-$N$ saddle-point methods \cite{Cuomo:2021kfm}. A complementary nonperturbative arena is the $3d$ Ising CFT, where fuzzy-sphere radial quantization has been used to access the $3d$ Ising spectrum and to study magnetic line defects, including the displacement operator and other defect primaries \cite{Zhu:2022gjc,Hu:2023ghk,Zhou:2024fgi}. It would be interesting to compute the corresponding defect spectral densities and retarded correlators in these models and compare the resulting relaxation mechanisms with the Liouville/holographic 2d CFT examples studied here.

\section*{Acknowledgments}

We thank Ahmed Abdalla, Tom Hartman, Luca Iliesiu, Zohar Komargodski, Mark Mezei, Martin Sasieta, Yifan Wang, Zixia Wei and Zhenbin Yang for helpful discussions. We also thank Tom Hartman and Zohar Komargodski for comments on a draft of this paper. We are grateful to the organizers of the workshop ``Random Geometry in Math and Physics" at the Simons Center for Geometry and Physics, Stony Brook for inviting us to attend and participate in the workshop. The discussions and talks by both mathematicians and physicists at this workshop were very helpful in developing this project. This work was supported in part by the Leinweber Institute for Theoretical Physics at UC Berkeley; and by the Department of Energy, Office of Science, Office of High Energy Physics through award DE-SC0025293 and QuantISED award DE-SC0019380.  BP is supported by the ERC Consolidator Grant GeoChaos No 101169611. This work is funded by the European Union. Views and opinions expressed are however those of the authors only and do not necessarily reflect those of the European Union or the European Research Council Executive Agency. Neither the European Union nor the granting authority can be held responsible for them. For the purpose of open access, the authors have applied a CC BY public copyright licence to any Author Accepted Manuscript (AAM) version arising from this submission.

\appendix

\section{Quadratic effective action for deformations}\label{sec:appA}

In this appendix, we expand the Liouville action to quadratic order around a background Liouville field, for both shape deformations and mass deformations.

\subsection{Effective action for shape deformations}

Consider the following shape deformation of the shell locus:
\begin{equation}
    \Sigma_\epsilon:\qquad y=\epsilon \xi(x), 
\end{equation}
where $\epsilon$ is an infinitesimal parameter.
Then we define the coefficient \(S_2[\xi]\) by
\begin{equation}
    S[\Sigma_\epsilon]
    =
    S[\Sigma_0]+\epsilon^2 S_2[\xi]+O(\epsilon^3).
\end{equation}
The total action, including the bulk Liouville term and the shell source term, is
\begin{equation}
    S=S_{\rm bulk}+S_{\rm shell},
\end{equation}
with
\begin{equation}
    S_{\rm bulk}
    =
    \frac{1}{4\pi}
    \left[
    \int_{\epsilon \xi(x)}^\infty \dd x\,\dd y\,\mathcal L_+
    +
    \int_{-\infty}^{\epsilon \xi(x)} \dd x\,\dd y\,\mathcal L_-
    \right],
\end{equation}
where
\begin{equation}
    \mathcal L_\pm
    =
    \frac14
    \left[
    \left(\partial_x\Phi_\pm\right)^2
    +
    \left(\partial_y\Phi_\pm\right)^2
    \right]
    +e^{\Phi_\pm}.
\end{equation}
The shell source term is
\begin{equation}
    S_{\rm shell}
    =
    -\frac{m_0}{4\pi}
    \int_0^{2\pi}\dd x\,
    \sqrt{1+\epsilon^2 \xi'(x)^2}\,
    \Phi_+\bigl(x,\epsilon \xi(x)\bigr).
\end{equation}
The proper-length factor can be expanded as
\begin{equation}
    \sqrt{1+\epsilon^2 \xi'(x)^2}
    =
    1+\frac{\epsilon^2}{2}\xi'(x)^2+O(\epsilon^4).
\end{equation}
We write
\begin{equation}
    \Phi_\pm(x,y)
    =
    \Phi_{0,\pm}(y)+\epsilon \varphi_\pm(x,y)+O(\epsilon^2),
\end{equation}
where
where $\Phi_0(y)$ is the background Liouville solution. Denote $\Phi_*$ as the value of the background Liouville field on the shell locus and denote the second derivatives as
\begin{equation}
    \Phi''_{0,+}(0)=\Phi''_{0,-}(0)\equiv 2s^2.
\end{equation}
Moreover, let
\begin{equation}
    d(x)\equiv \partial_y\varphi_+(x,0)=\partial_y\varphi_-(x,0).
\end{equation}
We first evaluate the bulk action. Expanding the bulk Lagrangian gives
\begin{equation}
    \mathcal L_\pm
    =
    \mathcal L_{0,\pm}
    +
    \epsilon \mathcal L_{1,\pm}
    +
    \epsilon^2 \mathcal L_{2,\pm}
    +O(\epsilon^3),
\end{equation}
where
\begin{equation}
    \mathcal L_{1,\pm}
    =
    \frac12 \Phi'_{0,\pm}\partial_y\varphi_\pm
    +
    e^{\Phi_{0,\pm}}\varphi_\pm,
\end{equation}
and
\begin{equation}
    \mathcal L_{2,\pm}
    =
    \frac14
    \left[
    \left(\partial_x\varphi_\pm\right)^2
    +
    \left(\partial_y\varphi_\pm\right)^2
    \right]
    +
    \frac12 e^{\Phi_{0,\pm}}\varphi_\pm^2.
\end{equation}
The moving integration domain gives
\begin{align}
    \int_{\epsilon \xi(x)}^\infty \dd y\,\mathcal L_+
    &+
    \int_{-\infty}^{\epsilon \xi(x)}\dd y\,\mathcal L_-
    =
    \int_0^\infty \dd y\,\mathcal L_+
    +
    \int_{-\infty}^0 \dd y\,\mathcal L_-  \nonumber\\
    &
    +
    \epsilon^2
    \left[
    \int_0^\infty \dd y\,\mathcal L_{2,+}
    +
    \int_{-\infty}^0 \dd y\,\mathcal L_{2,-}
    +
    \xi\left(\mathcal L_{1,-}-\mathcal L_{1,+}\right)_{y=0}
    +
    \frac{\xi^2}{2}
    \left(\mathcal L'_{0,-}-\mathcal L'_{0,+}\right)_{y=0}
    \right]
    +O(\epsilon^3).
\end{align}
At \(y=0\), we have
\begin{equation}
    \left(\mathcal L_{1,-}-\mathcal L_{1,+}\right)_{y=0}
    =
    m_0 d(x)-2m_0s^2\xi(x),
\end{equation}
and
\begin{equation}
    \frac12
    \left(\mathcal L'_{0,-}-\mathcal L'_{0,+}\right)_{y=0}
    =
    2m_0s^2.
\end{equation}
Therefore the explicit moving-boundary contribution reduces to
\begin{equation}
    \xi\left(\mathcal L_{1,-}-\mathcal L_{1,+}\right)_{y=0}
    +
    \frac{\xi^2}{2}
    \left(\mathcal L'_{0,-}-\mathcal L'_{0,+}\right)_{y=0}
    =
    m_0\xi d.
\end{equation}
Next, using the linearized equation
\begin{equation}
    \left(\partial_x^2+\partial_y^2\right)\varphi_\pm
    =
    2e^{\Phi_{0,\pm}}\varphi_\pm,
\end{equation}
we can reduce the fixed-domain quadratic term to a boundary term:
\begin{equation}
    \int_{\Omega_\pm} \dd x\,\dd y\,\mathcal L_{2,\pm}
    =
    \frac14
    \int_{\partial\Omega_\pm}\dd s\,\varphi_\pm\partial_n\varphi_\pm.
\end{equation}
The contribution from infinity vanishes. At \(y=0\), the outward normal of the
upper region is \(-\partial_y\), while the outward normal of the lower region is
\(+\partial_y\). Hence
\begin{align}
    \int_0^\infty \dd x\,\dd y\,\mathcal L_{2,+}
    +
    \int_{-\infty}^0 dx\,dy\,\mathcal L_{2,-}
    &=
    \frac14
    \int_0^{2\pi}\dd x\,
    \left[
    -\varphi_+(x,0)d(x)+\varphi_-(x,0)d(x)
    \right]  \nonumber\\
    &=
    \frac14
    \int_0^{2\pi}\dd x\,
    \left[
    \varphi_-(x,0)-\varphi_+(x,0)
    \right]d(x).
\end{align}
Using the linearized continuity condition
\begin{equation}
    \varphi_+(x,0)-\varphi_-(x,0)=2m_0\xi(x),
\end{equation}
we obtain
\begin{equation}
    \int_0^\infty \dd x\,\dd y\,\mathcal L_{2,+}
    +
    \int_{-\infty}^0 \dd x\,\dd y\,\mathcal L_{2,-}
    =
    -\frac{m_0}{2}
    \int_0^{2\pi}\dd x\,\xi(x)d(x).
\end{equation}
Combining this with the moving-boundary term gives
\begin{equation}
    S_{{\rm bulk},2}
    =
    \frac{1}{4\pi}
    \left[
    -\frac{m_0}{2}
    \int_0^{2\pi}\dd x\,\xi d
    +
    m_0
    \int_0^{2\pi}\dd x\,\xi d
    \right].
\end{equation}
Therefore
\begin{equation}
    S_{{\rm bulk},2}
    =
    \frac{m_0}{8\pi}
    \int_0^{2\pi}dx\,\xi(x)d(x).
\end{equation}
We now evaluate the shell source term. Along the deformed curve,
\begin{align}
    \Phi_+\bigl(x,\epsilon\xi(x)\bigr)
    &=
    \Phi_0(0)
    +
    \epsilon
    \left[
    \varphi_+(x,0)-m_0\xi(x)
    \right]  \nonumber\\[1em]
    &\quad
    +
    \epsilon^2
    \left[
    \xi(x)d(x)+s^2\xi(x)^2
    \right]
    +O(\epsilon^3),
\end{align}
where terms involving the second-order correction to the Liouville field cancel
against the first variation of the bulk action evaluated on the background
saddle. Including the proper-length factor, we obtain
\begin{align}
    \sqrt{1+\epsilon^2\xi'(x)^2}\,
    \Phi_+\bigl(x,\epsilon\xi(x)\bigr)
    &=
    \Phi_0(0)
    +
    \epsilon
    \left[
    \varphi_+(x,0)-m_0\xi(x)
    \right]  \nonumber\\[1em]
    &\quad
    +
    \epsilon^2
    \left[
    \xi d+s^2\xi^2
    +
    \frac12\Phi_0(0)(\xi')^2
    \right]
    +O(\epsilon^3).
\end{align}
Thus
\begin{equation}
    S_{{\rm shell},2}
    =
    -\frac{m_0}{4\pi}
    \int_0^{2\pi}\dd x\,
    \left[
    \xi d+s^2\xi^2
    +
    \frac12\Phi_0(0)(\xi')^2
    \right].
\end{equation}
Adding the bulk and shell contributions, we find the final expression for the action
\begin{align}\label{eq:appaction}
    S_2[\xi]
    &=
    S_{{\rm bulk},2}+S_{{\rm shell},2}  \nonumber\\
    &=
    -\frac{m_0}{8\pi}
    \int_0^{2\pi}\dd x\left(\,\xi \partial_y\varphi(x,0)
   +
   \Phi''_0(0)\xi^2
    +
    \Phi_0(0)(\xi')^2\right).
\end{align}
Written in this form, the action only has support on the shell locus and we can readily apply it to any of the setups being studied in the main text.

\subsection{Effective action for mass deformations}
Next, we study mass deformations $m(x) = m_0+\epsilon\mu(x)$.
The
Liouville action is
\begin{equation}
    S
    =
    S_{\rm bulk}+S_{\rm shell},
\end{equation}
where
\begin{equation}
    S_{\rm bulk}
    =
    \frac{1}{4\pi}
    \sum_{\pm}
    \int_{\Omega_\pm} \dd x\,\dd y
    \left[
    \frac14(\partial\Phi_\pm)^2+e^{\Phi_\pm}
    \right],
\end{equation}
and, since the shell is straight,
\begin{equation}
    S_{\rm shell}
    =
    -\frac{1}{4\pi}
    \int_0^{2\pi}\dd x\,m(x)\Phi(x,0).
\end{equation}
There is no arclength correction in this subsection because the shell locus is
not deformed.

Expanding the bulk action to quadratic order in \(\epsilon\), we get
\begin{equation}
    S_{{\rm bulk},2}
    =
    \frac{1}{4\pi}
    \sum_{\pm}
    \int_{\Omega_\pm} \dd x\,\dd y
    \left[
    \frac14(\partial\varphi_\pm)^2
    +
    \frac12 e^{\Phi_{0,\pm}}\varphi_\pm^2
    \right].
\end{equation}
Terms involving the second-order correction to the Liouville field cancel
against the first variation of the action evaluated on the background saddle,
so the quadratic on-shell action is determined entirely by the linearized field
\(\varphi_\pm\).

Using the linearized equation
\begin{equation}
    \left(\partial_x^2+\partial_y^2\right)\varphi_\pm
    =
    2e^{\Phi_{0,\pm}}\varphi_\pm,
\end{equation}
we can reduce the bulk quadratic action to a boundary term:
\begin{align}
    \int_{\Omega_\pm} \dd x\,\dd y
    \left[
    \frac14(\partial\varphi_\pm)^2
    +
    \frac12 e^{\Phi_{0,\pm}}\varphi_\pm^2
    \right]
    &=
    \frac14
    \int_{\partial\Omega_\pm}
    \dd s\,\varphi_\pm \partial_n\varphi_\pm .
\end{align}
The contributions from \(y=\pm\infty\) vanish for the nonzero modes. At
\(y=0\), the outward normal of the upper region \(y>0\) is
\begin{equation}
    \partial_n=-\partial_y,
\end{equation}
whereas the outward normal of the lower region \(y<0\) is
\begin{equation}
    \partial_n=+\partial_y.
\end{equation}
Therefore
\begin{align}
    S_{{\rm bulk},2}
    &=
    \frac{1}{16\pi}
    \int_0^{2\pi}\dd x 
    \Big[
    -\varphi_+(x,0)\partial_y\varphi_+(x,0)
    +
    \varphi_-(x,0)\partial_y\varphi_-(x,0)
    \Big]  \nonumber\\
    &=
    \frac{1}{16\pi}
    \int_0^{2\pi}\dd x\,
    \varphi(x,0)
    \left[
    \partial_y\varphi_-(x,0)-\partial_y\varphi_+(x,0)
    \right].
\end{align}
Using the linearized jump condition
\begin{equation}
    \partial_y\varphi_+(x,0)-\partial_y\varphi_-(x,0)
    =
    -2\mu(x),
\end{equation}
we obtain
\begin{equation}
    \partial_y\varphi_-(x,0)-\partial_y\varphi_+(x,0)
    =
    2\mu(x).
\end{equation}
Hence
\begin{equation}
    S_{{\rm bulk},2}
    =
    \frac{1}{8\pi}
    \int_0^{2\pi}\dd x\,\mu(x)\varphi(x,0).
\end{equation}
Next, consider the shell source term:
\begin{equation}
    S_{\rm shell}
    =
    -\frac{1}{4\pi}
    \int_0^{2\pi}\dd x\,
    \left(m_0+\epsilon\mu(x)\right)
    \left[
    \Phi_0(0)+\epsilon \varphi(x,0)+O(\epsilon^2)
    \right].
\end{equation}
At quadratic order, the term involving the product of the mass deformation and
the linearized Liouville field is
\begin{equation}
    S_{{\rm shell},2}
    =
    -\frac{1}{4\pi}
    \int_0^{2\pi}\dd x\,\mu(x)\varphi(x,0).
\end{equation}
Again, terms involving the second-order correction to \(\Phi\) cancel against
the first variation of the background action. Therefore, the total action at quadratic order is given by the simple compact form
\begin{equation}
    S_2=-\frac{1}{8\pi}\int_0^{2\pi}\dd x \mu(x)\varphi(x,0)\,.
\end{equation}
Written in this form, the above action has support only on the shell locus and we can apply to any of the setups being studied in the main text.

\section{More details on entropy correction for the one-sided and two-sided black holes}\label{app:entropy}

In this appendix, we will provide details on the solution to the second order Liouville equation that is used to compute the correction to the PETS entanglement entropy and to compute the correction to the apparent horizon entropy for the one-sided black hole. In particular, we compute the Green's functions used to solve the ODE for the zero mode of the second order Liouville field that is used to compute the correction to the entropy. For the PETS black hole, we need the Green's function on an interval with Neumann boundary condition at one endpoint and regularity at the other endpoint. For the one-sided black hole, we need the Green's function with Robin boundary condition at one endpoint and regularity at the other endpoint.

\subsection{Green's function on an interval with Neumann boundary condition at an endpoint}

The goal will be to solve for the Green's function of the differential operator,
\begin{equation} \label{eq:greensfunctiondefn}
    \left(\frac{d^2}{dt^2}-2\csc^2(t)\right)G_N(t,s)=\delta(t-s)
\end{equation}
with the domain being an interval from $t=0$ and $t=T$ ($T<\pi$), with the boundary conditions at the endpoints being regularity at $t=0$ and Neumann at $t=T$. Away from coincident points, the Green's function satisfies the homogeneous differential equation. A convenient basis for solving the homogeneous differential equation is the following,
\begin{equation}
    w(t)=\cot(t), \qquad v(t)= 1-t\cot(t)\,.
\end{equation}
Notice that $w(t)$ is not regular at $t=0$ while $v(t)$ is regular. Next, in order to satisfy the Neumann boundary condition at the other endpoint, we start with a linear combination of the above basis solutions,
\begin{equation}
    r(t)=w(t)+ \alpha v(t)
\end{equation}
and determine $\alpha$ by requiring $r'(T)=0$ to get
\begin{equation}
    \alpha= \frac{1}{T-\sin(T)\cos(T)} \implies r(t)=w(t)+\frac{1}{T-\sin(T)\cos(T)}v(t)\,.
\end{equation}
Therefore, we have shown that $v(t)$ is regular at $t=0$ and $r(t)$ satisfies the Neumann boundary condition at $y=T$. So, we can use a piecewise ansatz of the Green's function,
\begin{equation}
    G_N(t,s)=\begin{cases}
        A(s) v(t), \qquad & 0<t<s,\\
        B(s) r(t), \qquad & s<t<T\,.
    \end{cases}
\end{equation}
We determine the coefficients $A(s)$ and $B(s)$ by continuity of the Green's function at $t=s$, and jump in the first derivative across $t=s$ obtained by integrating (\ref{eq:greensfunctiondefn}) in a small interval around $t=s$,
\begin{equation}
    \partial_t G(s^+,s)-\partial_t G(s^-,s)=1\,.
\end{equation}
Solving the two conditions gives
\begin{equation}
    A(s)=-r(s), \qquad B(s)=-v(s)\,.
\end{equation}
Therefore, the Green's function can be written as
\begin{equation}
    G_N(t,s)=-(1-t_< \cot(t_<))\left[\cot(t_>)+\frac{1-t_>\cot(t_>)}{T-\sin(T)\cos(T)}\right]\,.
\end{equation}
where $t_< \equiv \text{min}(s,t)$ and $t_> \equiv \text{max}(s,t)$. We can use this Green's function to compute the solution to the differential equation in the presence of a source on the RHS,
\begin{equation}
    \left(\frac{d^2}{dt^2}-2\csc^2(t)\right) \phi(t)= J(t)
\end{equation}
to get
\begin{equation}
    \phi(t)=\int_0^{T}ds G_N(t,s) J(s)\,.
\end{equation}
This is the form of the solution that we will use whenever the second-order zero mode obeys a Neumann condition at the endpoint.

\subsection{Green's function on an interval with Robin boundary condition at an endpoint}\label{sec:robinbc}

Now, we compute the Green's function on the interval with Robin boundary condition at $t=T$,
\begin{equation}
    G_\gamma'(t,s)+\gamma G_\gamma(t,s)=0\,.
\end{equation}
We proceed in exactly the same way as we did for the Neumann boundary condition by noting that the regular solution at $t=0$ is $v(t)=1-t\cot(t)$. To find the solution satisfying the Robin boundary condition at the other endpoint, we form the linear combination,
 \begin{equation}
     r(t)=\cot(t)+A_\gamma (1-t\cot(t))\,.
 \end{equation}
We determine $A_\gamma$ by requiring $r'(T)+\gamma r(T)=0$ and obtain
\begin{equation}
    A_\gamma = \frac{1-\gamma \sin(T)\cos(T)}{T-\sin(T)\cos(T)+\gamma \sin(T)(\sin(T)-T\cos(T))}\,.
\end{equation}
Using a piecewise ansatz for the Green's function and by imposing continuity and the derivative jump, we obtain the final expression,
\begin{equation}
    G_\gamma(t,s)=-(1-t_< \cot(t_<))\left[\cot(t_>)+A_\gamma(1-t_>\cot(t_>))\right]\,.
\end{equation}
Note that when $\gamma=0$, the Robin boundary conditions reduces to the Neumann boundary condition and the Green's function reduces to the Neumann Green's function $G_N$. When $\gamma \to \infty$, the Robin boundary condition reduces to Dirichlet boundary condition and the corresponding Green's function is given by
\begin{equation}
    G_D(t,s)=-(1-t_< \cot(t_<))\left[\cot(t_>)-\frac{\cot(T)}{1-T\cot(T)}(1-t_>\cot(t_>))\right]\,.
\end{equation}
There is an exceptional value of the Robin parameter where the regular solution $v(t)$ itself satisfies the Robin boundary condition,
\begin{equation}
    \gamma_*= -\frac{T-\sin(T)\cos(T)}{\sin^2(T)(1-T\cot(T))}\,.
\end{equation}
At this value, the Green's function is ill-defined.

\subsection{Green's function on an interval with a source at the endpoint}

We have computed the Green's function when there is no endpoint source i.e., it satisfies Neumann or Robin boundary conditions at $t=T$. This Green's function is like a bulk-to-bulk propagator. In the presence of a boundary source, we also need to compute the bulk-to-boundary propagator $K_\gamma(t)$ which satisfies the homogeneous differential equation,
\begin{equation}
    \left(\frac{d^2}{dt^2}-2 \csc^2(t)\right)K_\gamma(t)=0\,,
\end{equation}
subject to regularity at $t=0$ and the normalized source condition, $K'(T)+\gamma K(T)=1$ at $t=T$. Since we have already identified that $v(t)=1-t\cot(t)$ is a regular solution to the above differential equation at $t=0$, we can work with the ansatz $K(t) =\alpha v(t)$ and determine the proportionality constant by imposing the source condition at $t=T$, giving
\begin{equation}
    K_\gamma(t)=\frac{1}{\gamma-(1+T\gamma)\cot(T)+T\csc^2(T)}(1-t\cot(t))\,.
\end{equation}
When $\gamma=0$, we get the Neumann bulk-to-boundary propagator,
\begin{equation}
     K_N(t)=\frac{1}{-\cot(T)+T\csc^2(T)}(1-t\cot(t))\,.
\end{equation}
If we also include a bulk source, then the general solution to 
\begin{equation}
     \left(\frac{d^2}{dt^2}-2\csc^2(t)\right) \phi(t)= J(t)
\end{equation}
subject to $\phi'(T)+\gamma \phi(T)=\eta$ and regularity at $t=0$
is given by 
\begin{equation}
     \phi(t)=\int_0^{T}ds G_\gamma(t,s) J(s) + \eta K_\gamma(t)\,.
\end{equation}
This is the form of the solution that we will use for the solution to the zero mode of the second order Liouville field for shape deformations.

\subsection{Boundary conditions for the second order Liouville field}

\subsubsection{Shape deformations}
Here we determine the boundary conditions for the second order Liouville field needed for the apparent-horizon calculation in section \ref{sec:ahshape}.
First, we determine the boundary condition obeyed by $c_n(y)$ at the shell. The
zero mode of the second-order Liouville equation on the cap side satisfies
\begin{equation}
\left({d^2\over ds^2}-2\csch^2(s)\right)\widetilde c_n(s)
=
\csch^2(s)\,\widetilde f_n(s)^2,
\qquad
s=y-\tau_0+A,
\end{equation}
where
\begin{equation}
\widetilde f_n(s)
=
2m_0\frac{\lambda_{-,n}}{\lambda_{+,n}+\lambda_{-,n}}
{\cal V}_n(s),
\qquad
{\cal V}_n(s):=
e^{-|n|(s-A)}
{|n|+\coth s\over |n|+\coth A}.
\end{equation}
The solution is required to be regular as $s\to\infty$. The homogeneous
solution that is regular at infinity is proportional to $\coth s$. Therefore,
if the value of the second-order field at the shell is fixed, the homogeneous
cap response gives
\begin{equation}
\partial_s\widetilde c_n(A)
=
-\lambda_{+,0}c_n(\tau_0)+\cdots ,
\qquad
\lambda_{+,0}={\csch^2A\over \coth A}
={r_0^2\over \sqrt{1+r_0^2}}.
\end{equation}
The omitted term is the particular solution sourced by $\widetilde f_n^2$.
Using the Wronskian with the homogeneous solution $\coth(s)$, this contribution
is
\begin{equation}\label{eq:Pshape}
P^{\rm shape}_n
=
-{1\over \coth A}
\int_A^\infty ds\,\coth s\,\csch^2s\,\widetilde f_n(s)^2 .
\end{equation}
The second-order continuity condition gives
\begin{equation}
\widetilde c_n(A)=c_n(\tau_0),
\end{equation}
while the second-order normal-derivative junction condition gives
\begin{equation}
\partial_s\widetilde c_n(A)-c'_n(\tau_0)
+
m_0(n^2+2r_0^2)=0.
\end{equation}
Combining these equations, we obtain a Robin boundary condition for the
second-order field on the black-hole side:
\begin{equation}
\partial_y c_n(\tau_0)+\lambda_{+,0}c_n(\tau_0)
=
m_0(n^2+2r_0^2)+P^{\rm shape}_n .
\end{equation}
Comparing this to the rescaled coordinate $t = r_H y$, we read off the $\gamma$-factor defined in section \ref{sec:robinbc} to be
\begin{equation}\label{eq:gammashape}
    \gamma = \lambda_{+,0}/r_H.
\end{equation}
 The endpoint is at $y= \tau_0$, so $T = r_H \tau_0$ in the notation of section \ref{sec:robinbc}, and 
\begin{equation}\label{eq:zetashape}
    \zeta_n^{\rm shape} = (m_0(n^2+2r_0^2)+P^{\rm shape}_n)/r_H.
\end{equation}
This allows us to apply the results of section \ref{sec:robinbc} to the calculation in section \ref{sec:ahshape}.

\subsubsection{Mass deformations}
For mass deformations, the cap-side second-order zero mode obeys
\begin{equation}
\left({d^2\over ds^2}-2\csch^2(s)\right)\widetilde d_n(s)
=
\csch^2(s)\,\widetilde g_n(s)^2,
\end{equation}
where
\begin{equation}
\widetilde g_n(s)=
\frac{2}{\lambda_{+,n}+\lambda_{-,n}}{\cal V}_n(s),
\qquad
{\cal V}_n(s) = 
e^{-|n|(s-A)}
{|n|+\coth s\over |n|+\coth A}.
\end{equation}
As before, solving the cap-side equation and imposing regularity as
$s\to\infty$ gives
\begin{equation}
\partial_s\widetilde d_n(A)
=
-\lambda_{+,0}d_n(\tau_0)
+
P^{\rm mass}_n,
\end{equation}
where
\begin{equation}\label{eq:Pn_mass}
P^{\rm mass}_n
=
-{1\over \coth A}
\int_A^\infty \dd s\,\coth s\,\csch^2s\,\widetilde g_n(s)^2 .
\end{equation}
For mass deformations, the shell is not displaced. Therefore the
second-order continuity and derivative junction conditions are simply
\begin{equation}
\widetilde d_n(A)=d_n(\tau_0),
\qquad
\partial_s\widetilde d_n(A)=d'_n(\tau_0).
\end{equation}
It follows that the second-order field on the black-hole side satisfies the
Robin boundary condition
\begin{equation}
\partial_y d_n(\tau_0)+\lambda_{+,0}d_n(\tau_0)
=
P^{\rm mass}_n .
\end{equation}
Comparing to the notation of section \ref{sec:robinbc}, we read off
\begin{equation}\label{eq:zetamass}
    \gamma = \lambda_{+,0}/r_H, \quad \zeta^{\rm mass}_n = P_n^{\rm mass}/r_H.
\end{equation}

\bibliographystyle{ourbst}
\bibliography{draft/ref.bib}

\providecommand{\href}[2]{#2}\begingroup\raggedright\begin{thebibliography}{10}

\bibitem{Anous:2016kss}
T.~Anous, T.~Hartman, A.~Rovai and J.~Sonner, {{Black Hole Collapse in the 1/c Expansion}}, \href{http://dx.doi.org/10.1007/JHEP07(2016)123}{JHEP {\bf 07}, 123, 2016}, [\href{http://arxiv.org/abs/arXiv:1603.04856}{{arXiv:1603.04856 [hep-th]}}].

\bibitem{Chandra:2022fwi}
J.~Chandra and T.~Hartman, {{Coarse graining pure states in AdS/CFT}}, \href{http://dx.doi.org/10.1007/JHEP10(2023)030}{JHEP {\bf 10}, 030, 2023}, [\href{http://arxiv.org/abs/arXiv:2206.03414}{{arXiv:2206.03414 [hep-th]}}].

\bibitem{Chandra:2024vhm}
J.~Chandra, T.~Hartman and V.~Meruliya, {{Statistics of three-dimensional black holes from Liouville line defects}}, \href{http://dx.doi.org/10.1007/JHEP11(2024)090}{JHEP {\bf 11}, 090, 2024}, [\href{http://arxiv.org/abs/arXiv:2404.15183}{{arXiv:2404.15183 [hep-th]}}].

\bibitem{Sasieta:2022ksu}
M.~Sasieta, {{Wormholes from heavy operator statistics in AdS/CFT}}, \href{http://dx.doi.org/10.1007/JHEP03(2023)158}{JHEP {\bf 03}, 158, 2023}, [\href{http://arxiv.org/abs/arXiv:2211.11794}{{arXiv:2211.11794 [hep-th]}}].

\bibitem{Bah:2022uyz}
I.~Bah, Y.~Chen and J.~Maldacena, {{Estimating global charge violating amplitudes from wormholes}}, \href{http://dx.doi.org/10.1007/JHEP04(2023)061}{JHEP {\bf 04}, 061, 2023}, [\href{http://arxiv.org/abs/arXiv:2212.08668}{{arXiv:2212.08668 [hep-th]}}].

\bibitem{ZamoRecursion}
A.~B. Zamolodchikov, {{Conformal symmetry two-dimensional space: Recursion representation of conformal block}}, \href{http://dx.doi.org/10.1007/BF01022967}{Theor. Math. Phys. {\bf 73}, 1088--1093, 1987}.

\bibitem{Verlinde:1989ua}
H.~L. Verlinde, {{Conformal Field Theory, 2-$D$ Quantum Gravity and Quantization of Teichmuller Space}}, \href{http://dx.doi.org/10.1016/0550-3213(90)90510-K}{Nucl. Phys. B {\bf 337}, 652--680, 1990}.

\bibitem{Witten:1988hc}
E.~Witten, {{(2+1)-Dimensional Gravity as an Exactly Soluble System}}, \href{http://dx.doi.org/10.1016/0550-3213(88)90143-5}{Nucl. Phys. B {\bf 311}, 46, 1988}.

\bibitem{Brown:1986nw}
J.~D. Brown and M.~Henneaux, {{Central Charges in the Canonical Realization of Asymptotic Symmetries: An Example from Three-Dimensional Gravity}}, \href{http://dx.doi.org/10.1007/BF01211590}{Commun. Math. Phys. {\bf 104}, 207--226, 1986}.

\bibitem{Hartman:2013mia}
T.~Hartman, {{Entanglement Entropy at Large Central Charge}},  2013, [\href{http://arxiv.org/abs/arXiv:1303.6955}{{arXiv:1303.6955 [hep-th]}}].

\bibitem{Chandra:2022bqq}
J.~Chandra, S.~Collier, T.~Hartman and A.~Maloney, {{Semiclassical 3D gravity as an average of large-c CFTs}}, \href{http://dx.doi.org/10.1007/JHEP12(2022)069}{JHEP {\bf 12}, 069, 2022}, [\href{http://arxiv.org/abs/arXiv:2203.06511}{{arXiv:2203.06511 [hep-th]}}].

\bibitem{Collier:2023fwi}
S.~Collier, L.~Eberhardt and M.~Zhang, {{Solving 3d gravity with Virasoro TQFT}}, \href{http://dx.doi.org/10.21468/SciPostPhys.15.4.151}{SciPost Phys. {\bf 15}, 151, 2023}, [\href{http://arxiv.org/abs/arXiv:2304.13650}{{arXiv:2304.13650 [hep-th]}}].

\bibitem{Billo:2016cpy}
M.~Billo, V.~Goncalves, E.~Lauria and M.~Meineri, {Defects in conformal field theory}, \href{http://dx.doi.org/10.1007/JHEP04(2016)091}{JHEP {\bf 04}, 091, 2016}, [\href{http://arxiv.org/abs/arXiv:1601.02883}{{arXiv:1601.02883 [hep-th]}}].

\bibitem{Cuomo:2021rkm}
G.~Cuomo, Z.~Komargodski and A.~Raviv-Moshe, {{Renormalization Group Flows on Line Defects}}, \href{http://dx.doi.org/10.1103/PhysRevLett.128.021603}{Phys. Rev. Lett. {\bf 128}, 021603, 2022}, [\href{http://arxiv.org/abs/arXiv:2108.01117}{{arXiv:2108.01117 [hep-th]}}].

\bibitem{Green1954}
M.~S. Green, {Markoff random processes and the statistical mechanics of time-dependent phenomena. {II}. irreversible processes in fluids}, \href{http://dx.doi.org/10.1063/1.1740082}{The Journal of Chemical Physics {\bf 22}, 398--413, 1954}.

\bibitem{Kubo:1957mj}
R.~Kubo, {Statistical-mechanical theory of irreversible processes. i. general theory and simple applications to magnetic and conduction problems}, \href{http://dx.doi.org/10.1143/JPSJ.12.570}{Journal of the Physical Society of Japan {\bf 12}, 570--586, 1957}.

\bibitem{Matsubara1955}
T.~Matsubara, {A new approach to quantum-statistical mechanics}, \href{http://dx.doi.org/10.1143/PTP.14.351}{Progress of Theoretical Physics {\bf 14}, 351--378, 1955}.

\bibitem{Martin:1959jp}
P.~C. Martin and J.~Schwinger, {Theory of many-particle systems. i}, \href{http://dx.doi.org/10.1103/PhysRev.115.1342}{Physical Review {\bf 115}, 1342--1373, 1959}.

\bibitem{KadanoffMartin1963}
L.~P. Kadanoff and P.~C. Martin, {Hydrodynamic equations and correlation functions}, \href{http://dx.doi.org/10.1016/0003-4916(63)90078-2}{Annals of Physics {\bf 24}, 419--469, 1963}.

\bibitem{Birmingham:2002ph}
D.~Birmingham, I.~Sachs and S.~N. Solodukhin, {Relaxation in conformal field theory, hawking--page transition, and quasinormal/normal modes}, \href{http://dx.doi.org/10.1103/PhysRevD.67.104026}{Phys. Rev. D {\bf 67}, 104026, 2003}, [\href{http://arxiv.org/abs/arXiv:hep-th/0212308}{{arXiv:hep-th/0212308}}].

\bibitem{Wang:2025eow}
D.~Wang, Z.~Wang and Z.~Wei, {Wormholes with ends of the world}, \href{http://dx.doi.org/10.1007/JHEP09(2025)166}{JHEP {\bf 2025}, 166, 2025}, [\href{http://arxiv.org/abs/arXiv:2504.12278}{{arXiv:2504.12278 [hep-th]}}].

\bibitem{Deutsch1991ETH}
J.~M. Deutsch, {Quantum statistical mechanics in a closed system}, \href{http://dx.doi.org/10.1103/PhysRevA.43.2046}{Phys. Rev. A {\bf 43}, 2046--2049, 1991}.

\bibitem{Srednicki1994ETH}
M.~Srednicki, {Chaos and quantum thermalization}, \href{http://dx.doi.org/10.1103/PhysRevE.50.888}{Phys. Rev. E {\bf 50}, 888--901, 1994}, [\href{http://arxiv.org/abs/arXiv:cond-mat/9403051}{{arXiv:cond-mat/9403051}}].

\bibitem{Belin:2020hea}
A.~Belin and J.~de~Boer, {{Random statistics of OPE coefficients and Euclidean wormholes}}, \href{http://dx.doi.org/10.1088/1361-6382/ac1082}{Class. Quant. Grav. {\bf 38}, 164001, 2021}, [\href{http://arxiv.org/abs/arXiv:2006.05499}{{arXiv:2006.05499 [hep-th]}}].

\bibitem{Belin:2021ryy}
A.~Belin, J.~de~Boer and D.~Liska, {{Non-Gaussianities in the statistical distribution of heavy OPE coefficients and wormholes}}, \href{http://dx.doi.org/10.1007/JHEP06(2022)116}{JHEP {\bf 06}, 116, 2022}, [\href{http://arxiv.org/abs/arXiv:2110.14649}{{arXiv:2110.14649 [hep-th]}}].

\bibitem{Chandra:2024bqz}
J.~Chandra, {{Euclidean wormholes in holographic RG flows}}, \href{http://dx.doi.org/10.1007/JHEP11(2024)096}{JHEP {\bf 11}, 096, 2024}, [\href{http://arxiv.org/abs/arXiv:2407.15630}{{arXiv:2407.15630 [hep-th]}}].

\bibitem{Chandra:2023rhx}
J.~Chandra, {{Euclidean wormholes for individual 2d CFTs}}, \href{http://dx.doi.org/10.1007/JHEP04(2024)051}{JHEP {\bf 04}, 051, 2024}, [\href{http://arxiv.org/abs/arXiv:2305.07183}{{arXiv:2305.07183 [hep-th]}}].

\bibitem{deBoer:2023vsm}
J.~de~Boer, D.~Liska, B.~Post and M.~Sasieta, {{A principle of maximum ignorance for semiclassical gravity}}, \href{http://dx.doi.org/10.1007/JHEP02(2024)003}{JHEP {\bf 2024}, 003, 2024}, [\href{http://arxiv.org/abs/arXiv:2311.08132}{{arXiv:2311.08132 [hep-th]}}].

\bibitem{Collier:2024mgv}
S.~Collier, L.~Eberhardt and M.~Zhang, {{3d gravity from Virasoro TQFT: Holography, wormholes and knots}},  2024, [\href{http://arxiv.org/abs/arXiv:2401.13900}{{arXiv:2401.13900 [hep-th]}}].

\bibitem{deBoer:2024znb}
J.~de~Boer, D.~Li{\v{s}}ka and B.~Post, {Multiboundary wormholes and {OPE} statistics}, \href{http://dx.doi.org/10.1007/JHEP10(2024)207}{JHEP {\bf 10}, 207, 2024}, [\href{http://arxiv.org/abs/arXiv:2405.13111}{{arXiv:2405.13111 [hep-th]}}].

\bibitem{Saad:2019pqd}
P.~Saad, {{Late Time Correlation Functions, Baby Universes, and ETH in JT Gravity}},  2019, [\href{http://arxiv.org/abs/arXiv:1910.10311}{{arXiv:1910.10311 [hep-th]}}].

\bibitem{Chandra:2025fef}
J.~Chandra, {{Statistics in 3d gravity from knots and links}}, \href{http://dx.doi.org/10.1007/JHEP12(2025)139}{JHEP {\bf 12}, 139, 2025}, [\href{http://arxiv.org/abs/arXiv:2508.10864}{{arXiv:2508.10864 [hep-th]}}].

\bibitem{Jafferis:2024random}
D.~L. Jafferis, L.~Rozenberg and G.~Wong, {{3d} gravity as a random ensemble}, \href{http://dx.doi.org/10.1007/JHEP02(2025)208}{JHEP {\bf 02}, 208, 2025}, [\href{http://arxiv.org/abs/arXiv:2407.02649}{{arXiv:2407.02649 [hep-th]}}].

\bibitem{Belin:2026pko}
A.~Belin, S.~Collier, L.~Eberhardt, D.~Liska and B.~Post, {{A universal sum over topologies in 3d gravity}},  2026, [\href{http://arxiv.org/abs/arXiv:2601.07906}{{arXiv:2601.07906 [hep-th]}}].

\bibitem{Goel:2018ubv}
A.~Goel, H.~T. Lam, G.~J. Turiaci and H.~Verlinde, {{Expanding the Black Hole Interior: Partially Entangled Thermal States in SYK}}, \href{http://dx.doi.org/10.1007/JHEP02(2019)156}{JHEP {\bf 02}, 156, 2019}, [\href{http://arxiv.org/abs/arXiv:1807.03916}{{arXiv:1807.03916 [hep-th]}}].

\bibitem{Bekenstein1973Entropy}
J.~D. Bekenstein, {Black holes and entropy}, \href{http://dx.doi.org/10.1103/PhysRevD.7.2333}{Phys. Rev. D {\bf 7}, 2333--2346, 1973}.

\bibitem{Hawking1975Radiation}
S.~W. Hawking, {Particle creation by black holes}, \href{http://dx.doi.org/10.1007/BF02345020}{Commun. Math. Phys. {\bf 43}, 199--220, 1975}. [Erratum: Commun. Math. Phys. 46, 206 (1976)].

\bibitem{RyuTakayanagi2006}
S.~Ryu and T.~Takayanagi, {Holographic derivation of entanglement entropy from the anti-de sitter space/conformal field theory correspondence}, \href{http://dx.doi.org/10.1103/PhysRevLett.96.181602}{Phys. Rev. Lett. {\bf 96}, 181602, 2006}, [\href{http://arxiv.org/abs/arXiv:hep-th/0603001}{{arXiv:hep-th/0603001}}].

\bibitem{HubenyRangamaniTakayanagi2007}
V.~E. Hubeny, M.~Rangamani and T.~Takayanagi, {A covariant holographic entanglement entropy proposal}, \href{http://dx.doi.org/10.1088/1126-6708/2007/07/062}{JHEP {\bf 07}, 062, 2007}, [\href{http://arxiv.org/abs/arXiv:0705.0016}{{arXiv:0705.0016 [hep-th]}}].

\bibitem{Penington:2019kki}
G.~Penington, S.~H. Shenker, D.~Stanford and Z.~Yang, {{Replica wormholes and the black hole interior}}, \href{http://dx.doi.org/10.1007/JHEP03(2022)205}{JHEP {\bf 03}, 205, 2022}, [\href{http://arxiv.org/abs/arXiv:1911.11977}{{arXiv:1911.11977 [hep-th]}}].

\bibitem{Balasubramanian:2022gmo}
V.~Balasubramanian, A.~Lawrence, J.~M. Magan and M.~Sasieta, {{Microscopic Origin of the Entropy of Black Holes in General Relativity}}, \href{http://dx.doi.org/10.1103/PhysRevX.14.011024}{Phys. Rev. X {\bf 14}, 011024, 2024}, [\href{http://arxiv.org/abs/arXiv:2212.02447}{{arXiv:2212.02447 [hep-th]}}].

\bibitem{Balasubramanian:2025zey}
V.~Balasubramanian and T.~Yildirim, {{The nonperturbative Hilbert space of quantum gravity with one boundary}}, \href{http://dx.doi.org/10.1007/JHEP03(2026)040}{JHEP {\bf 03}, 040, 2026}, [\href{http://arxiv.org/abs/arXiv:2506.04319}{{arXiv:2506.04319 [hep-th]}}].

\bibitem{Krasnov:2005dm}
K.~Krasnov and J.-M. Schlenker, {Minimal surfaces and particles in 3-manifolds}, \href{http://dx.doi.org/10.1007/s10711-007-9132-1}{Geometriae Dedicata {\bf 126}, 187--254, 2007}, [\href{http://arxiv.org/abs/arXiv:math/0511441}{{arXiv:math/0511441 [math.DG]}}].

\bibitem{Maldacena:2016upp}
J.~Maldacena, D.~Stanford and Z.~Yang, {{Conformal symmetry and its breaking in two dimensional Nearly Anti-de-Sitter space}}, \href{http://dx.doi.org/10.1093/ptep/ptw124}{PTEP {\bf 2016}, 12C104, 2016}, [\href{http://arxiv.org/abs/arXiv:1606.01857}{{arXiv:1606.01857 [hep-th]}}].

\bibitem{Stanford:2017thb}
D.~Stanford and E.~Witten, {{Fermionic Localization of the Schwarzian Theory}}, \href{http://dx.doi.org/10.1007/JHEP10(2017)008}{JHEP {\bf 10}, 008, 2017}, [\href{http://arxiv.org/abs/arXiv:1703.04612}{{arXiv:1703.04612 [hep-th]}}].

\bibitem{Schneider:2014convex}
R.~Schneider, \emph{Convex Bodies: The Brunn--Minkowski Theory}, vol.~151 of \emph{Encyclopedia of Mathematics and its Applications}.
\newblock Cambridge University Press, Cambridge, 2~ed., 2014, \href{http://dx.doi.org/10.1017/CBO9781139003858}{10.1017/CBO9781139003858}.

\bibitem{Bers:1960Simultaneous}
L.~Bers, {Simultaneous uniformization}, \href{http://dx.doi.org/10.1090/S0002-9904-1960-10413-2}{Bulletin of the American Mathematical Society {\bf 66}, 94--97, 1960}.

\bibitem{Bers:1960BoundedDomains}
L.~Bers, {Spaces of {Riemann} surfaces as bounded domains}, \href{http://dx.doi.org/10.1090/S0002-9904-1960-10415-6}{Bulletin of the American Mathematical Society {\bf 66}, 98--103, 1960}.

\bibitem{Witten:1988hf}
E.~Witten, {Coadjoint orbits of the {Virasoro} group}, \href{http://dx.doi.org/10.1007/BF01218287}{Communications in Mathematical Physics {\bf 114}, 1--53, 1988}.

\bibitem{Balog:1997zz}
J.~Balog, L.~Feh{\'e}r and L.~Palla, {Coadjoint orbits of the {Virasoro} algebra and the global {Liouville} equation}, \href{http://dx.doi.org/10.1142/S0217751X98000147}{International Journal of Modern Physics A {\bf 13}, 315--362, 1998}, [\href{http://arxiv.org/abs/arXiv:hep-th/9703045}{{arXiv:hep-th/9703045 [hep-th]}}].

\bibitem{Alekseev:1989ce}
A.~Alekseev and S.~L. Shatashvili, {Path integral quantization of the coadjoint orbits of the {Virasoro} group and 2d gravity}, \href{http://dx.doi.org/10.1016/0550-3213(89)90130-2}{Nuclear Physics B {\bf 323}, 719--733, 1989}.

\bibitem{Gao:2016bin}
P.~Gao, D.~L. Jafferis and A.~C. Wall, {Traversable wormholes via a double trace deformation}, \href{http://dx.doi.org/10.1007/JHEP12(2017)151}{JHEP {\bf 12}, 151, 2017}, [\href{http://arxiv.org/abs/arXiv:1608.05687}{{arXiv:1608.05687 [hep-th]}}].

\bibitem{Maldacena:2018lmt}
J.~Maldacena and X.-L. Qi, {Eternal traversable wormhole},  2018, [\href{http://arxiv.org/abs/arXiv:1804.00491}{{arXiv:1804.00491 [hep-th]}}].

\bibitem{Radnell:2005sewing}
D.~Radnell and E.~Schippers, {Quasisymmetric sewing in rigged {Teichm{\"u}ller} space}, \href{http://dx.doi.org/10.1142/S0219199706002210}{Communications in Contemporary Mathematics {\bf 8}, 481--534, 2006}, [\href{http://arxiv.org/abs/arXiv:math-ph/0507031}{{arXiv:math-ph/0507031 [math-ph]}}].

\bibitem{MaibachPeltola:2026complex}
S.~Maibach and E.~Peltola, {Complex deformations of the circle: Group cohomology and {Virasoro} uniformization},  2026, [\href{http://arxiv.org/abs/arXiv:2605.20175}{{arXiv:2605.20175 [math-ph]}}].

\bibitem{Srednicki_1994}
M.~Srednicki, {Chaos and quantum thermalization}, \href{http://dx.doi.org/10.1103/physreve.50.888}{Physical Review E {\bf 50}, 888--901, 1994}.

\bibitem{PhysRevA.43.2046}
J.~M. Deutsch, {Quantum statistical mechanics in a closed system}, \href{http://dx.doi.org/10.1103/PhysRevA.43.2046}{Phys. Rev. A {\bf 43}, 2046--2049, 1991}.

\bibitem{DAlessio:2015qtq}
L.~D'Alessio, Y.~Kafri, A.~Polkovnikov and M.~Rigol, {{From quantum chaos and eigenstate thermalization to statistical mechanics and thermodynamics}}, \href{http://dx.doi.org/10.1080/00018732.2016.1198134}{Adv. Phys. {\bf 65}, 239--362, 2016}, [\href{http://arxiv.org/abs/arXiv:1509.06411}{{arXiv:1509.06411 [cond-mat.stat-mech]}}].

\bibitem{Kondo:1964}
J.~Kondo, {Resistance minimum in dilute magnetic alloys}, \href{http://dx.doi.org/10.1143/PTP.32.37}{Progress of Theoretical Physics {\bf 32}, 37--49, 1964}.

\bibitem{Erdmenger:2013dpa}
J.~Erdmenger, C.~Hoyos, A.~O'Bannon and J.~Wu, {A holographic model of the kondo effect}, \href{http://dx.doi.org/10.1007/JHEP12(2013)086}{JHEP {\bf 12}, 086, 2013}, [\href{http://arxiv.org/abs/arXiv:1310.3271}{{arXiv:1310.3271 [hep-th]}}].

\bibitem{Erdmenger:2016msd}
J.~Erdmenger, M.~Flory, M.-N. Newrzella, M.~Strydom and J.~M.~S. Wu, {Quantum quenches in a holographic kondo model}, \href{http://dx.doi.org/10.1007/JHEP04(2017)045}{JHEP {\bf 04}, 045, 2017}, [\href{http://arxiv.org/abs/arXiv:1612.06860}{{arXiv:1612.06860 [hep-th]}}].

\bibitem{Drukker:2006xg}
N.~Drukker and S.~Kawamoto, {Small deformations of supersymmetric wilson loops and open spin-chains}, \href{http://dx.doi.org/10.1088/1126-6708/2006/07/024}{JHEP {\bf 07}, 024, 2006}, [\href{http://arxiv.org/abs/arXiv:hep-th/0604124}{{arXiv:hep-th/0604124}}].

\bibitem{Cooke:2017qgm}
M.~Cooke, A.~Dekel and N.~Drukker, {The wilson loop cft: insertion dimensions and structure constants from wavy lines}, {J. Phys. A {\bf 50}, 335401, 2017}, [\href{http://arxiv.org/abs/arXiv:1703.03812}{{arXiv:1703.03812 [hep-th]}}].

\bibitem{Giombi:2018qox}
S.~Giombi and S.~Komatsu, {Exact correlators on the wilson loop in \(\mathcal{N}=4\) sym: Localization, defect cft, and integrability}, \href{http://dx.doi.org/10.1007/JHEP05(2018)109}{JHEP {\bf 05}, 109, 2018}, [\href{http://arxiv.org/abs/arXiv:1802.05201}{{arXiv:1802.05201 [hep-th]}}].

\bibitem{Liendo:2018ukf}
P.~Liendo, C.~Meneghelli and V.~Mitev, {Bootstrapping the half-bps line defect}, \href{http://dx.doi.org/10.1007/JHEP10(2018)077}{JHEP {\bf 10}, 077, 2018}, [\href{http://arxiv.org/abs/arXiv:1806.01862}{{arXiv:1806.01862 [hep-th]}}].

\bibitem{Giombi:2017cqn}
S.~Giombi, R.~Roiban and A.~A. Tseytlin, {Half-bps wilson loop and \(ads_2/cft_1\)}, {Nucl. Phys. B {\bf 922}, 499--527, 2017}, [\href{http://arxiv.org/abs/arXiv:1706.00756}{{arXiv:1706.00756 [hep-th]}}].

\bibitem{Correa:2012at}
D.~Correa, J.~Henn, J.~Maldacena and A.~Sever, {An exact formula for the radiation of a moving quark in \(\mathcal{N}=4\) super yang mills}, \href{http://dx.doi.org/10.1007/JHEP06(2012)048}{JHEP {\bf 06}, 048, 2012}, [\href{http://arxiv.org/abs/arXiv:1202.4455}{{arXiv:1202.4455 [hep-th]}}].

\bibitem{Bianchi:2018zpb}
L.~Bianchi, M.~Lemos and M.~Meineri, {Line defects and radiation in \(\mathcal{N}=2\) conformal theories}, \href{http://dx.doi.org/10.1103/PhysRevLett.121.141601}{Phys. Rev. Lett. {\bf 121}, 141601, 2018}, [\href{http://arxiv.org/abs/arXiv:1805.04111}{{arXiv:1805.04111 [hep-th]}}].

\bibitem{Barrat:2024vwu}
J.~Barrat, B.~Fiol, E.~Marchetto, A.~Miscioscia and E.~Pomoni, {Conformal line defects at finite temperature}, {SciPost Phys. {\bf 18}, 018, 2025}, [\href{http://arxiv.org/abs/arXiv:2407.14600}{{arXiv:2407.14600 [hep-th]}}].

\bibitem{Giombi:2026line}
S.~Giombi, Y.-Z. Li and J.~Shan, {Bouncing singularities and thermal correlators on line defects},  2026, [\href{http://arxiv.org/abs/arXiv:2603.11012}{{arXiv:2603.11012 [hep-th]}}].

\bibitem{Antonini:2023hdh}
S.~Antonini, M.~Sasieta and B.~Swingle, {{Cosmology from random entanglement}}, \href{http://dx.doi.org/10.1007/JHEP11(2023)188}{JHEP {\bf 11}, 188, 2023}, [\href{http://arxiv.org/abs/arXiv:2307.14416}{{arXiv:2307.14416 [hep-th]}}].

\bibitem{Antonini:2024mci}
S.~Antonini and P.~Rath, {{Do holographic CFT states have unique semiclassical bulk duals?}}, \href{http://dx.doi.org/10.1142/S0218271825440250}{Int. J. Mod. Phys. D {\bf 34}, 2544025, 2025}, [\href{http://arxiv.org/abs/arXiv:2408.02720}{{arXiv:2408.02720 [hep-th]}}].

\bibitem{Antonini:2025ioh}
S.~Antonini, P.~Rath, M.~Sasieta, B.~Swingle and A.~Vilar~L{\'o}pez, {{The baby universe is fine and the CFT knows it: on holography for closed universes}}, \href{http://dx.doi.org/10.1007/JHEP12(2025)159}{JHEP {\bf 12}, 159, 2025}, [\href{http://arxiv.org/abs/arXiv:2507.10649}{{arXiv:2507.10649 [hep-th]}}].

\bibitem{Kudler-Flam:2025cki}
J.~Kudler-Flam and E.~Witten, {{Emergent mixed states for baby universes and black holes}}, \href{http://dx.doi.org/10.1007/JHEP05(2026)090}{JHEP {\bf 05}, 090, 2026}, [\href{http://arxiv.org/abs/arXiv:2510.06376}{{arXiv:2510.06376 [hep-th]}}].

\bibitem{Liu:2025ikq}
H.~Liu, {{''Filtering'' CFTs at large N: Euclidean Wormholes, Closed Universes, and Black Hole Interiors}},  2025, [\href{http://arxiv.org/abs/arXiv:2512.13807}{{arXiv:2512.13807 [hep-th]}}].

\bibitem{Iliesiu:2025ias}
L.~V. Iliesiu, ``On the hilbert space of quantum gravity from the gravitational path integral.'' Talk at the workshop ``Quantum Aspects of Black Holes and Spacetime,'' Institute for Advanced Study, 2025.

\bibitem{DiUbaldo:2026rly}
G.~Di~Ubaldo, L.~V. Iliesiu, H.~W. Lin and C.~Yan, {{Positivity of the gravitational path integral implies the axionic weak gravity conjecture}},  2026, [\href{http://arxiv.org/abs/arXiv:2605.05305}{{arXiv:2605.05305 [hep-th]}}].

\bibitem{Meineri:2019ycm}
M.~Meineri, J.~Penedones and A.~Rousset, {{Colliders and conformal interfaces}}, \href{http://dx.doi.org/10.1007/JHEP02(2020)138}{JHEP {\bf 02}, 138, 2020}, [\href{http://arxiv.org/abs/arXiv:1904.10974}{{arXiv:1904.10974 [hep-th]}}].

\bibitem{Bachas:2024hcb}
C.~Bachas and Z.~Chen, {Invariant tensions from holography}, \href{http://dx.doi.org/10.1007/JHEP08(2024)028}{JHEP {\bf 08}, 028, 2024}, [\href{http://arxiv.org/abs/arXiv:2404.14998}{{arXiv:2404.14998 [hep-th]}}].

\bibitem{Bachas:2020yxv}
C.~Bachas, S.~Chapman, D.~Ge and G.~Policastro, {Energy reflection and transmission at {$2D$} holographic interfaces}, \href{http://dx.doi.org/10.1103/PhysRevLett.125.231602}{Physical Review Letters {\bf 125}, 231602, 2020}, [\href{http://arxiv.org/abs/arXiv:2006.11333}{{arXiv:2006.11333 [hep-th]}}].

\bibitem{Bachas:2021fqo}
C.~Bachas and V.~Papadopoulos, {Phases of holographic interfaces}, \href{http://dx.doi.org/10.1007/JHEP04(2021)262}{JHEP {\bf 04}, 262, 2021}, [\href{http://arxiv.org/abs/arXiv:2101.12529}{{arXiv:2101.12529 [hep-th]}}].

\bibitem{Bachas:2022cba}
C.~Bachas, S.~Baiguera, S.~Chapman, G.~Policastro and T.~Schwartzman, {Energy transport for thick holographic branes}, \href{http://dx.doi.org/10.1103/PhysRevLett.131.021601}{Physical Review Letters {\bf 131}, 021601, 2023}, [\href{http://arxiv.org/abs/arXiv:2212.14058}{{arXiv:2212.14058 [hep-th]}}].

\bibitem{Bak:2011ga}
D.~Bak, M.~Gutperle and R.~A. Janik, {Janus black holes}, \href{http://dx.doi.org/10.1007/JHEP10(2011)056}{JHEP {\bf 10}, 056, 2011}, [\href{http://arxiv.org/abs/arXiv:1109.2736}{{arXiv:1109.2736 [hep-th]}}].

\bibitem{Abdalla:2026wdx}
A.~I. Abdalla, J.~Chandra and Y.~Wang, {{Line Defects in Liouville Conformal Field Theory: Localized Cosmological Constants and Decohered Hyperbolic Geometries}},  2026, [\href{http://arxiv.org/abs/arXiv:2603.02166}{{arXiv:2603.02166 [hep-th]}}].

\bibitem{Cuomo:2021kfm}
G.~Cuomo, Z.~Komargodski and M.~Mezei, {{Localized magnetic field in the O(N) model}}, \href{http://dx.doi.org/10.1007/JHEP02(2022)134}{JHEP {\bf 02}, 134, 2022}, [\href{http://arxiv.org/abs/arXiv:2112.10634}{{arXiv:2112.10634 [hep-th]}}].

\bibitem{Zhu:2022gjc}
W.~Zhu, C.~Han, E.~Huffman, J.~S. Hofmann and Y.-C. He, {Uncovering conformal symmetry in the {$3D$} {Ising} transition: State-operator correspondence from a fuzzy sphere regularization}, \href{http://dx.doi.org/10.1103/PhysRevX.13.021009}{Phys. Rev. X {\bf 13}, 021009, 2023}, [\href{http://arxiv.org/abs/arXiv:2210.13482}{{arXiv:2210.13482 [cond-mat.stat-mech]}}].

\bibitem{Hu:2023ghk}
L.~Hu, Y.-C. He and W.~Zhu, {Solving conformal defects in {$3D$} conformal field theory using fuzzy sphere regularization}, \href{http://dx.doi.org/10.1038/s41467-024-47978-y}{Nature Communications {\bf 15}, 3659, 2024}, [\href{http://arxiv.org/abs/arXiv:2308.01903}{{arXiv:2308.01903 [cond-mat.stat-mech]}}].

\bibitem{Zhou:2024fgi}
Z.~Zhou, D.~Gaiotto, Y.-C. He and Y.~Zou, {The {$g$}-function and defect changing operators from wavefunction overlap on a fuzzy sphere}, \href{http://dx.doi.org/10.21468/SciPostPhys.17.1.021}{SciPost Phys. {\bf 17}, 021, 2024}, [\href{http://arxiv.org/abs/arXiv:2401.00039}{{arXiv:2401.00039 [hep-th]}}].

\end{thebibliography}\endgroup

\end{document}